\documentclass[fleqn,usenatbib]{mnras}
\usepackage{newtxtext,newtxmath,lscape}
\usepackage[T1]{fontenc}
\usepackage{lineno}

\DeclareRobustCommand{\VAN}[3]{#2}
\let\VANthebibliography\thebibliography
\def\thebibliography{\DeclareRobustCommand{\VAN}[3]{##3}\VANthebibliography}

\usepackage{graphicx}	
\usepackage{amsmath}	

\newcommand{\oiii}{O\,{\sc iii}} 
\newcommand{\oi}{O\,{\sc i}} 
\newcommand{\feii}{Fe\,{\sc ii}} 
\newcommand{\civ}{C\,{\sc iv}} 
\newcommand{\heii}{He\,{\sc ii}} 
\newcommand{\hei}{He\,{\sc i}}
\newcommand{\mgii}{Mg\,{\sc ii}}
\newcommand{\ciii}{C\,{\sc iii}}

\title[Long-term Reverberation Mapping of NGC 4151]{\bf Broad-line region in NGC 4151 monitored by two decades of reverberation mapping campaigns. I. Evolution of structure and kinematics}
	
\author[Chen et al.]
{Yong-Jie Chen$^{1,2}$, 
Dong-Wei Bao$^{1,2}$,
Shuo Zhai$^{1,2}$,
Feng-Na Fang$^{1,2}$,
Chen Hu$^{1}$\thanks{E-mail: huc@ihep.ac.cn},
Pu Du$^{1}$,
Sen Yang$^{1,2}$,
\newauthor
Zhu-Heng Yao$^{1,2}$,
Yan-Rong Li$^{1}$\thanks{E-mail: liyanrong@mail.ihep.ac.cn},	
Michael S. Brotherton$^{3}$,
Jacob N. McLane$^{3}$,
T.\,E. Zastrocky$^{3}$,
\newauthor
Kianna A. Olson$^{3}$,
Edi Bon$^{4}$,
Hua-Rui Bai$^{1,2}$,
Yi-Xin Fu$^{1,2}$,
Jun-Rong Liu$^{1,2}$,
Yi-Lin Wang$^{1,2}$,
Jaya Maithil$^{3}$,
\newauthor
H.\,A. Kobulnicky$^{3}$,
D.\,A. Dale$^{3}$,
C. Adelman$^{3, 5}$,
M.\,J. Caradonna$^{3, 6}$,
Z. Carter$^{3, 7}$,
J. Favro$^{3, 8}$,
A.\,J. Ferguson$^{3}$,
\newauthor
I.\,M. Gonzalez$^{3, 9}$,
L.\,M. Hadding$^{3, 10}$,
H.\,D. Hagler$^{3, 11}$,
G. Murphree$^{3, 12}$,
M. Oeur$^{3, 13}$,
C.\,J. Rogers$^{3, 14}$,
\newauthor
T. Roth$^{3, 15}$,
S. Schonsberg$^{3, 16}$,
T.\,R. Stack$^{3, 17}$,
Jian-Min Wang$^{1,18,19}$\thanks{E-mail: wangjm@mail.ihep.ac.cn}
\\
$^{1}$Key Laboratory for Particle Astrophysics, Institute of High 
Energy Physics, Chinese Academy of Sciences, 19B Yuquan Road, Beijing 100049, China\\
$^{2}$School of Physical Science, University of Chinese Academy of Sciences, 19A Yuquan Road, Beijing 100049, People’s Republic of China \\
$^{3}$Department of Physics and Astronomy, University of Wyoming, Laramie, WY 82071, USA\\
$^{4}$Astronomical Observatory Belgrade, Volgina 7, 11060 Belgrade, Serbia\\
$^{5}$Department of Physics \& Astronomy, Cal Poly Pomona, Pomona, CA 91768, USA\\
$^{6}$Department of Physics, University of North Texas, Denton, Texas 76203, USA\\
$^{7}$Department of Physics and Astronomy, Trinity University, San Antonio, TX 78212, USA\\
$^{8}$Department of Astronomy, Diablo Valley College, Pleasant Hill, CA 94523, USA\\
$^{9}$Department of Physics, Oklahoma Baptist University, Shawnee, OK 74804, USA\\
$^{10}$Department of Astronomy, Grinnell College, Grinnell, IA 50112, USA\\
$^{11}$Department of Astronomy, Whitman College, Walla Walla, WA 99362, USA\\
$^{12}$Department of Physics, Rhodes College, Memphis, TN 38112, USA\\
$^{13}$Department of Physics and Astronomy, State Long Beach, Long Beach, CA 90840, USA\\
$^{14}$Department of Physics and Astronomy, Cal Poly Humboldt, Arcata, CA 95521, USA\\
$^{15}$Department of Physics \& Astronomy, California State University, Sacramento, CA 95747, USA\\
$^{16}$Department of Physics and Astronomy, University of Montana, Missoula, MT 59812, USA\\
$^{17}$Department of Astronomy, Illinois Institute of Technology, Chicago, IL 60616, USA\\
$^{18}$National Astronomical Observatories of China, Chinese Academy of Sciences, A20 Datun Road, Beijing 100012, China\\
$^{19}$School of Astronomy and Space Sciences, University of Chinese Academy of Sciences, 19A Yuquan Road, Beijing 100049, People’s Republic of China
}
  
\begin{document}	
    \label{firstpage}
    \pagerange{\pageref{firstpage}--\pageref{lastpage}}
    \maketitle
    \begin{abstract}
		We report the results of long-term reverberation mapping (RM) campaigns of the nearby active galactic nuclei (AGN) NGC 4151, spanning from 1994 to 2022, based on archived observations of the FAST Spectrograph Publicly Archived Programs and our new observations with the 2.3m telescope at the Wyoming Infrared Observatory. We reduce and calibrate all the spectra in a consistent way,
		and derive light curves of the broad H$\beta$ line and 5100\,{\AA} continuum. Continuum light curves are also constructed using public archival photometric data to increase sampling cadences. We subtract the host galaxy contamination using {\it HST} imaging to correct fluxes of the calibrated light curves. Utilizing the long-term archival photometric data, we complete the absolute flux-calibration of the AGN continuum. We find that the H$\beta$ time delays are correlated with the 5100\,{\AA} luminosities as $\tau_{\rm H\beta}\propto L_{5100}^{0.46\pm0.16}$. This is remarkably consistent with Bentz et al. (2013)'s global size-luminosity relationship of AGNs. Moreover, the data sets for five of the seasons allow us to obtain the velocity-resolved delays of the H$\beta$ line, showing diverse structures (outflows, inflows and disks). 
		Combining our results with previous independent measurements, we find the measured dynamics of the H$\beta$ broad-line region (BLR) are possibly related to the long-term trend of the luminosity. There is also a possible additional $\sim$1.86 years time lag between the variation in BLR radius and luminosity. These results suggest that dynamical changes in the BLR may be driven by the effects of radiation pressure. 
	\end{abstract}


        \section{introduction} \label{sec:intro}
        Over the past four decades, the technique of reverberation mapping (\citealt{ Blandford1982, Peterson1993}) has been well established as a powerful tool for measuring the masses of supermassive black holes (SMBHs) and investigating the geometry and dynamics of broad line regions (BLRs) in AGNs. There are now more than 100 AGNs with RM observations from different campaigns, e.g., \cite{peterson2004,bentz2006,bentz2009lick,bentz2010,Denny2009,Denney2010,dupu2014,dupu2015,dupu2016,dupu2018,rosa2018,huchen2021,lu2021,lishasha2021} and \cite{Bao2022}. These efforts led to the discovery of the relationship between the characteristic radius of the BLR and optical 5100\,{\AA} luminosity (hereafter $R_{\rm H\beta}-L_{5100}$\footnote{For convince, we have marked $\lambda L_{\lambda, \rm AGN}(5100\,\text{\AA})$ as $L_{5100}$.} relationship; \citealt{Kaspi2000,bentz2009,bentz2013}), which provides an economical way to estimate black hole (BH) masses from single-epoch spectra. However, recent RM campaigns discovered that the measurement of AGNs with high accretion rate show significant deviation from the traditional $R_{\rm H\beta}-L_{5100}$ relationship by \cite{dupu2015,Du2016a,Du2018b} and \cite{lishasha2021}. Therefore, \citet{Du2019} established a new scaling relationship including the flux ratio between \feii~ and H$\beta$ emission lines (i.e., $\mathcal{R}_{\rm Fe}=F_{\rm Fe}/F_{\rm H\beta}$).
	
	Furthermore, an intrinsic $R_{\rm H\beta}-L_{5100}$ relationship can be investigated for individual AGNs using multiple observing seasons that cover a sufficiently broad luminosity range. Such a relationship provides information about the scatter of the global relationship.
	NGC 5548 is the first AGN studied in such a way, with 23 RM campaigns (e.g., \citealt{peterson2002,bentz2007,Denney2009,lu2016,pei2017,rosa2018,Lu2022}). The resulting relationship has a power law index of ${0.57\pm 0.3}$, which is in agreement with the value of $0.53^{+0.04}_{-0.03}$ for the ensemble RM sample obtained by \cite{bentz2013} (see also \citealt{Lu2022}). Nevertheless, the $R_{\rm H\beta}-L_{5100}$ relationship for NGC 5548 has a larger intrinsic scatter of 0.26 dex.
	In particular, the campaign conducted by \cite{pei2017} found that the H$\beta$ time lag was significantly shorter than expected (see also Figure 13 in~\citealt{Lu2022}). 
	
	\begin{figure*}
		\centering
		\includegraphics[width=0.9\textwidth]{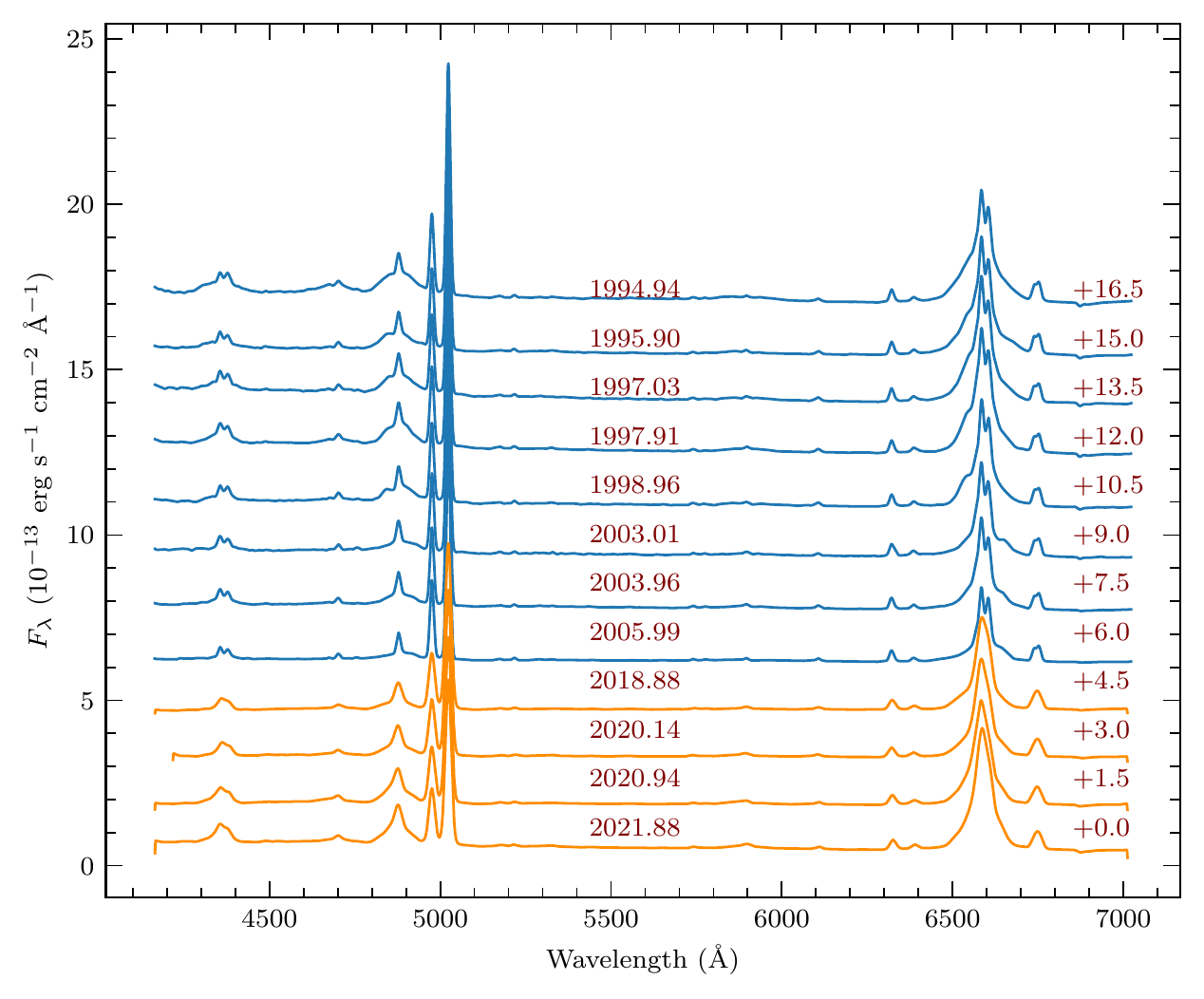}
		\caption{Some examples of flux-calibrated spectra in the observed frame. Blue: the spectra from the sixth project in FAST (AGN Watch). Orange: the spectra from MAHA. The spectra are shifted for clarity.} 
		\label{figure:spectra_example} 
	\end{figure*}
	
	At a distance of $\sim$15.8 Mpc (e.g., \citealt{yuan2020}), NGC 4151 is also one of the most intensively studied nearby AGNs across the electromagnetic spectrum (e.g., \citealt{Crenshaw1996, Edelson1996, kaspi1996, Warwick1996, Kraemer2005, Wang2011, Bon2012, Williams2017, Lyu2021}). 
	NGC 4151 is bright, highly variable and has been previously monitored by a number of RM campaigns (\citealt{clavel1990,ulrich1996,maoz1991,kaspi1996,bentz2006,rosa2018}). Meanwhile, due to the proximity of NGC 4151, the bulk kinematics of stars (\citealt{onken2007,Onken2014,roberts2021}) and gas (\citealt{hicks2008}) in the nucleus of the host galaxy can be spatially resolved and thereby the mass of the central SMBH can be directly constrained. This provides a direct way to verify the BH mass measurement by the RM method, which is dependent on the $f$-factor determined by the $M_\bullet-\sigma_\ast$ relation (e.g., \citealt{onken2004, graham2011, park2012, grier2013, woo2013, ho2014, woo2015, batiste2017, yu2019}), where $M_\bullet$ is the BH mass and $\sigma_{\ast}$ is the velocity dispersion of the bulge or spheroid. 
	
	The spectra of NGC 4151 show the asymmetric broad H$\beta$ profile, strongly varying with time (\citealt{Sergeev2001,Shapovalova2010}).  
	Since November 2018, the Monitoring AGNs with H$\beta$ Asymmetry (MAHA, hereafter) program (\citealt{dupu2018, Brotherton2020}) has been spectroscopically monitoring NGC 4151 in the optical and has obtained a wealth of high-quality spectra for four observing seasons.
	In addition, the AGN Watch Project from the FAST Spectrograph Publicly Archive Programs provides a database for long-term spectroscopic monitoring of a sample of bright AGNs, among which NGC 4151 was observed most and has a thousand optical spectra over 17 seasons from 1994 to 2011. These spectroscopic databases allow us to study the reverberation properties of the H$\beta$ broad emission line, geometry and kinematic of the H$\beta$ BLR and investigate the intrinsic $R_{\rm H\beta}-L_{5100}$ relationship for NGC 4151. As the first paper in a series, this work reports our data compilation and reduction and the results from basic analysis.
	
	The paper is organized as follows. In Section \ref{sec2}, we describe our data collection and reduction, flux calibration, light curve measurements and starlight correction. In Section \ref{analysis}, we perform data analysis and present the measured time lags, line width measurement, velocity-resolved delays, virial testing and BH mass. In Section \ref{discussions}, we discuss the $R_{\rm H\beta}-L_{5100}$ relationship, the evolution of BLR kinematics and the relationship of BLR radius versus luminosity. Conclusions are summarized in Section \ref{sec:summary}. 
	We adopt the redshift $z=0.00332$ and the luminosity distance 15.8 Mpc~\citep{yuan2020}.

	\section{Data Compilation and Reduction}
	\label{sec2}
	\subsection{The FAST Spectrograph Publicly Archived Program: AGN Watch}
	Table~\ref{table:info} summarizes the sampling information of optical spectroscopic data for NGC 4151 we analyze. The public archive spectra mainly come from the AGN Watch project (PI: Belinda Wilkes), which is the sixth project of the FAST Spectrograph Publicly Archive Programs\footnote{\url{http://tdc-www.harvard.edu/instruments/fast/progs/}} (FAST hereafter). These data can be accessed through the Smithsonian Astrophysical Observatory Telescope Data Center. The AGN Watch project is a long-running program aiming at developing a database for the reverberation mapping of bright AGNs. The project provided spectral data for a total of 17 years from February 1994 to March 2011 for more than 10 AGNs. Moreover, the Redshift Survey (PI: John Huchra) and AGN Emission Line Regions project (PI: Hermine Landt) of FAST also targeted NGC 4151 and provided several spectroscopic data. In total, there are 1027 spectra over 621 nights for NGC 4151 in FAST.
	
	All spectroscopic observations were made with the FAST spectrograph \citep{Fabricant1998} on the 1.5 m Tillinghast telescope at the Fred L. Whipple Observatory (FLWO), using a 300 line/mm grating and $3\arcsec$ wide long slit. The spectral coverage is 3400 to 7200\,{\AA} with a dispersion of $\sim$1.47~{\AA} pixel$^{-1}$. Most of the spectroscopic data have an exposure time of 30s, yielding an averaged signal-to-noise ratio (SNR) of $\sim$29. The spectrophotometric standard stars were observed each night and can be retrieved from the Spectrophotometric Standards project in the public archive of FAST. These data were processed by the FAST data reduction pipeline \citep{Tokarz1997}. However, the pipeline does not include the procedures of flux calibration because most of the FAST spectra suffered a wavelength-dependent flux loss (see \citealt{Mink2020} for the details).
	Therefore, we just use flux standards to correct the spectral shapes based on the pipeline preprocessed data. The further procedures of flux-calibration are conducted with Section~\ref{sec:calibration} and Section~\ref{sec:flux_correction}.
	
	We check these spectra by using flux ratios of [\oiii] doublet ($R( \text{[\oiii]})=F({\text{[\oiii]}\lambda5007})/F({\text{[\oiii]}\lambda4959})$). We integrate the fluxes of [\oiii]$\lambda$4959,5007 separately by subtracting a local continuum underneath determined by two nearby background windows. For [\oiii]$\lambda$4959 line, we choose 4950-4950~{\AA}, 4995-5000~{\AA} and 4950-4995~{\AA} as the blue and red background, and the line integration windows, respectively. For [\oiii]$\lambda$5007 line, the three windows separately are 4995-5000~{\AA}, 5045-5055~{\AA} and 5000-5045~{\AA}. We tabulate $R(\text{[\oiii]})$ in Appendix~\ref{appendixa}. The average of flux ratios $R(\text{[\oiii]})\sim$3.05 in our measurement is consistent with 3.01 for normal AGNs~\citep{Storey2000}.
	However, there are some spectra with $R( \text{[\oiii]})$ being much lower than 3 (about fifteen spectra). According to the information recorded in the headers of fits files and the observation logs, the problematic spectra usually have longer exposure time than others, which may cause the [\oiii]$\lambda$5007 flux saturation. We, therefore, discard them in our following analysis.
	
	For NGC 4151, the narrow line region and host galaxy are extensive with non-homogeneous surface brightness. A uniform aperture of spectral extraction and position angle of the slit is needed during monitoring periods of RM campaigns. This is important for the accuracy of the spectral flux-calibration ([\oiii]-based method, Section~\ref{sec:calibration}) and obtaining light variations of the AGN continuum (Section~\ref{sec:Spectroscopic_Lightcurves}). However, after checking observational and spectra reducing information recorded in the header of fits files, we find that the spectroscopic data were obtained by using different slit position angles with 3{\arcsec} wide long slit. The different extracted apertures also were applied in the process of data reduction for these spectra. We summarize them in Appendix ~\ref{appendixa}. We present a simple test for the light loss of [\oiii] emission when using different extracted apertures and position angles in Appendix~\ref{appendixb}. Fortunately, there is no significant light loss for measurements of [\oiii] fluxes in our simple test. Most of them are less than 5 percent. This indicates the accuracy of spectral flux-calibration utilizing the [\oiii]-based technique (Section~\ref{sec:calibration}).
	\label{sec:fast}

 	\subsection{MAHA Program}
	\label{sec:maha}
	The details of the MAHA program have been described by \citet{dupu2018}. In this section, we briefly introduce the observations for NGC 4151. The spectroscopic data were obtained by using the 2.3m telescope at the Wyoming Infrared Observatory (WIRO) with the long slit spectrograph in remote mode~\citep{findlay2016}. We used the 900 line/mm grating and $5\arcsec$ wide long slit. The wavelength coverage of spectra was $\rm 4000-7000\,\AA$ with a dispersion of 1.49\,{\AA} pixel$^{-1}$. In each night, we took CuAr lamp exposures to calibrate the wavelength and observed the spectrophotometric standard stars (Feige 34, G191B2B, HZ44, or BD+28d4211) for the correction of the spectral shape. We reduced the spectra with the standard \texttt{IRAF v2.16} procedures and extracted them using a uniform aperture of $\pm$6\arcsec.18.

        \begin{figure*}
            \centering
            \includegraphics[width=0.9\textwidth]{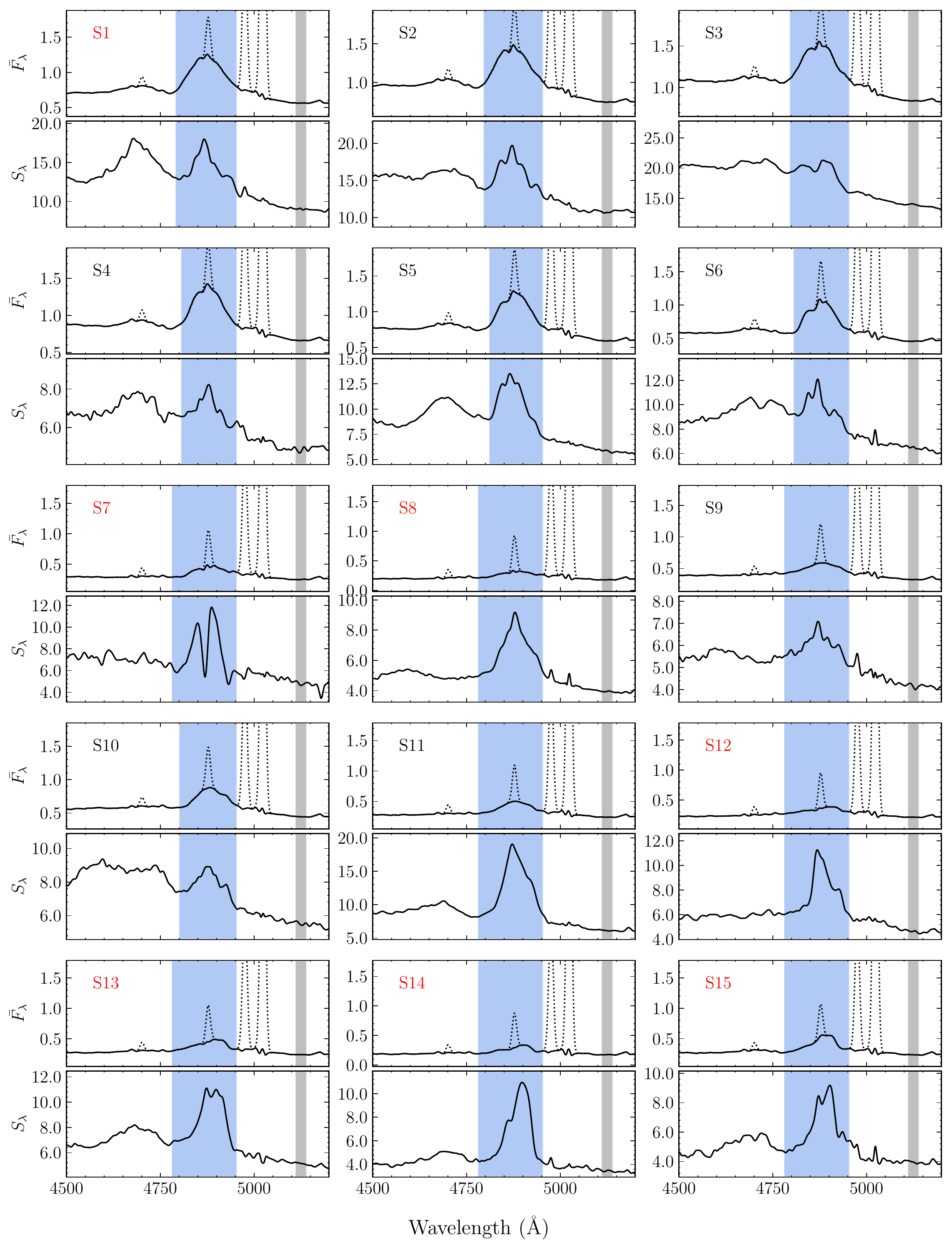}
            \caption{Mean and rms spectra in the observed frame for each season. The top panels are the mean spectra with the rms spectra shown underneath. The units of the mean and rms spectra are $\rm 10^{-13}~erg\ s^{-1}\ cm^{-1}\ \text{\AA}^{-1}$ and $10^{-15}~\rm erg\ s^{-1}\ cm^{-1}\ \text{\AA}^{-1} $, respectively. The solid lines are the rms and the narrow-line-subtracted mean spectra. The dotted lines are the narrow H$\beta$ and [\oiii]$\lambda$4959, 5007 lines. The blue regions are the windows for integrating the H$\beta$ line. The 5100$\rm \AA$ continuum windows are marked in gray. The red/black season labels in the top left corners represents that the data are used/not used to calculate the time lags, respectively.}
            \label{figure:mean_rms}
        \end{figure*}
    
        \begin{figure*}
            \centering
            \includegraphics[width=0.9\textwidth]{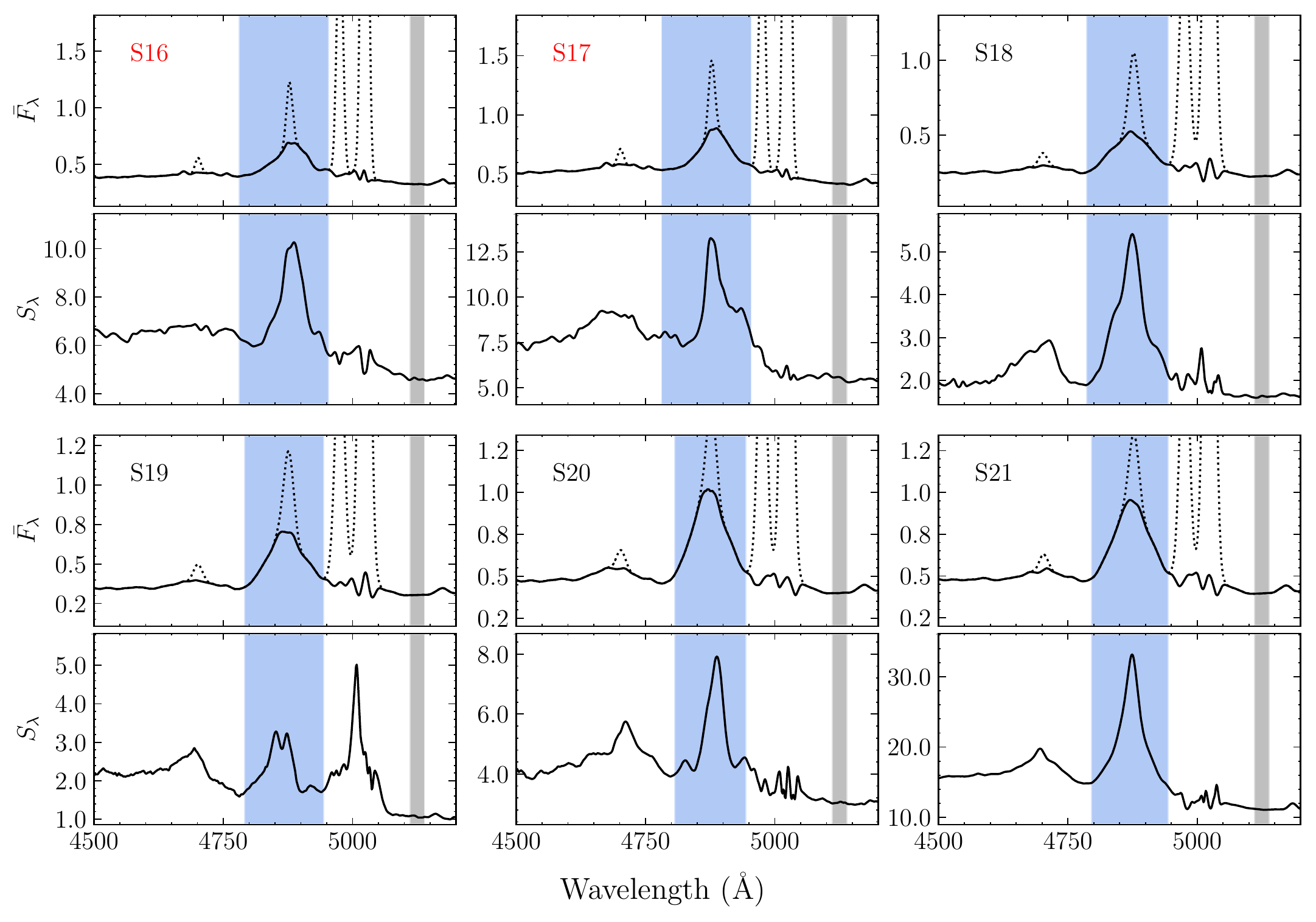}
                \addtocounter{figure}{-1}
            \caption{Continued.}
            \label{figure:mean_rms}
        \end{figure*}

      \subsection{Relative Flux Calibration of Spectra}
      \label{sec:calibration}
      The flux-calibration for the spectra from FAST and WIRO was initially done by using the spectrophotometric standard stars on corresponding nights. However, additional procedures for relative flux-calibration are necessary because of variable atmospheric extinction during the night and potential slit losses. Narrow emission lines are known to originate from an extended region and vary over a long timescale from years to decades \citep{pterson2013}. We thus assume that the fluxes of [\oiii]$\lambda$5007 line are constant during the FAST projects and MAHA campaign, and scale the observed spectra to match its standard flux. We stress that this assumption does not affect our time lag measurement in each season (Section~\ref{sec:lag_measurements}) because the variations of the [\oiii] flux can be neglected in such a short timescale ($\sim$200 days); However, this assumption might bias the estimation of the 5100~{\AA} luminosity and therefore affect the analysis of the $R_{\rm H\beta}-L_{5100}$ relationship (Section~\ref{sec:rl relationship}).
      In Section~\ref{sec:flux_correction}, we will examine whether the [\oiii]$\lambda$5007 flux changes over the time span of two campaigns and perform an extra correction to the 5100\,{\AA} continuum fluxes based on photometric data.
      
      The variable seeing and sometimes inaccurate telescope tracking would result in variable broadening to the observed spectra. Therefore, we artificially broaden the spectra from FAST and WIRO to achieve the same spectral resolution throughout (see more details in \citealt{dupu2014}). Due to the different flux losses of [\oiii] lines in the FAST spectra (see Section~\ref{sec:fast}), we determine the standard [\oiii] flux using the spectra from WIRO taken in photometric nights, which is $1.00\times10^{-11}\rm\ erg\ s^{-1}\ cm^{-2}$ in our measurement. Finally, we scale these spectra to match the standard [\oiii]$\lambda$5007 flux by employing the [\oiii]-based technique from \citet{van1992}. Some examples of relative flux-calibrated spectra are shown in Figure~\ref{figure:spectra_example}. 

        \subsection{Mean and RMS Spectra}
    	\label{sec:mean_rms_spectra}
        To investigate variability amplitudes and changes of H$\beta$ profile in the different seasons, we calculate the mean and root-mean-square (rms) spectra by the definition of the mean spectrum  (e.g., \citealt{peterson2004})
    	
    	\begin{equation}
    		\Bar{F}_\lambda = \frac{1}{N}\sum_{i=1}^{N}F_{\lambda}^{i}
    	\end{equation}
    	and the rms spectrum
    	\begin{equation}
    		S_{\lambda} = \left[\frac{1}{N-1}\sum_{i=1}^{N}\left(F_{\lambda}^{i}-\Bar{F}_\lambda\right)^2\right]^{1/2},
    		\label{eqn_rms}
    	\end{equation}
    	where $F_{\lambda}^{i}$ is the $i$th spectrum, and $N$ is the number of spectra in each season. The mean and rms spectra are shown in Figure~\ref{sec:mean_rms_spectra}. The rms spectra defined as above do not take into account the differences in the errors of the spectra. Alternatively, we can also calculate S/N-weighted rms spectra (e.g., see  \citealt{Park2012b,Barth2015}). In Appendix~\ref{appendixc}, we present a comparison between the two types of rms spectra for our data. We find that the two types of rms spectra are almost the same and the resulting line widths differ no more than 1\%. Therefore, we, by default, measure line widths of the rms spectra  with Equation~(\ref{eqn_rms}) in the subsequent analysis. 
    	
    	Both the mean and rms spectra show complicated profiles with significant changes over the 21 monitoring seasons. Interestingly, there is apparent blue absorption in the H$\beta$ broad line in S3, S6 and S7, which has been reported by~\citet{Hutchings2002} and \citet{Shapovalova2010}. The investigation of the evolution of the broad emission line profiles will be featured in Paper II of this series.
    	Additionally, the [\oiii] and other narrow emission lines are very weak in rms spectra, which indicates that our relative flux-calibration procedures in Section \ref{sec:calibration} work well. There is a stronger residual of [\oiii] in the rms spectrum of S19. This causes by some spectra taken in poor weather conditions. For these instances, the procedures of broadening the spectra do not work very well.

    	\begin{figure*}
    		\centering
    		\includegraphics[ width=0.9\textwidth]{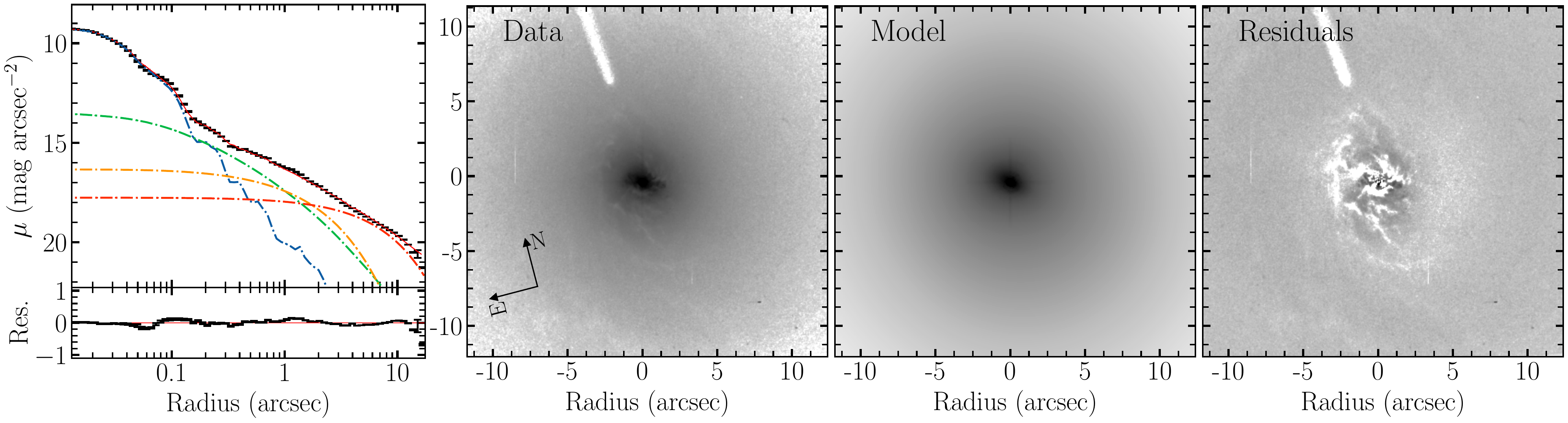}
    		\caption{\texttt{GALFIT} decomposition for NGC 4151. From left to right, each panel shows the 1D profile, 2D image of the data, the best-fit model for the host and the residual. The 1D brightness profiles show the data (black dots), the best-fit total model (red solid line) and the subcomponents (the blue dash line for PSF, the green, orange and red dotted–dashed lines for the bulge, the inner and outer disk, respectively).}
    		\label{figure:galfit}
    	\end{figure*}

	\subsection{Flux Contribution from the Host Galaxy}
        \label{sec:host}
        For reducing the flux deviations in the light curves introduced by the different spectral extraction apertures and position angles (Section~\ref{sec:fast}), it is necessary to subtract the different starlight contamination for each spectrum. We estimate the fluxes from the host galaxy by adopting the following procedures. First, we employ the two-dimensional (2D) image-decomposition program \texttt{GALFIT}\footnote{\texttt{GALFIT} is a nonlinear least-squares 2D image-fitting algorithm, and Version 3.0.5 is used in this work.} (\citealt{peng2002,Peng2010}) to model the image with typical galaxy parameters for exactly subtracting the contribution from the center point source. The image data of NGC 4151 was observed by the \emph{Hubble Space Telescope (HST)} Advanced Camera for Surveys (ACS) on 2004-03-28 \citep{bentz2009}. 
        We separately use a central PSF and a series of Sersic profiles to model the nucleus region and the host components. Sersic profile~\citep{Sersic1968} is given by
        \begin{equation}
            \Sigma{(r)} = {\Sigma}_{e}e^{-\kappa\left[\left(r/r_{e}\right)^{(1/n)}-1\right]}
        \end{equation}
        where $\Sigma_{e}$ is the surface brightness at effective radius, $r_e$ is the effective radius and $\kappa$ is a factor coupled with n to ensure the $R_e$ is the effective radius. The Sersic profile is very flexible and typically represents exponential, Gaussian and de Vaucouleurs profiles when n = 0.5, 1, and 4 respectively. We set free parameters to achieve the best 2D decomposition and reasonable host galaxy components. Figure~\ref{figure:galfit} shows the best models and their corresponding residuals.
        
        After that, we use the decomposition of the \emph{HST} image to estimate the contribution of starlight with the appropriate spectral extraction aperture and position angle. \cite{ho2014} classified the bulge type of NGC 4151 as a classical bulge. We thus adopt the spectrum of a model bulge~\citep{Kinney1996} for the host galaxy of NGC 4151. The host fluxes at 5100\,{\AA} are transformed from F550M magnitude with \texttt{pyphot}\footnote{\url{https://pypi.org/project/pyphot/}} package, after correcting for redshift ($z\approx0.00332$) and Galactic reddening ($E(B-V)=0.071$ mag; \citealt{Schlafly2011}) adopting the extinction curve of \citet{Cardelli1989}. We list the host fluxes in each spectrum obtained in FAST and WIRO observations in Table~\ref{table:log}.

        \subsection{Spectroscopic Light Curves}
	\label{sec:Spectroscopic_Lightcurves}
\begin{figure*}
        \centering
        \includegraphics[width=0.9\textwidth]{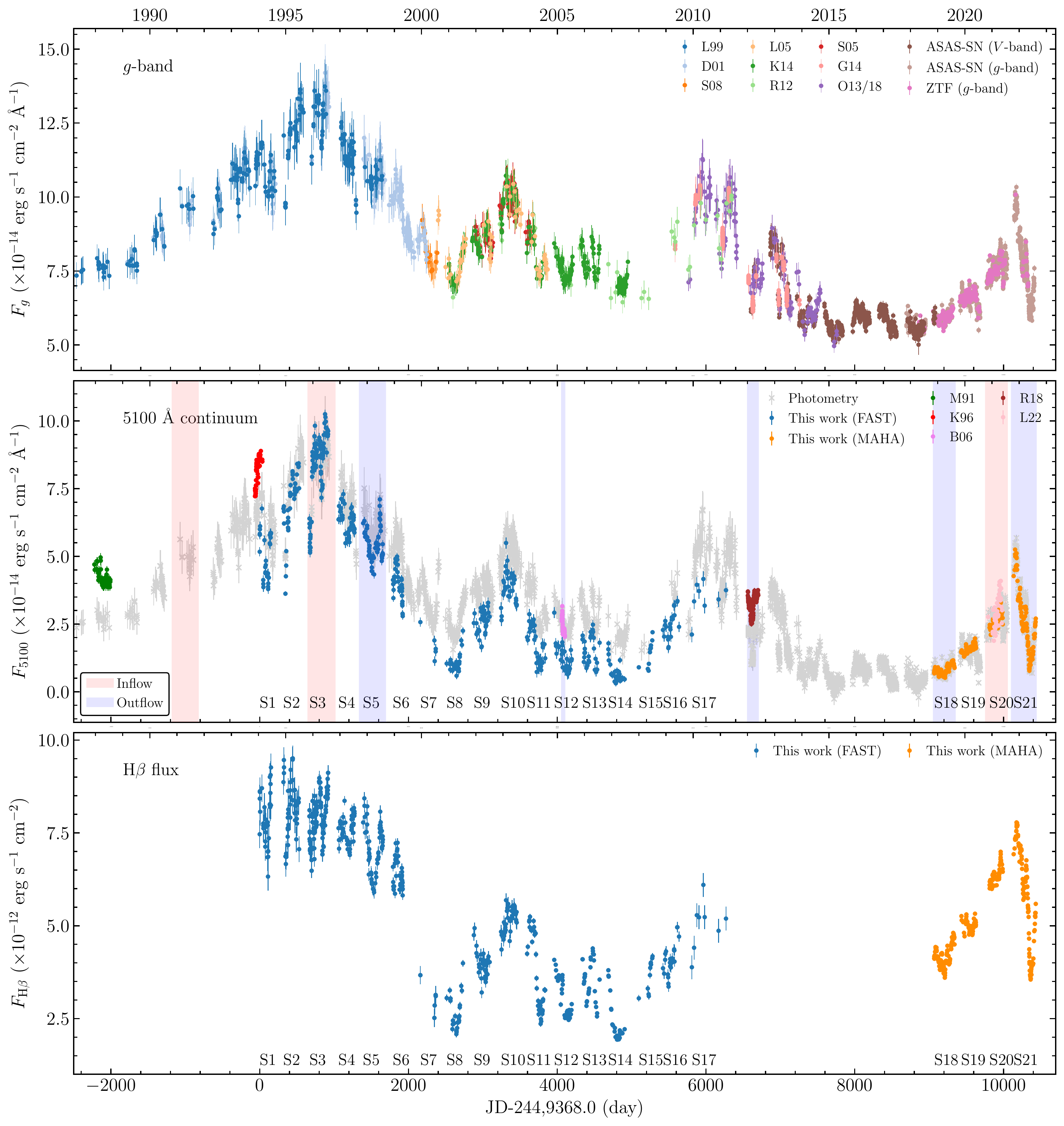}
        \caption{Panels from top to bottom are the light curves of photometry, the continuum at rest-frame 5100~{\AA} and the H$\beta$ emission line, respectively. In the middle panel, the gray points are the flux-corrected photometric light curve, and the red/blue regions represent the BLR kinematics of the corresponding season are the virial motion with a contribution from inflow/outflow, respectively.}
        \label{figure:lc_all}
        \end{figure*}

    	Light curves of the AGN continuum and H$\beta$ emission are shown in the middle and bottom panels in Figure~\ref{figure:lc_all}, respectively.
        Generally, there are two methods for flux measurements of emission lines, one of which is the direct integration (e.g., \citealt{dupu2014,dupu2018,rosa2018,Bao2022}) and the other is the spectral fitting (e.g., \citealt{lu2021,huchen2021,Li2022}).
        For the first method, the emission fluxes are measured by a simple integration after subtracting a local background defined by two continuum windows. This method is suitable for the strong and single emission lines (e.g., H$\beta$ and H$\alpha$). For the second method, the fluxes of emission lines are obtained by fitting the spectra in a wide wavelength range with multiple spectral components. This method is suitable for measuring the fluxes of highly blended lines (e.g., \heii~and \feii). For the objects with strong contributions from the host galaxy, the local continuum would deviate considerably from a simple straight line. Considering the weak host contamination for NGC 4151, we employ the integration method to measure the H$\beta$ light curves.

        The measurement windows of the H$\beta$ line are listed in Table~\ref{table:info} and shown in Figure~\ref{figure:mean_rms}. As described in Section~\ref{sec:mean_rms_spectra}, the rms variations of H$\beta$ line are significant from year to year (see also Figure~\ref{figure:mean_rms}). Therefore, we adopt different windows for H$\beta$ flux measurements based on rms residual variation. This can mitigate flux contamination from \heii. 
        For two continuum windows, we determine them using the mean flux at the wavelength from 4510-4520\,{\AA} and 5090-5140\,{\AA} in observer-frame for blue- and red-ward, respectively. We measure the 5100\,{\AA} continuum light curves in the red-ward windows, in which there is only very weak \feii~emission but contamination from the host.
        At last, we subtract the starlight contribution for the light curves at 5100\,{\AA}. For the spectra from FAST, the different aperture sizes and position angles would cause non-uniform fluxes from the host galaxy for each spectrum. Therefore, the starlight fluxes are diverse values for these spectra (see details in Section~\ref{sec:fast}). 

        The conditions of the weather and telescope (e.g., the poor weather condition, telescope-tracking inaccuracies and slit positioning) would introduce an additional influence on the measurements of fluxes. Hence, except for the statistical errors, we also include the systematic uncertainties in the fluxes. The estimation of systematic error is estimated by the same technique in \citet{dupu2014}, and is included in the following analysis.

        \begin{table*}
        \renewcommand{\arraystretch}{1.3}
        \setlength{\tabcolsep}{9.5pt}
        \caption{Datasets for the photometric light curves of NGC 4151}
        \label{table:photometry}
        \begin{tabular}{lcccccccc}
        \hline
        Dataset & Filter & JD & Date & Number of & $\varphi$ & $G$ & Ref. \\
        &    & (-2,400,000) & (yyyy/mm-yyyy/mm) & Observations & & ($\rm 10^{-14}\ erg\ s^{-1}\ cm^{-2}\ \text{\AA}^{-1}$) & \\
        \hline
        L99 & $B$ & 941-12,015 & 1968/03-1998/07 & 789 & 0.873$\pm$0.064 & 4.972$\pm$0.376 & 1\\
        D01 & $B$ & 8,591-12,617 & 1989/03-2000/07 & 440 & 0.868$\pm$0.066 & 4.999$\pm$0.370 & 2\\
        S08$^a$ & 5100~\AA & 12,552-12,755 & 2000/01-2000/07 & 18 & 0.761$\pm$0.123 & 6.081$\pm$0.357 & 3\\
        L05 & $B$ & 12,621-14,240 & 2000/03-2004/08 & 67 & 0.974$\pm$0.079 & 5.406$\pm$0.325 & 4\\
        K14 & $V$ & 12,915-15,319 & 2001/01-2007/08 & 234 & 1.080$\pm$0.078 & 3.923$\pm$0.326 & 5\\
        R12 & $B$ & 12,967-16,715 & 2001/02-2011/06 & 57 & 1.018$\pm$0.079 & 4.365$\pm$0.362 & 6\\
        S05 & $B$ & 13,270-14,022 & 2001/12-2004/01 & 149 & 1.136$\pm$0.083 & 4.978$\pm$0.319 & 7\\
        G14 & $B$ & 15,954-17,626 & 2009/05-2013/11 & 600 & 1.950$\pm$0.183 & 1.086$\pm$0.445 & 8\\
        O13/18 & $B$ & 16,134-18,127 & 2009/10-2015/04 & 163 & 0.837$\pm$0.077 & 4.240$\pm$0.146 & 9, 10\\
        ASAS-SN$^b$ & $V$ & 16,974-19,449 & 2012/02-2018/11 & 905 & 1.423$\pm$0.117 & 4.784$\pm$0.850 & 11, 12\\
        ASAS-SN$^b$ & $g$ & 19,064-20,783 & 2017/11-2022/07 & 1285 & 1.000$\pm$0.000 & 0.000$\pm$0.000 & 11, 12\\
        ZTF$^c$ & $g$ & 19,202-20,701 & 2018/03-2022/05 & 401 & 0.992$\pm$0.008 & 3.331$\pm$0.027 & 13, 14\\
        \hline
        \end{tabular}
        \begin{list}{}{}
        \item[$^a$] {Only the part of the continuum light curve that has temporal overlap with S6 is used.}
        \item[$^b$] {All-Sky Automated Survey for SuperNovae.}
        \item[$^c$] {Zwicky Transient Facility project.}
        \item[Ref:] {(1)~\citet{Lyuty1999}, (2)~\citet{Doroshenko2001}, (3)~\citet{shapovalova2008}, (4)~\citet{Lyuty2005}, (5)~\citet{Koshida2014}, (6)~\citet{Roberts2012}, (7)~\citet{Sergeev2005}, (8)~\citet{Guo2014}, (9)~\citet{Oknyanskij2013}, (10)~\citet{Oknyanskij2018}, (11)~\citet{Shappee2014}, (12)~\citet{Kochanek2017}, (13)~\citet{Bellm2019}, (14)~\citet{Masci2019}.}
        \end{list}
        \end{table*}

        \subsection{Photometry and Flux Correction of the Continuum}
        
        Based on the assumption of the constant [\oiii] flux over the monitoring periods of the FAST and MAHA program, we have finished the relative flux-calibration for the spectra in Section~\ref{sec:calibration}. However, these two campaigns span a total of $\sim 28$ years, and the [\oiii] flux may have undergone significant variation between different observation seasons years apart. 
        The long-term photometric light curves can be used to test this and further correct the fluxes in the AGN continuum measured in different seasons.
        
        \subsubsection{The Long-term Photometric Light Curves}
        Table~\ref{table:photometry} summarizes photometric datasets from published literature and time-domain photometric surveys. 
        These data are often measured with different aperture sizes and filters making it necessary to inter-calibrate them to construct an accurate light curve. 
        We convert these photometric data ($B$-, $V$- and $g$-band) to flux densities by adopting the zero-point $F_{\lambda}(B,\ V\ {\rm and}\ g=0)=3.63\times 10^{-9}\rm\ erg\ s^{-1}\ cm^{-2}\ \text{\AA}^{-1}$. With the fluxes of a dataset of ASAS-SN ($g$-band) as a reference, we calibrate the other datasets with the Bayesian-based package \texttt{PyCALI}\footnote{\url{https://github.com/LiyrAstroph/PyCALI}} \citep{li2014}. This package is based on the assumption that light curves can be described by a damped random walk process and determines the best multiplicative scale factor $\varphi$ and additive flux adjustment $G$ (i.e. $F=\varphi F_{\rm obs}-G$) by exploring the posterior distribution with a diffusive nested sampling algorithm \citep{Brewer2011}. 
        Table~\ref{table:photometry} lists the values of $\varphi$ and $G$ obtained for all datasets and the upper panel in Figure~\ref{figure:lc_all} shows the inter-calibration light curves.

        There are two things needed to be noted in the above procedure of inter-calibration. First, for the datasets L99, D01, L05, G14 and O13/18, the measurement error of each photometric data point is not specified in the corresponding literature, but most of them are in the range from 0.01 mag to 0.1 mag (about 1\% to 10\% of flux). We adopt 5\%  of fluxes as photometric errors. Moreover, the systematic uncertainties are also considered by using the same procedures as Section~\ref{sec:Spectroscopic_Lightcurves}. Second, there is no temporal overlap between the datasets from \citet{Doroshenko2001} and \citet{Lyuty2005}. Therefore, the continuum light curve from \citet{shapovalova2008} is also included for inter-calibration. Their spectral flux calibration was based on [\oiii]$\lambda$5007 and [\oi]$\lambda$6300 (see more details in \citealt{shapovalova2008}). Given the possibility of flux variations of narrow lines, only the part of the data of S08 that has temporal overlap with S6 is used.
	    \label{sec:photometry}

        \subsubsection{Correction for Fluxes in Continuum Light Curves}
        Figure~\ref{o3imag} shows the continuum-subtracted image of the [\oiii] region and the aperture size that we used in the observations of the WIRO spectra (red rectangle). This demonstrates that the loss of [\oiii] emission (about 3.4 percent) during the period of the MAHA program can be ignored. Therefore, we assume that WIRO spectra have accurate absolute flux. Adopting the starlight-subtracted AGN fluxes in S18 as the standard, we inter-calibrate the fluxes in the photometric light curve using the same method as Section~\ref{sec:photometry}. The comparison between flux-corrected long-term variation and AGN continuum in the other seasons is shown in Figure~\ref{figure:lc_all}. 
        The fluxes of corrected photometry and AGN continuum in S19-S21 are substantially consistent. However, the photometric fluxes in S1-S17 are slightly higher than that of the AGN continuum.
        This implies that the [\oiii] emission underwent a change during the period of two RM campaigns but does not change significantly in the short term. On the other hand, \citet{Li2022} calibrated their spectra from Lijiang observations by using the comparison-star-based technique. Moreover, they measured the AGN continuum by adopting the method of spectral decomposition with some necessary components such as the contribution of the host galaxy. As shown in the middle panel of Figure~\ref{figure:lc_all}, the fluxes in the part temporal overlapping between the calibrated photometric light curve and the data from \citet{Li2022} are consistent. This indicates that our absolute calibration procedures for WIRO spectra work well.
        
        Finally, considering the change of the [\oiii] emission over the 28 years of RM monitoring, we calibrate the light curves for each season to match the variation of the flux-corrected photometric data. The calibration method is the same as described in Section~\ref{sec:photometry}. Moreover, the fluxes of the 5100\,{\AA}~continuum from previous H$\beta$ RM campaigns (M91, K96, B06, R18 and L22; see the details in Section~\ref{sec:rl relationship}) also are corrected using the above procedure. We list the average of the corrected fluxes for each H$\beta$ RM campaign in Table~\ref{table:summary}. As shown in Figure~\ref{mapping1}, the variation of the 5100\,{\AA} continuum is well matched to the temporally overlapping photometry, which indicates that the deviation of the [\oiii]-based calibration introduced by the different aperture sizes and position angles for the FAST spectra (Section~\ref{sec:fast}) can be largely ignored. 
        \label{sec:flux_correction}

        \begin{table*}
        \renewcommand{\arraystretch}{1.3}
  \setlength{\tabcolsep}{3pt}
            \caption{Light Curve Statistics and H$\beta$ Line Integration Windows}
            \label{table:info}
            \begin{tabular}{lccccccccccccc}
            \hline
            Campaign & JD & Date & $N^a$& \multicolumn{2}{c}{Sampling Interval (day)}  & Integration Window (\AA) & \multicolumn{2}{c}{$F_{\rm var}\ (\%)$}  & & \multicolumn{2}{c}{$P_{\rm var}\ (\%)$}  & \\
            \cline{5-6}
            \cline{8-9}
            \cline{11-12}
             & (-2,449,000) & (yyyy/mm-yyyy/mm) & & median & $\left<T\right>^b$ & H$\beta$ & Continuum & $\rm H\beta$ & & Continuum & $\rm H\beta$\\
             \hline
             \multicolumn{4}{l}{The FAST Spectrograph Publicly Archived Programs}\\
             \hline
             S1$^c$ & 368-520 & 1994/01-1994/06 & 33 & 1.5 & 4.7 & 4790-4955 & 16.4 $\pm$ 2.1 & 8.0 $\pm$ 1.3 & & 56.3 & 37.7\\
             S2 & 688-898 & 1994/12-1995/06 & 44 & 2.0 & 4.9 & 4795-4955 & 15.4 $\pm$ 1.7 & 7.9 $\pm$ 1.1 & & 79.9 & 35.2\\
             S3 & 1,038-1,288 & 1995/11-1996/07 & 92 & 1.0 & 2.8 & 4795-4955 & 15.9 $\pm$ 1.2 & 7.3 $\pm$ 0.6 & & 66.5 & 33.8\\
             S4 & 1,429-1,643 & 1996/12-1997/07 & 48 & 2.0 & 4.6 & 4805-4955 & 6.9 $\pm$ 0.8 & 4.5 $\pm$ 0.5 & & 28.4 & 19.1\\
             S5 & 1,760-2,024 & 1997/11-1998/07 & 54 & 2.0 & 5.0 & 4810-4955 & 10.7 $\pm$ 1.1 & 8.7 $\pm$ 0.9 & & 48.4 & 35.3\\
             S6 & 2,161-2,291 & 1998/12-1999/04 & 39 & 2.0 & 3.4 & 4805-4955 & 14.6 $\pm$ 1.7 & 6.0 $\pm$ 0.7 & & 56.5 & 21.7\\
             S7$^c$ & 2,529-2,735 & 1999/12-2000/07 & 5 & 8.5 & 51.6 & 4780-4955 & 29.6 $\pm$ 10.5 & 11.2 $\pm$ 5.4 & & 77.9 & 37.2\\
             S8$^c$ & 2,883-3,101 & 2000/12-2001/07 & 26 & 5.9 & 8.7 & 4780-4955 & 32.3 $\pm$ 5.1 & 17.3 $\pm$ 2.5 & & 116.7 & 62.8\\
             S9 & 3,248-3,460 & 2001/12-2002/07 & 33 & 3.9 & 6.6 & 4780-4955 & 14.7 $\pm$ 2.2 & 7.5 $\pm$ 1.2 & & 66.6 & 42.6\\
             S10 & 3,614-3,824 & 2002/12-2003/07 & 37 & 3.0 & 5.9 & 4800-4955 & 13.9 $\pm$ 1.9 & 5.1 $\pm$ 0.9 & & 70.2 & 26.4\\
             S11 & 3,964-4,207 & 2003/11-2004/07 & 38 & 4.0 & 6.6 & 4780-4955 & 36.1 $\pm$ 4.6 & 27.6 $\pm$ 3.2 & & 118.6 & 75.8\\
             S12$^c$ & 4,329-4,563 & 2004/11-2005/07 & 32 & 3.9 & 7.6 & 4780-4955 & 35.0 $\pm$ 5.3 & 17.0 $\pm$ 2.2 & & 126.7 & 49.4\\
             S13$^c$ & 4,709-4,938 & 2005/12-2006/07 & 30 & 3.0 & 7.9 & 4780-4955 & 32.2 $\pm$ 4.7 & 15.1 $\pm$ 2.0 & & 108.6 & 52.7\\
             S14$^c$ & 5,056-5,275 & 2006/11-2007/06 & 27 & 5.4 & 8.4 & 4780-4955 & 49.2 $\pm$ 7.4 & 25.4 $\pm$ 3.5 & & 140.7 & 65.0\\
             S15$^c$ & 5,466-5,650 & 2007/12-2008/07 & 11 & 5.0 & 18.5 & 4780-4955 & 32.8 $\pm$ 7.0 & 11.6 $\pm$ 2.6 & & 95.4 & 31.3\\
             S16$^c$ & 5,791-6,006 & 2008/11-2009/06 & 24 & 3.9 & 9.4 & 4780-4955 & 18.3 $\pm$ 3.1 & 9.3 $\pm$ 1.5 & & 70.1 & 38.5\\
             S17$^{c,d}$ & 6,179-6,637 & 2009/12-2011/03 & 8 & 36.9 & 65.5 & 4780-4955 & 16.3 $\pm$ 5.1 & 11.5 $\pm$ 3.8 & & 65.3 & 44.4\\
             \hline
             \multicolumn{4}{l}{MAHA Program}\\
             \hline
             S18 & 9,437-9,710 & 2018/11-2019/08 & 106 & 1.9 & 2.6 & 4785-4945 & 22.1 $\pm$ 1.6 & 7.1 $\pm$ 0.5 & & 76.5 & 28.7\\
             S19 & 9,799-9,998 & 2019/11-2020/05 & 75 & 1.1 & 2.7 & 4790-4945 & 6.8 $\pm$ 0.6 & 2.7 $\pm$ 0.2 & & 40.8 & 12.1\\
             S20 & 10,182-10,358 & 2020/11-2021/05 & 58 & 1.9 & 3.1 & 4805-4945 & 11.6 $\pm$ 1.1 & 3.7 $\pm$ 0.4 & & 44.3 & 15.4\\
             S21 & 10,505-10,802 & 2021/12-2022/08 & 108 & 2.0 & 2.8 & 4795-4945 & 45.8 $\pm$ 3.1 & 22.2 $\pm$ 1.5 & & 148.4 & 74.6\\
             \hline
             \multicolumn{8}{l}{$^a$ The number of observational nights in each season.}\\
             \multicolumn{8}{l}{$^b$ The mean sampling interval for different seasons.}\\
             \multicolumn{8}{l}{$^c$ The data in these seasons are not used to analyze in Section~\ref{analysis} because of their low quality.}\\
             \multicolumn{8}{l}{$^d$ These two years taken together are regarded as one observational season due to the small number of epochs in each.}
            \end{tabular}
        \end{table*}
        
	\section{Analysis}
	\label{analysis}
 
	\subsection{Variability}
        We adopt two methods to describe the light curve characteristics of the continuum and H$\beta$ emission line. The first represents the AGN intrinsic variability, which is defined \citep{rodr1997} as 
        \begin{equation}
            F_{\rm var} = \frac{(\sigma^2-\Delta^2)^{1/2}}{\left<F\right>}
        \end{equation}
        where 
        \begin{equation}
            \sigma = \sum_{i=1}^{N}\frac{(F_i-\left<F\right>)^2}{N-1},\ \Delta^2 = \sum_{i=1}^{N}\frac{\Delta_i^2}{N},\ \left<F\right>=\sum_{i=1}^{N}\frac{F_i}{N}
        \end{equation}
        
        and the uncertainty in $F_{\rm var}$ is defined \citep{edelson2002} as 
        \begin{equation}
            \sigma_{\rm var}=\frac{1}{(2N)^{1/2}F_{\rm var}}\frac{\sigma^2}{\left<F\right>^2}
        \end{equation}
        where $N$ is the number of epochs, $F_i$ is the flux of the $i$th observation and $\Delta_i$ is the error of $F_i$. The other is the variability amplitude of the peak-to-peak variation, which is given by 
        \begin{equation}
            P_{\rm var}=\frac{F_{\rm max}- F_{\rm min}}{\Bar{F}}
        \end{equation}
        where $\Bar{F} = (F_{\rm max}+ F_{\rm min})/2$, $F_{\rm max}$ and $F_{\rm min}$ are the maximum and minimum flux in the light curve, respectively. We summarize the variability amplitudes of the light curves in different periods in Table~\ref{table:info}. The larger $P_{\rm var}$ in the two campaigns indicates that the relatively smaller effect of different aperture sizes and position angles for flux calibration can be neglected.

        \begin{figure*}
            \centering
            \includegraphics[width=0.82\textwidth]{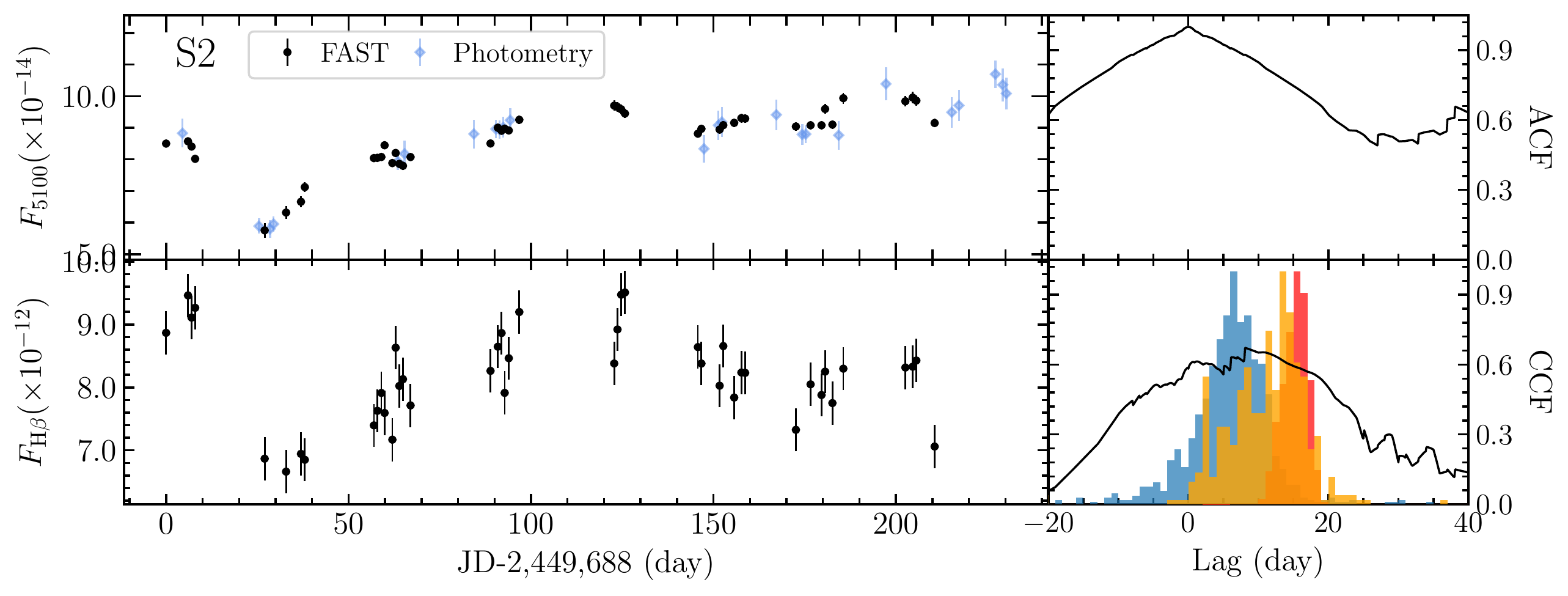}
            \includegraphics[width=0.82\textwidth]{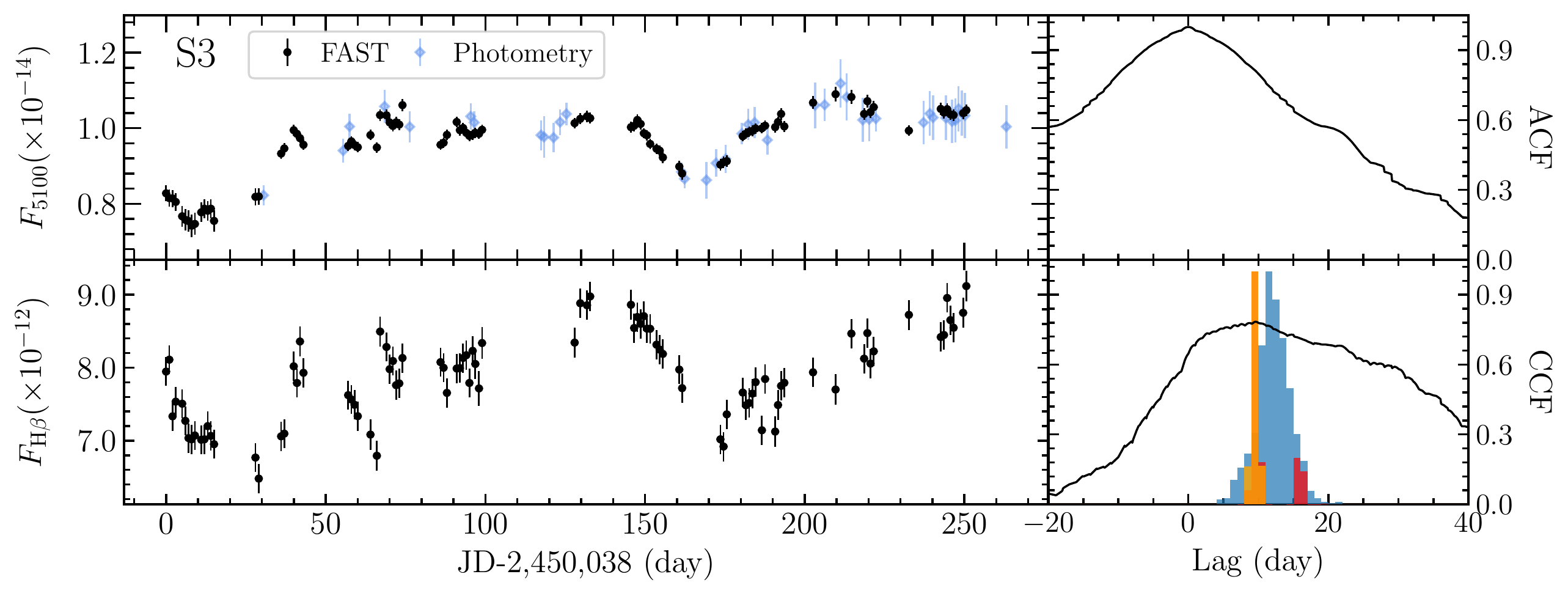}
            \includegraphics[width=0.82\textwidth]{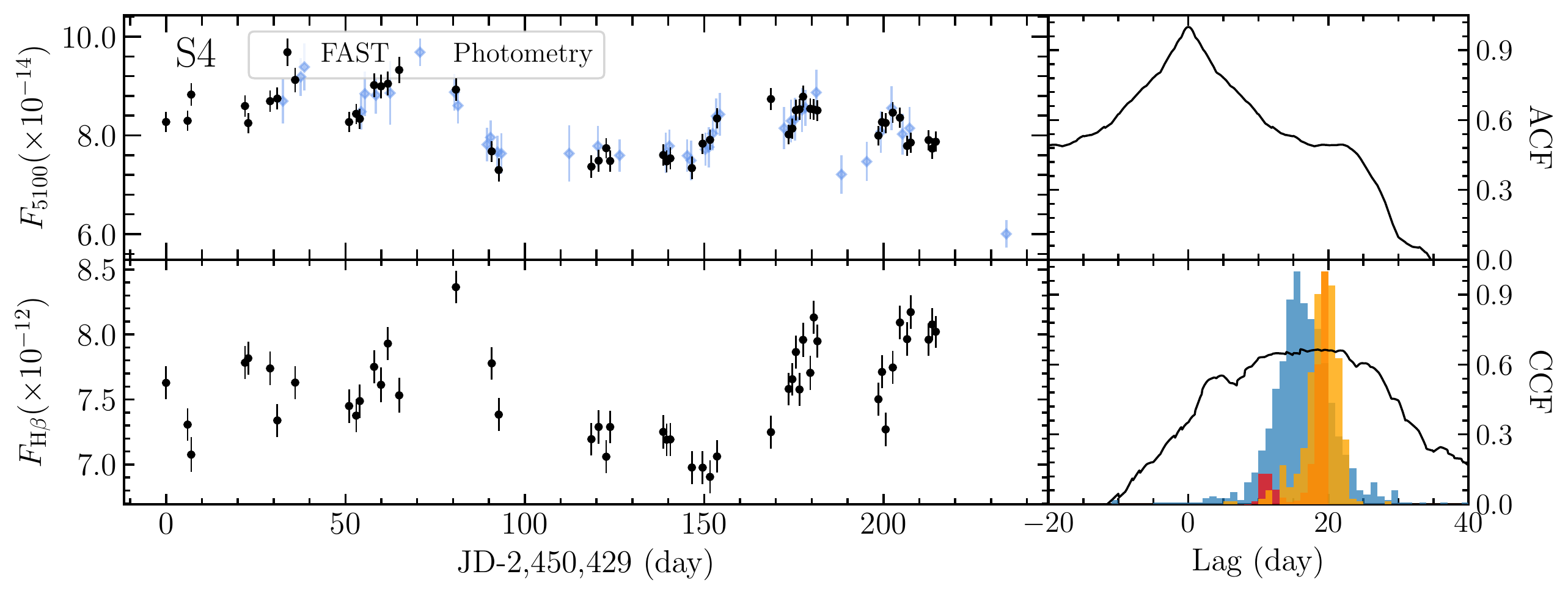}
            \includegraphics[width=0.82\textwidth]{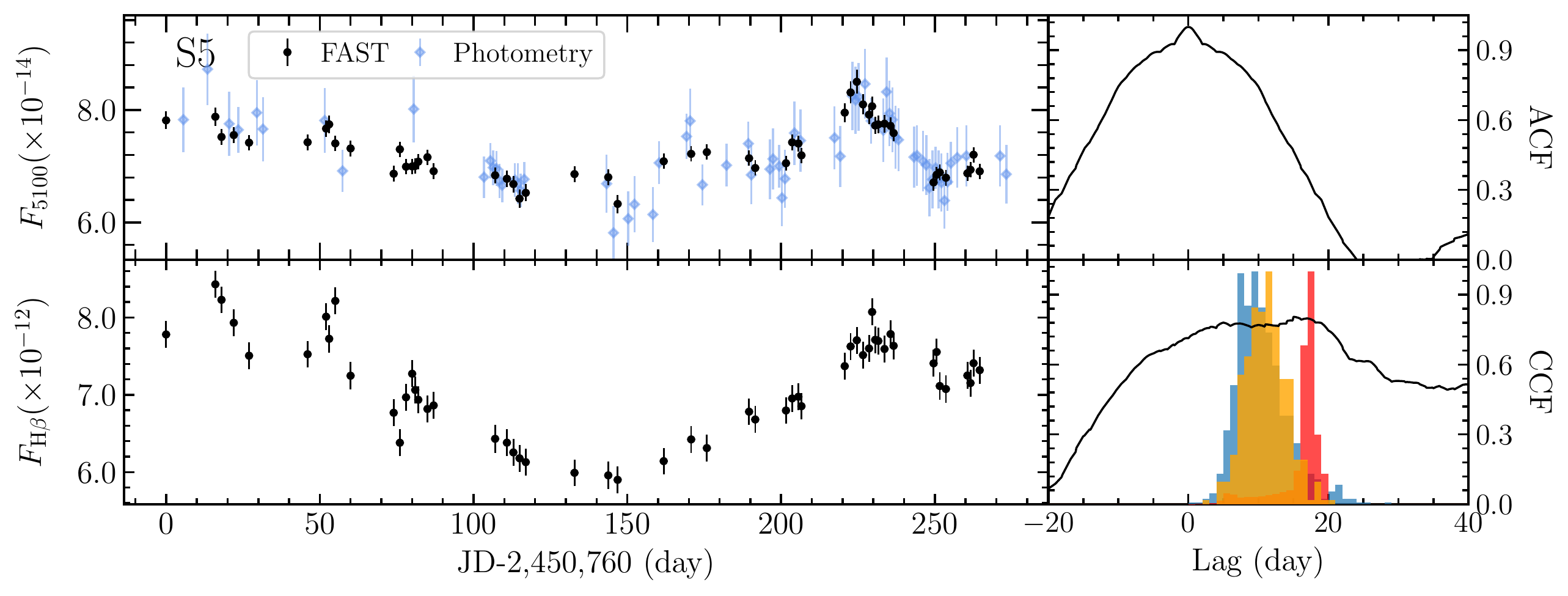}
            \caption{(Left) the corrected light curves of the continuum at 5100~{\AA}, calibrated photometry and the integrated H$\beta$ fluxes. (Right) results of the correlation analysis. The top panel shows the ACFs of the continua. The bottom panel shows CCFs between the broad emission lines and continuum light curves. The blue histogram is the cross-correlation centroid distribution. The red and orange histograms show the posterior distribution of time lags from JAVELIN and MICA, respectively. }
            \label{mapping1}
       \end{figure*}
		
        \begin{figure*}
            \centering
            \includegraphics[width=0.82\textwidth]{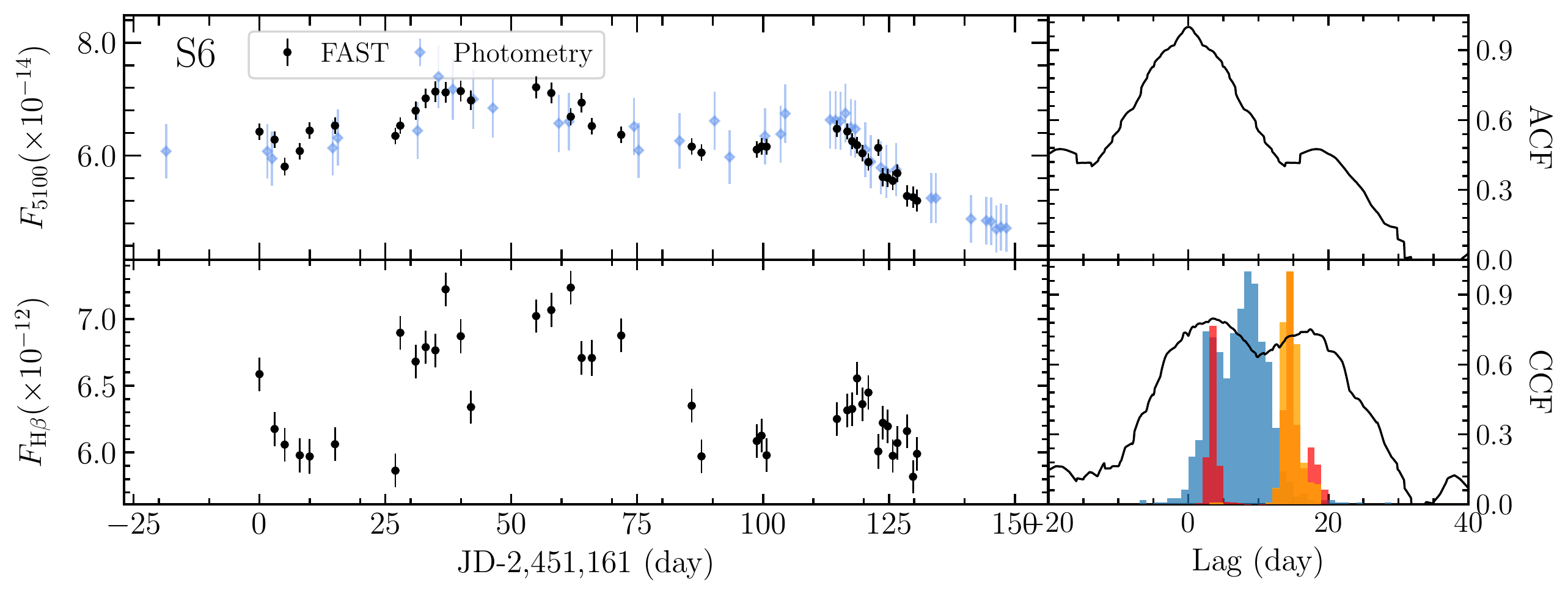}
                \includegraphics[width=0.82\textwidth]{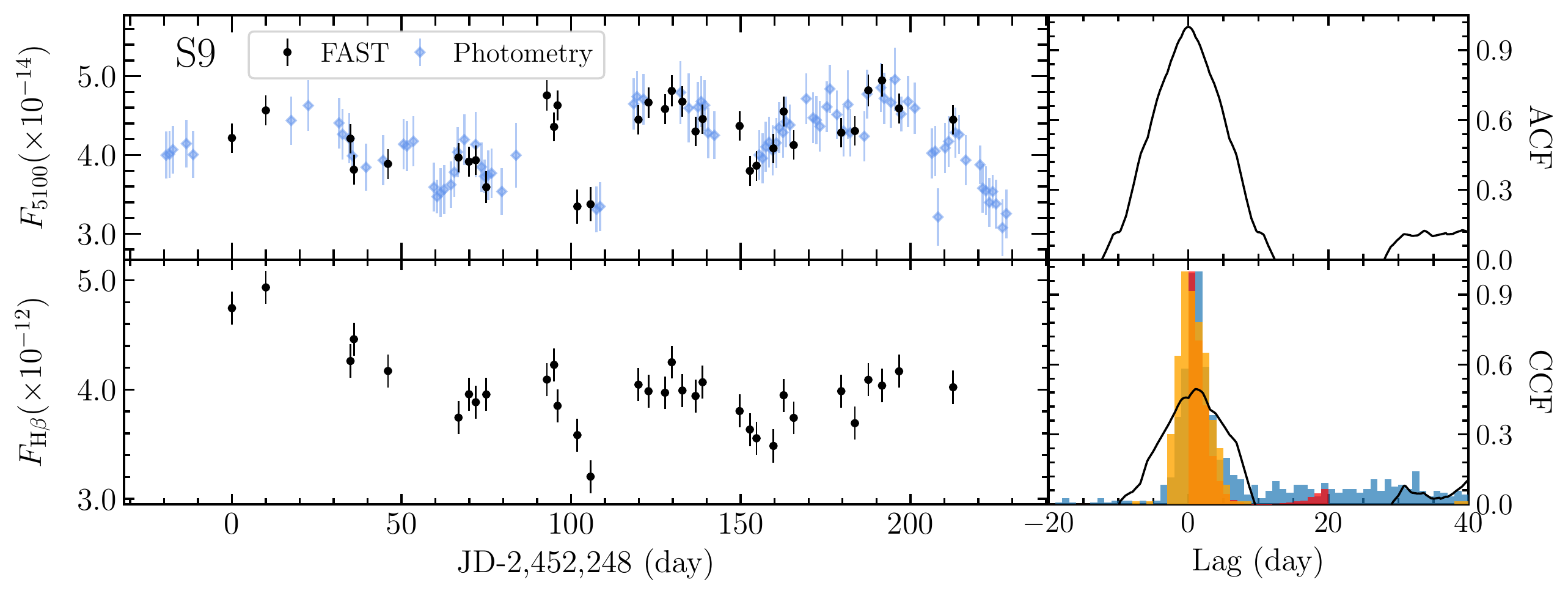}
                \includegraphics[width=0.82\textwidth]{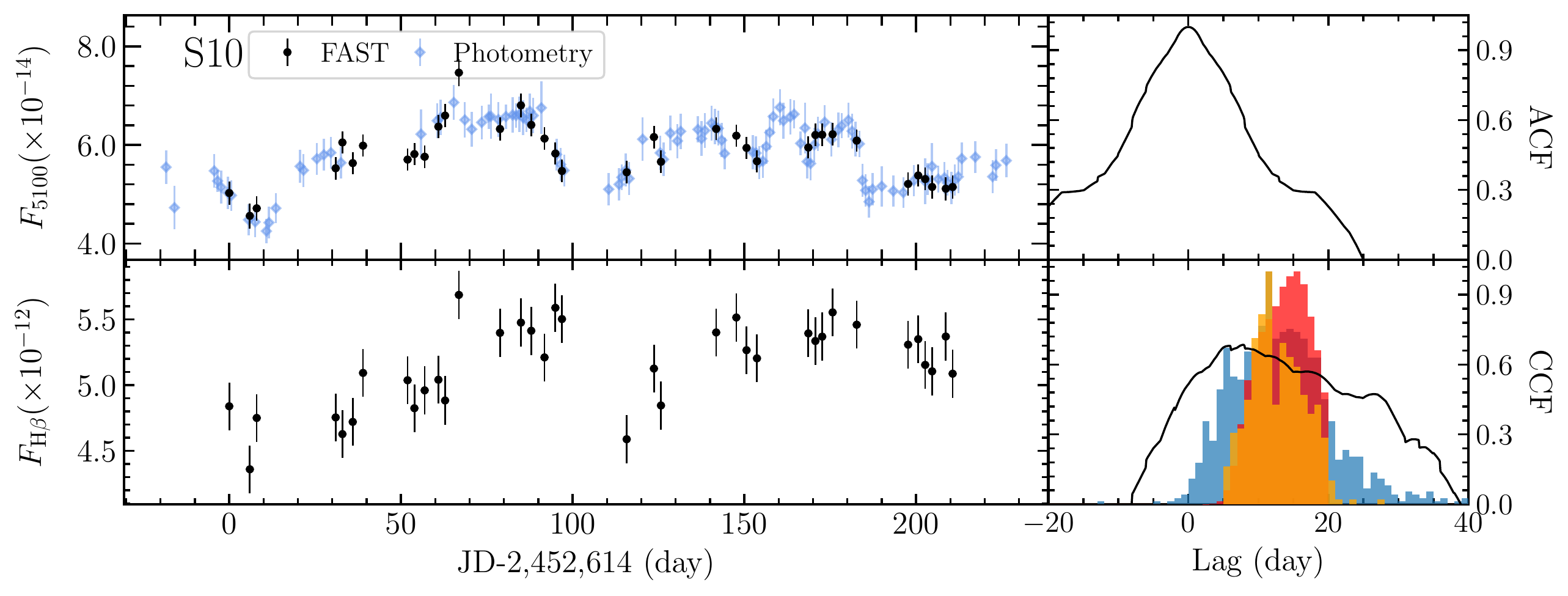}
            \includegraphics[width=0.82\textwidth]{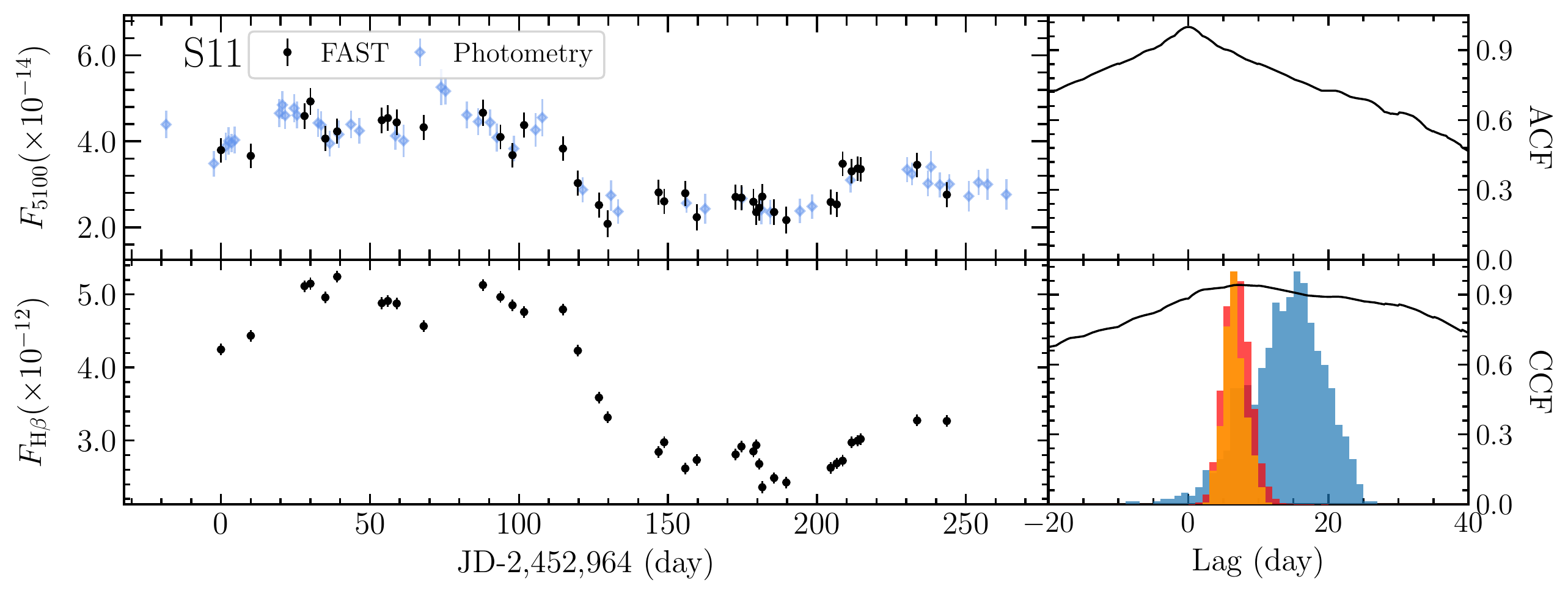}
            \addtocounter{figure}{-1}
            \caption{(Continued).}
            \label{mapping2}
        \end{figure*}
        
        \begin{figure*}
            \centering
                \includegraphics[width=0.82\textwidth]{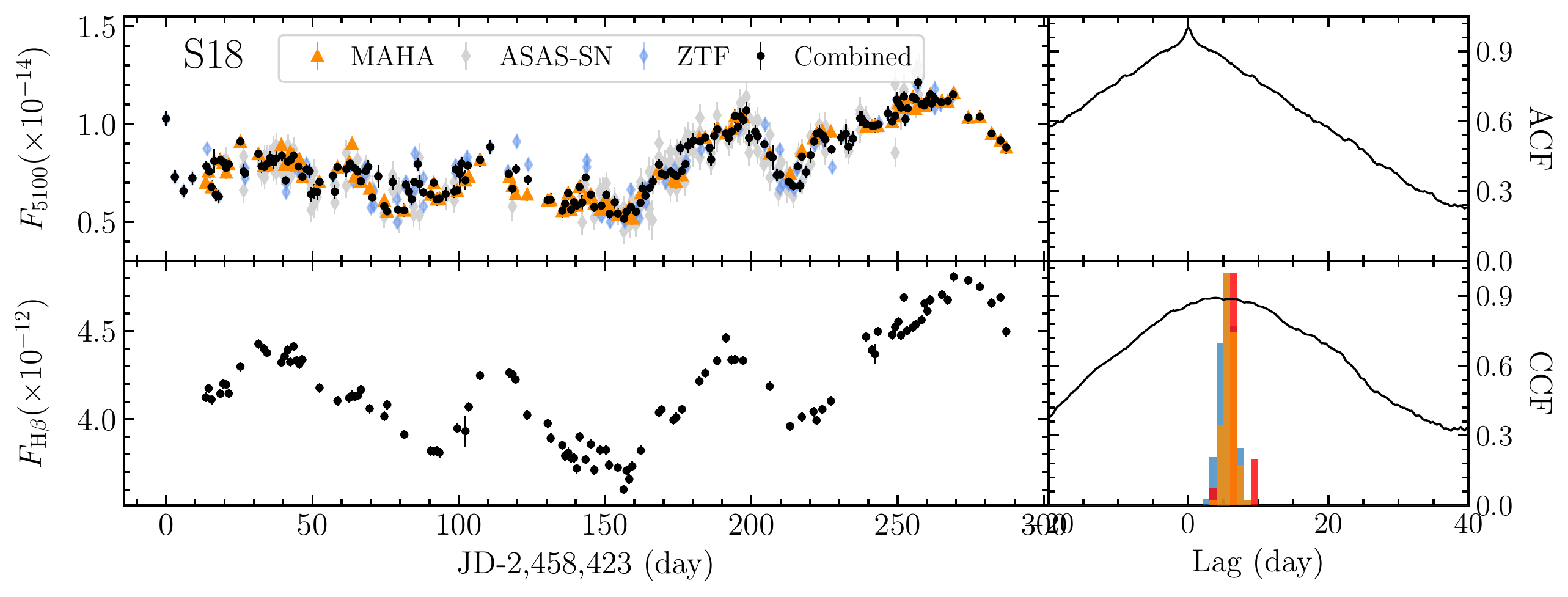}
            \includegraphics[width=0.82\textwidth]{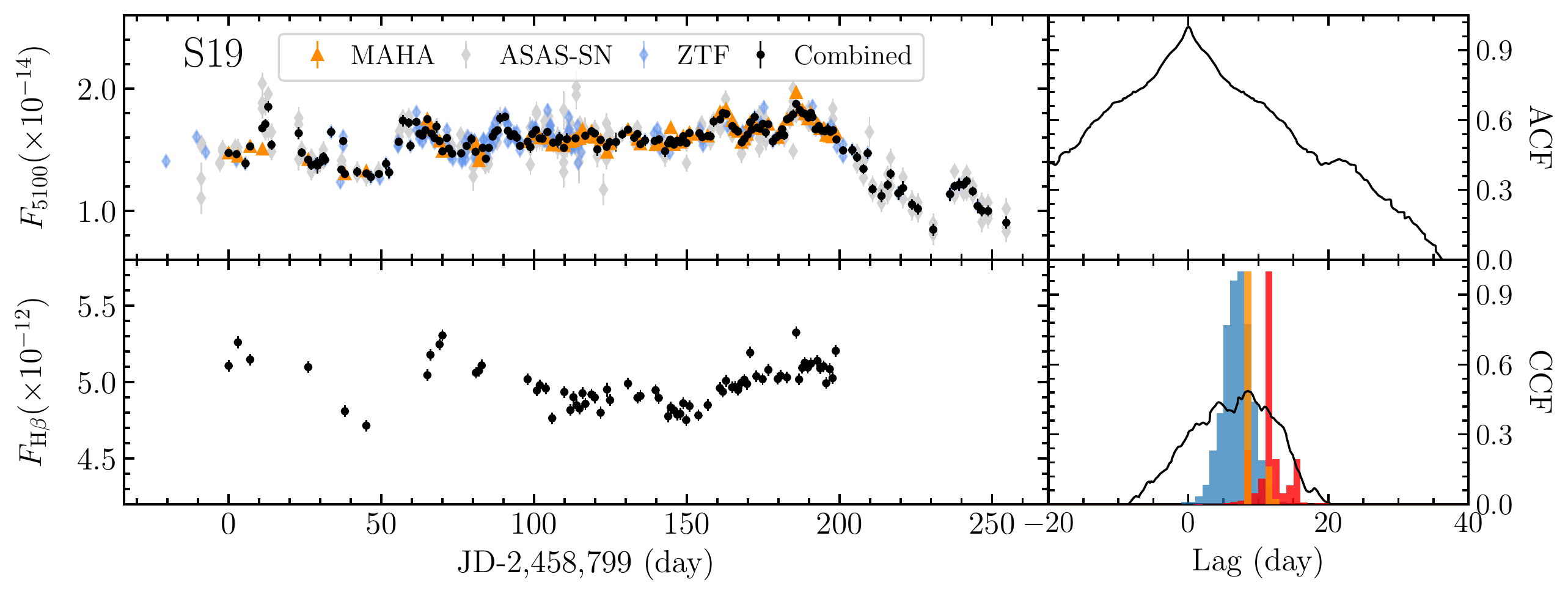}
            \includegraphics[width=0.82\textwidth]{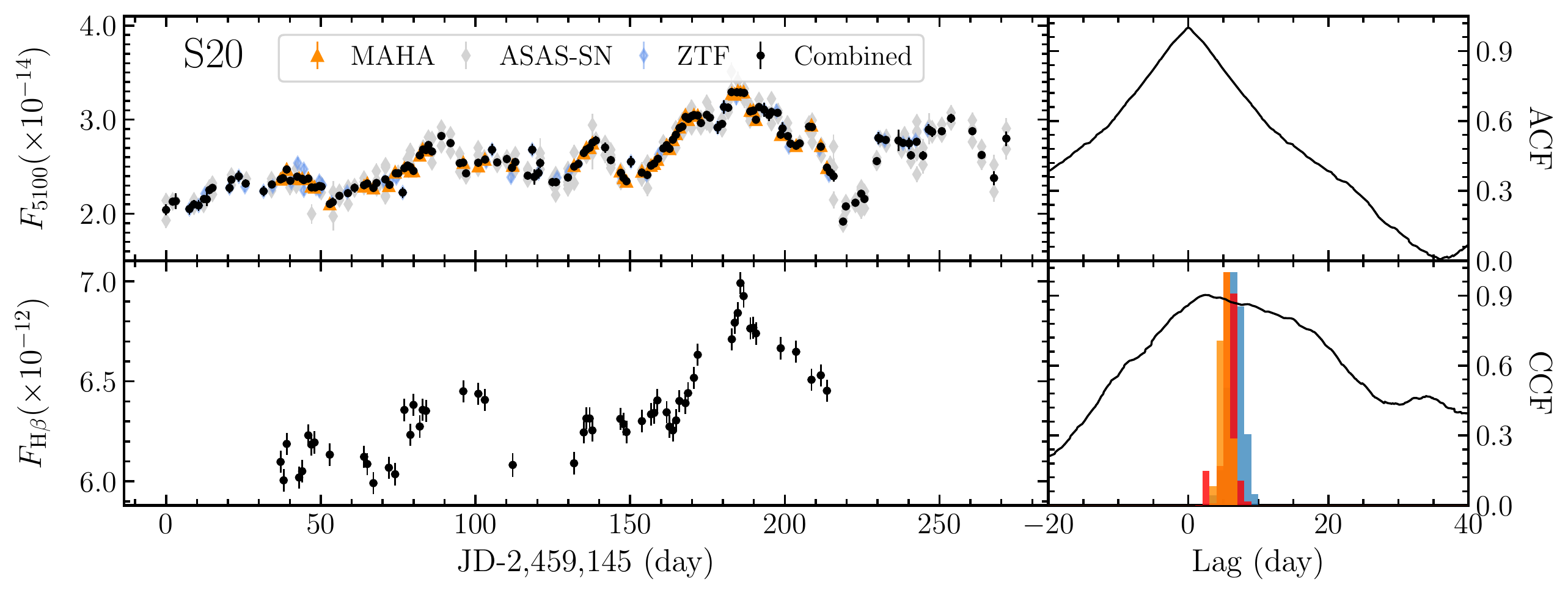}
                \includegraphics[width=0.82\textwidth]{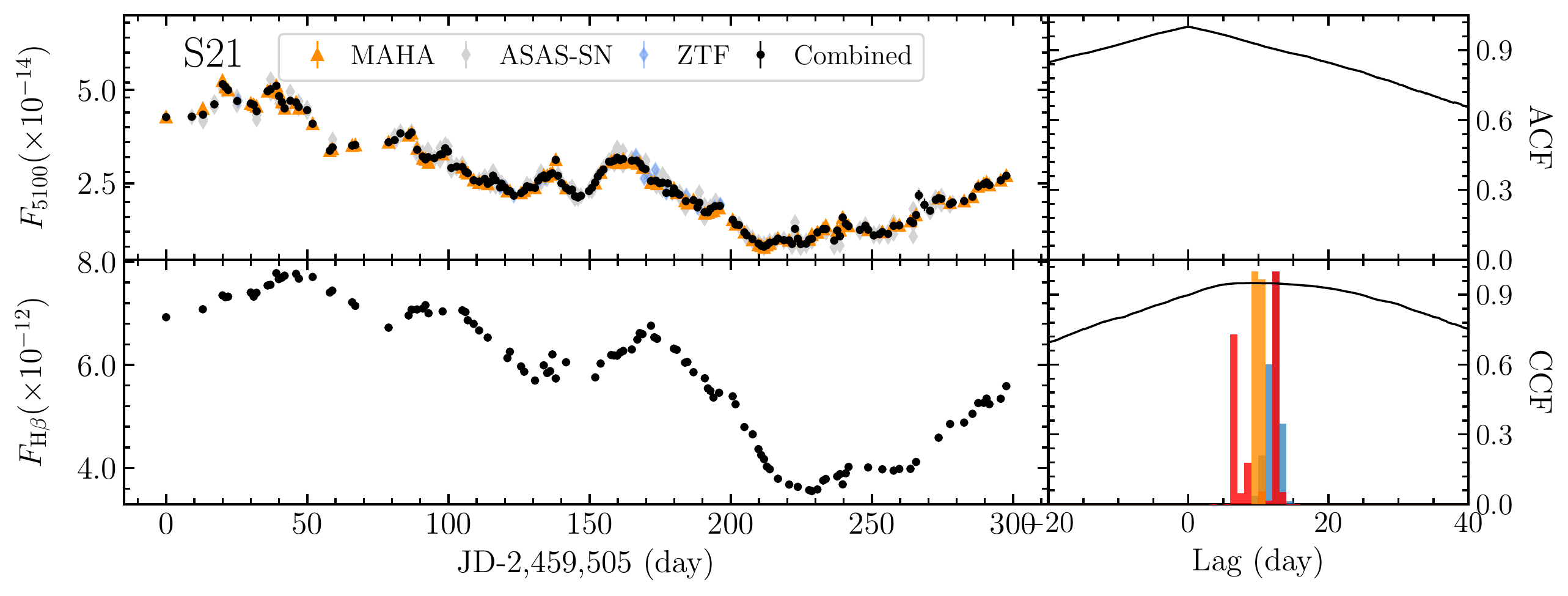}
            \addtocounter{figure}{-1}
            \caption{(Continued).}
            \label{mapping3}
        \end{figure*}
  
	\subsection{Time Lag Measurements}
        \label{sec:lag_measurements}
        As listed in Table~\ref{table:info}, the sampling is sparse and highly uneven for the data in some seasons. Generally, this may cause the measured lags to be over- or under-estimated \citep{Hu2021}. Therefore, the data with uneven sampling or undersampling (the sampling intervals are larger than 7 days) in S1, S7-9 and S12-17 are not included in our following analysis. 
        Furthermore, the photometric light curves usually can be used to improve the sampling and extend the temporal coverage of the continuum light curves. 
        Therefore, for S18-21, we combine the 5100\,{\AA} and the well-sampled ASAS-SN and ZTF light curves by weight averaging all the observations on the same nights. The inter-calibration between the three light curves has been done in Section~\ref{sec:flux_correction}. 
        For other seasons (S2-6 and S9-11), their photometric data have unknown uncertainties and uneven sampling (see the details in Section~\ref{sec:photometry}), thus we do not include them in our following lag measurements. 
        
        We use three different methods to measure the time lag of the H$\beta$ emission line with respect to the flux variations of the continuum: the interpolated cross-correlation function (ICCF; \citealt{Gaskell1986,Gaskell1987,White1994}), the stochastic process modeling algorithm JAVELIN\footnote{\url{https://github.com/nye17/javelin}} \citep{zu2011}, and the Multiple and Inhomogeneous Component Analysis (MICA\footnote{\url{https://github.com/LiyrAstroph/MICA2}}; \citealt{li2016}). 
        The method of ICCF is a very common technique used to calculate the time lag. We measure the time lag by using the centroid of the CCF above 80\% of peak ($r_{\rm max}$). The uncertainties of the time lag are estimated by the 15.87\% and 84.13\% quantiles of the cross-correlation function distribution (CCCD), which is given by the flux randomization and the random subset selection (FR/RSS, \citealt{Peterson1998,peterson2004}). 
        The other two methods, JAVELIN and MICA, both use a damped random walk (DRW) process to rebuild the continuum light curve and infer a specific transfer function. While JAVELIN adopts a top-hat function as the transfer function, MICA employs a sum of Gaussians. In this work, we use one Gaussian in MICA. Thus, the time delays are given at the center of the top-hat/Gaussian function, and their associated uncertainties are estimated from the 15.87\% and 84.13\% quantiles of their corresponding posterior samples generated by Markov Chain Monte Carlo (MCMC). For convenience, we separately mark the time lags measured by ICCF, JAVELIN and MICA as $\tau_{\rm cent}$, $\tau_{\rm JAV}$ and $\tau_{\rm MICA}$.

        The right column of Figure~\ref{mapping1} shows the lag distributions using ICCF, JAVELIN and MICA. The time delays and their corresponding uncertainties are listed in Table~\ref{table:delays_widths_mass} and the comparison between the lag measurements of the three methods is plotted in Figure~\ref{figure:LagMethodComparison}. The lag measurements of the three methods are generally consistent with each other within errors. The light curves are sparse and have uneven sampling in the early part of S9 and S19, while their variability amplitudes in fluxes in the late part are too small (as shown in Figure~\ref{mapping1}). This may cause unreliable time lag measurements. Also, the lower $r_{\rm max}$ ($\leqslant$0.50) in S9 and S19 indicate poor response between the variations of the 5100\,{\AA} continuum and the H$\beta$ emission line ( see a detailed discussion in Appendix~\ref{appedixd}). 
        Finally, we average the time lag measurements of three methods in order to reduce the possible bias caused by their different treatments.
	
	\subsection{Line Width Measurements}
	We list the line widths in Table~\ref{table:delays_widths_mass}.
        For mean and rms spectra, we adopt the following procedures to determine the broad H$\beta$ line widths and their corresponding uncertainties.
        First, for a set of $N$ spectra, we randomly select $N$ spectra and remove the $n$ duplicated ones. The left ($N-n$) spectra are used to construct a new mean and rms spectrum. For the newly constructed mean spectra, we subtract the continuum and the narrow \heii, H$\beta$ and $\rm [\text{\oiii}] \lambda 4959, 5007$ lines by using spectral decomposition. 
        The spectral components include: (1) a single power law for the AGN continuum; (2) double Gaussians for the broad H$\beta$ line; (3) a single Gaussian for the broad {\heii} line; (4) a single Gaussian for the narrow lines. The iron blends are not included because they are weak and degenerate with the AGN continuum components.
        For the newly constructed rms spectra, the narrow lines are very weak as shown in Figure~\ref{figure:mean_rms}. We directly interpolate between two continuum windows to subtract the local background continuum flux of the H$\beta$ line. We measure both FWHM and $\rm \sigma_{line}$ as characteristic parameters of the line width. After that, the above procedures are repeated 1000 times. We utilize the median values and standard deviations of the line width distribution as our measurements and the corresponding uncertainties. Finally, we correct the line widths by subtracting the instrumental broadening contribution in quadrature. 
        The instrumental broadening is estimated by comparing the [\oiii]$\lambda$5007 widths in the mean spectra with the intrinsic one (425 $\rm km\ s^{-1}$ measured by \citealt{Whittle1992}). We obtain the instrumental broadening of $\sim$ 700 $\rm km\ s^{-1}$ and 1200 $\rm km\ s^{-1}$ for the FAST and WIRO spectra, respectively.
        \label{sec:width}

    	\begin{figure*}
    		\centering
    		\includegraphics[width=0.45\textwidth]{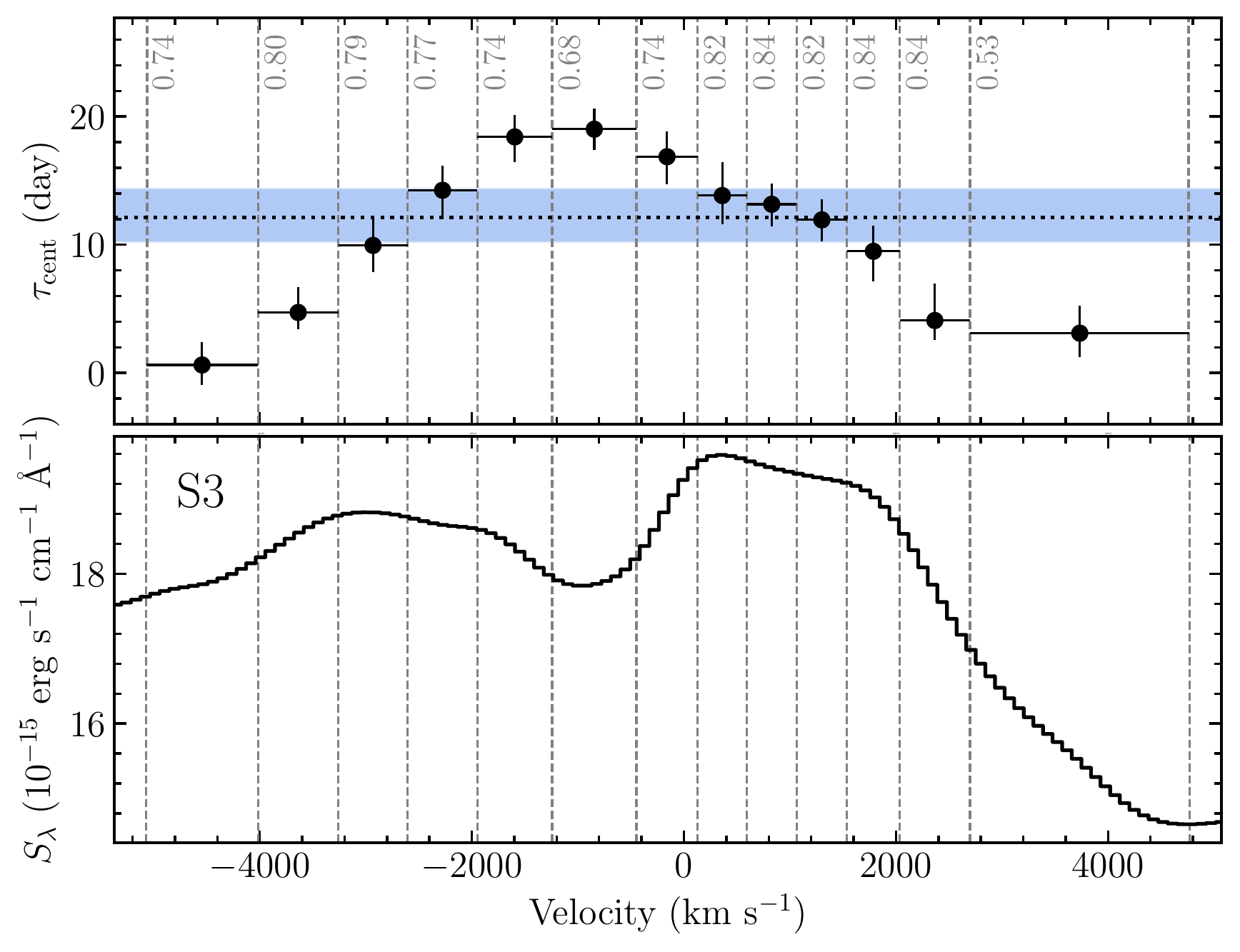}
    		\includegraphics[width=0.45\textwidth]{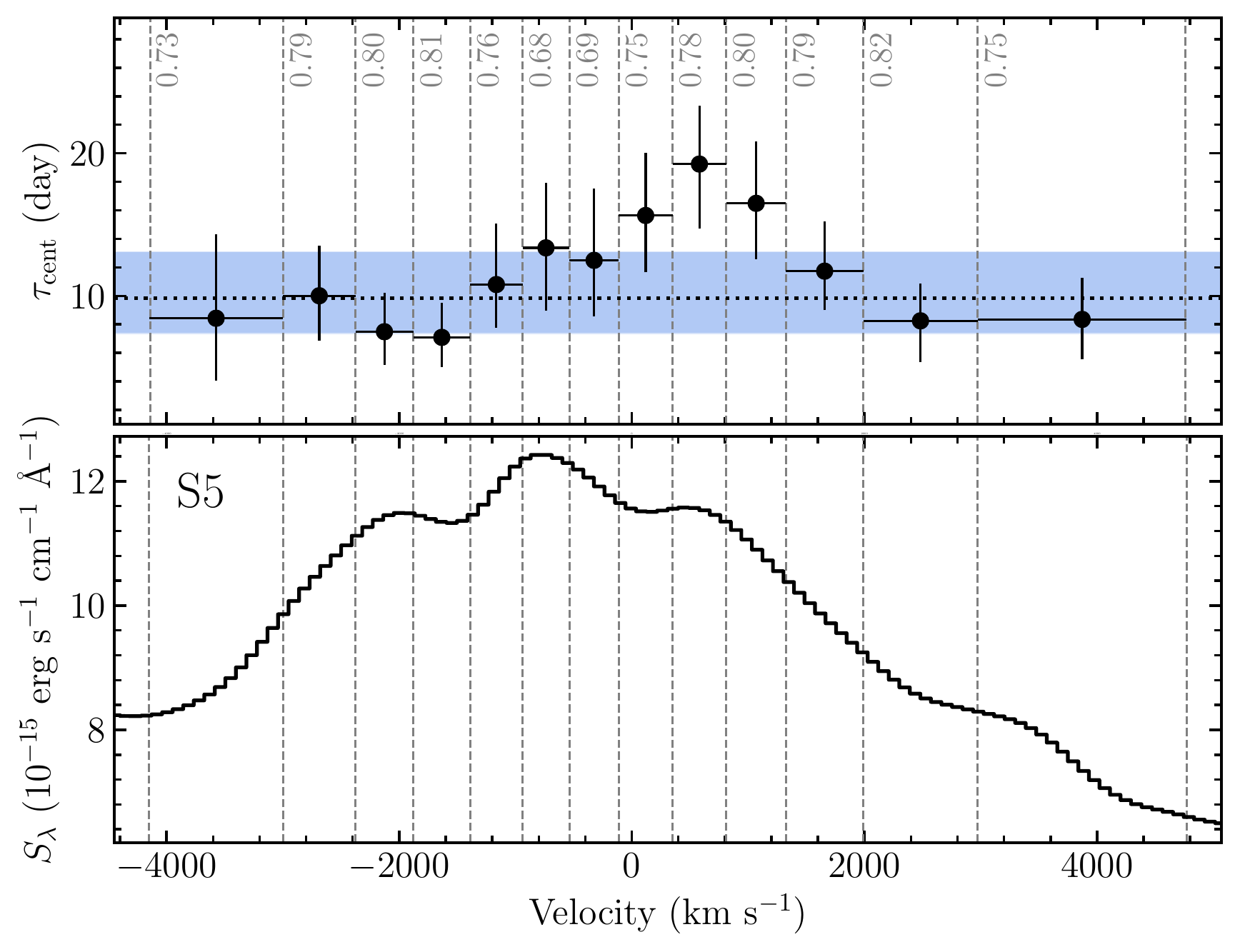}
    		\includegraphics[width=0.45\textwidth]{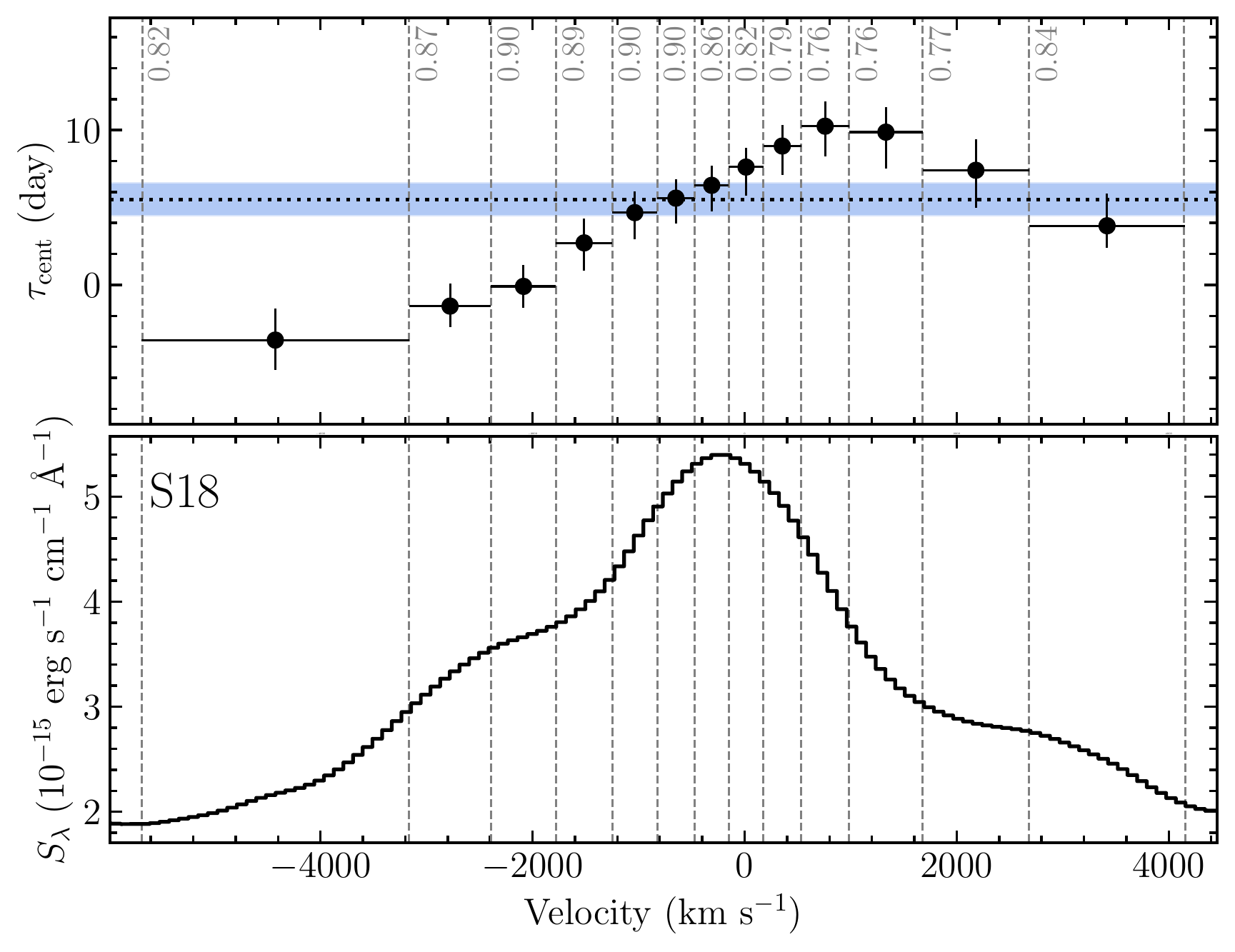}
    		\includegraphics[width=0.45\textwidth]{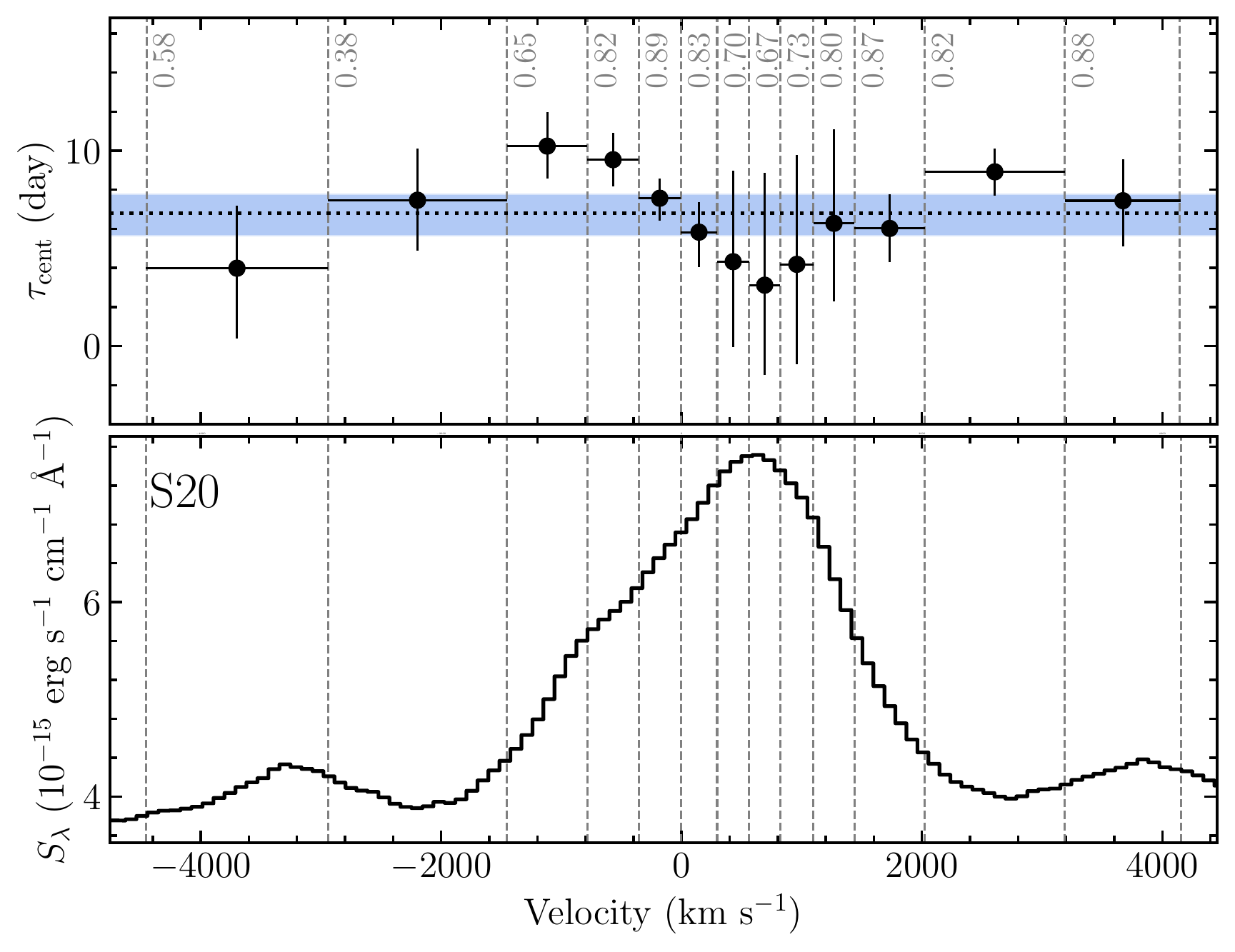}
    		\includegraphics[width=0.45\textwidth]{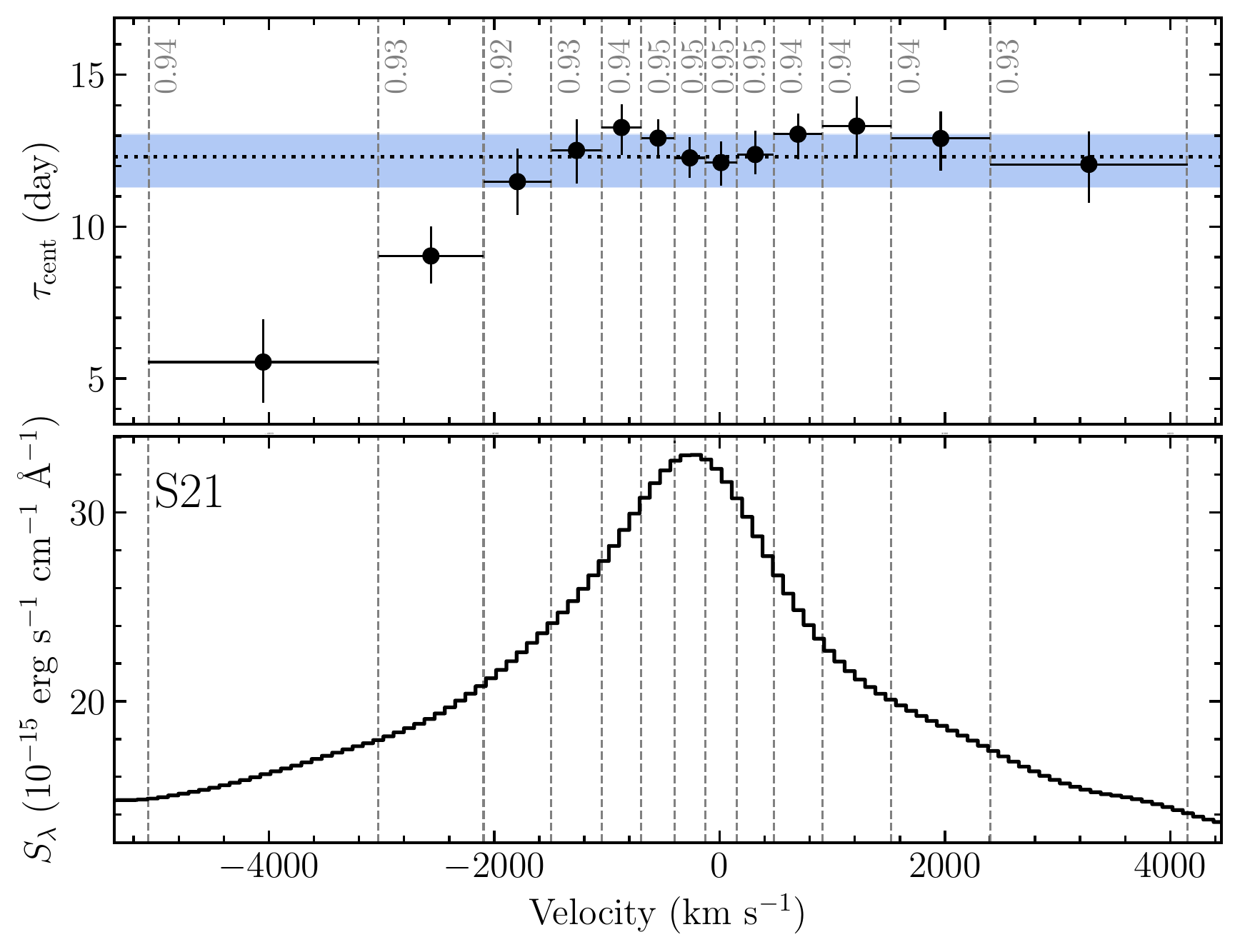}
    		\caption{Velocity-resolved delays in the rest frame and the corresponding rms spectra. The upper panel in each plot shows the centroid lags in different velocity bins and the rms spectrum is shown underneath. The vertical gray dash lines show the velocity bins used to produce velocity-resolved light curves. The vertical number inside each bin is the value of their peak correlation coefficient. The horizontal dotted lines and the blue regions are the H$\beta$ centroid lags and their uncertainties for the total H$\beta$ emission line in Table~\ref{table:delays_widths_mass}.}
    		\label{figure:velocity_resolved}
    	\end{figure*}

        \subsection{Velocity-resolved Delays}
    	\label{sec:velocity_resolved}
    	 The lags of the total flux of broad emission lines with respect to the varying 5100\,{\AA} continuum represent the emissivity-averaged radius of the BLR. For those with high-enough data quality, velocity-resolved delays have been used to investigate the geometry and kinematics of BLR for many AGNs (e.g., \citealt{dupu2018,Brotherton2020,Bao2022}). 
    	
        Higher cadence and more uniform sampling for the spectroscopic data were obtained in S3, S5, S18, S20 and S21. This allows us to recover velocity-resolved delays.
    	We first divide the H$\beta$ lines into several bins, that have equal integrated fluxes measured in the rms spectra. Then, we obtain the light curves of the H$\beta$ emission in each bin by integrating the continuum-subtracted fluxes. We perform the measurements of time lags between the continuum and the line for every bin by employing the ICCF method. The velocity-resolved delays are demonstrated in Figure~\ref{figure:velocity_resolved}. The structures in S3, S5 and S18 are similar while the ones in S20 and S21 display complicated double peaks. This implies the BLR kinematics and/or geometry are likely undergoing significant changes between the five seasons. A further discussion of the kinematics of BLR is described in Section~\ref{sec:kinematics}. 

		\begin{figure*}
			\centering
			\includegraphics[width=0.9\textwidth]{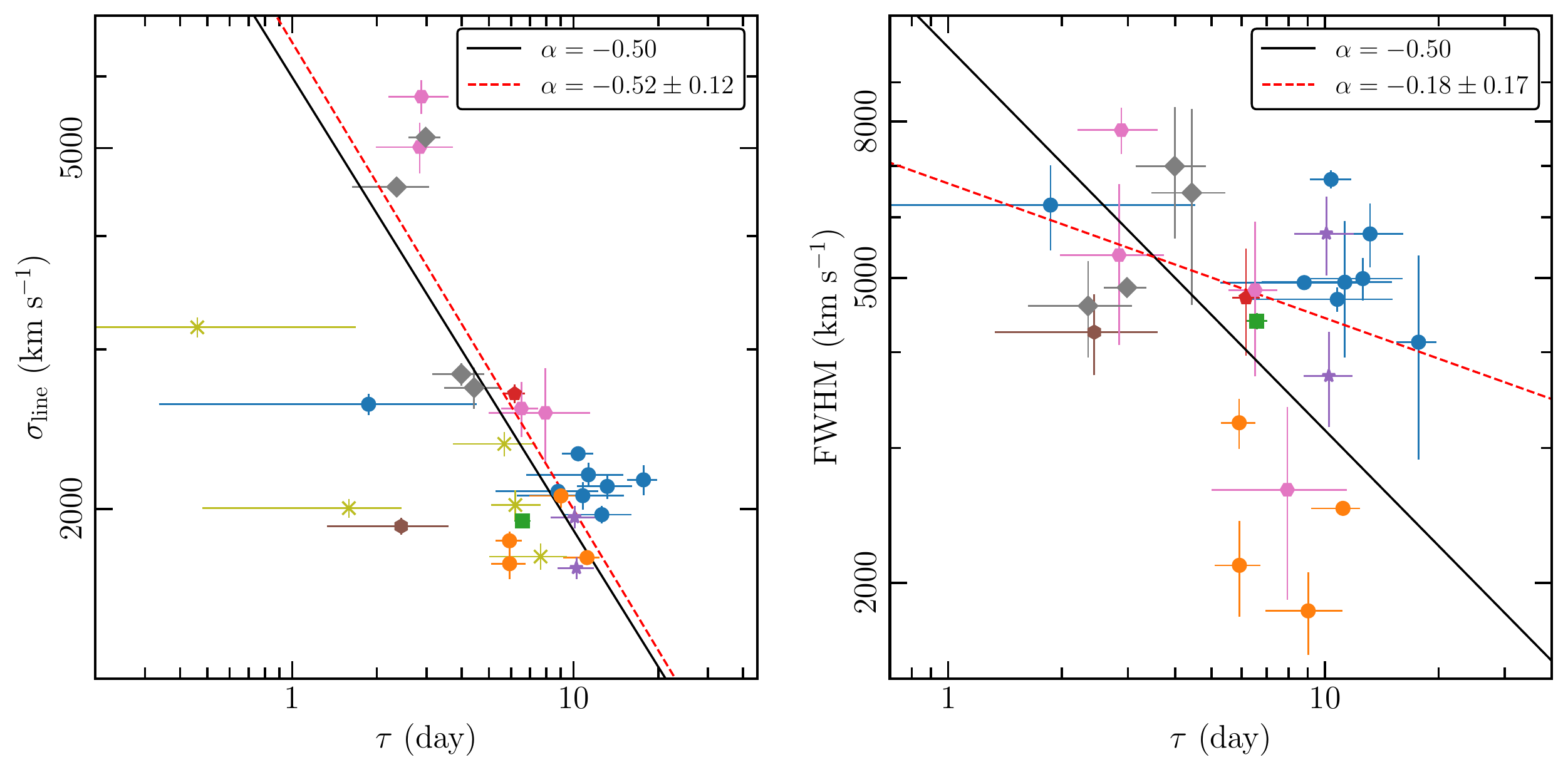}
			\includegraphics[width=0.9\textwidth]{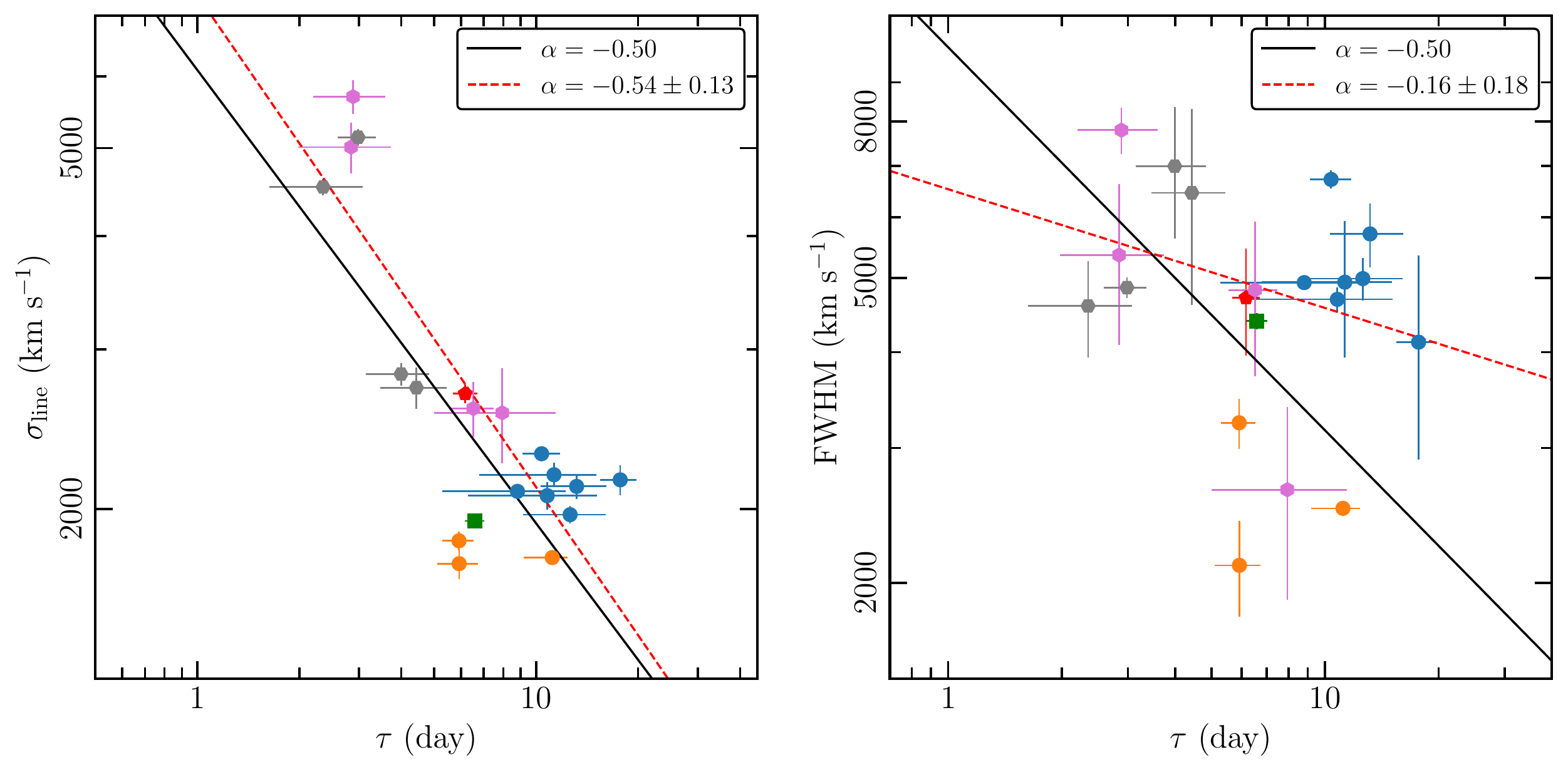}
			\caption{The relationships of line dispersion and FWHM versus H$\beta$ lags for the rms spectra of NGC 4151. The top panels show the fit with all measurements and the bottom panels with the measurements from M91, K96 and L22 removed. The red dashed lines are the best linear fit slopes. The black solid lines have a fixed slope of $-$0.5. The grey rhombus and pink points are the results for the UV emission lines from U96 and C90. The other points are the results for the optical lines from M91 (purple star), K96 (brown diamond), B06 (red pentagon), R18 (green square), L22 (gold asterisks), FAST (blue points) and MAHA (orange points).}
			\label{virial_relationship}
		\end{figure*}

        \subsection{Testing the Virial Relation}
        \label{sec:virialtesting}
       
\begin{figure}
            \centering
            \includegraphics[width=0.47\textwidth]{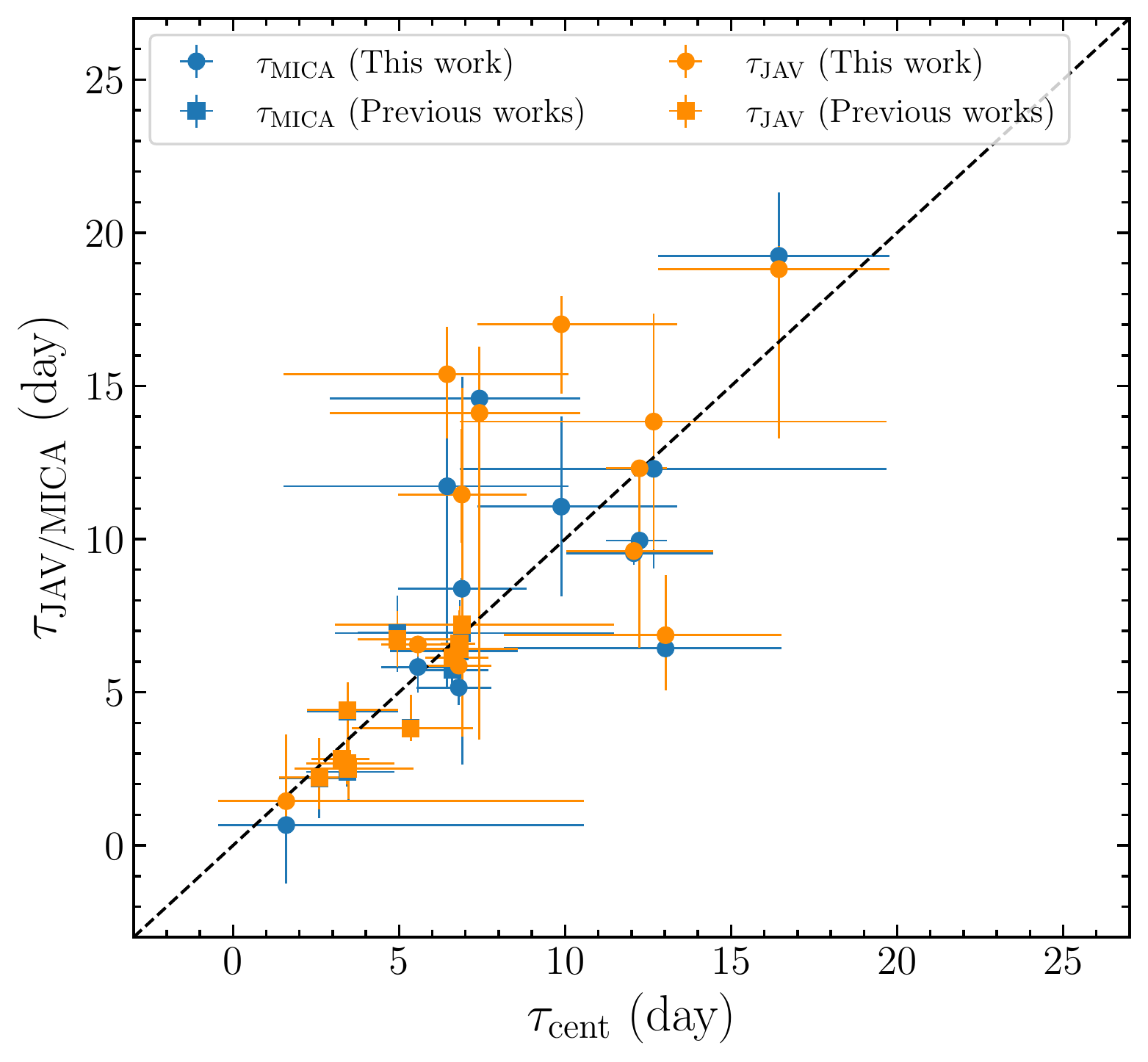}
            \caption{Comparison between the lag measurements of three methods.}
            \label{figure:LagMethodComparison}
        \end{figure}

        As found in previous investigations for the other AGNs (e.g., \citealt{peterson2000,peterson2004,lu2016,lu2021,Lu2022}), the dynamics of the BLR is dominated by the gravitational potential of the central SMBH (i.e., the virial relation between the widths of broad emission lines and the time lags follow $\Delta V\propto\tau^{-\frac{1}{2}}$). For NGC 4151, \citet{Li2022} collected the previous RM results and combined their measurements. They found that the observed time lag and velocity basically follow a virial relation.
        In this work, there are twelve new RM measurements, which allow better examination. In order to avoid the possible discrepancy of the time lag measurements caused by different techniques, we reanalyzed the data from \citet{maoz1991}, \citet{kaspi1996}, \citet{ulrich1996}, \citet{Metzroth2006}, \citet{bentz2006} and \citet{rosa2018} using JAVELIN and MICA. As shown in Figure~\ref{figure:LagMethodComparison}, the lags measured by JAVELIN and MICA are generally consistent with those measured by ICCF in previous works within the errors. Then, we averaged the time lags of the three methods, as we did for our new measurements in Section~\ref{sec:lag_measurements} above. We list the results of reanalysis and the line widths for all available campaigns in Table~\ref{table:delays_widths_mass}. 
        In previous RM campaigns, most have used the measurements of line dispersion and FWHM from rms spectra, we thus similarly use line dispersions and FWHM in combination with the time lags to verify the BLR virial relation, respectively.

	    The relations between the time lags and the velocity widths for all datasets are plotted in Figure~\ref{virial_relationship} (the upper panels). We fit the relation $\rm \log\Delta V = \alpha\log\tau+\beta$ by employing \texttt{linmix} algorithm\footnote{\url{https://github.com/jmeyers314/linmix}} \citep{kelly2007}, which adopts a hierarchical Bayesian approach to linear regression with measurement errors in both X and Y and a component of intrinsic random scatter. For the relation of the time lag versus the line dispersion, the best-fit slope $\alpha=-0.52\pm0.12$ indicates that the BLR follows virial motion. For the relation between the time lag versus FWHM, the best fitting slope of $\alpha=-0.18\pm0.17$ significantly deviates from the expected slope of $\alpha=-0.5$. This deviation may be caused by the uncertainty of the FWHM measurements due to the complicated H$\beta$ profiles (see Figure~\ref{figure:mean_rms}).
	    
	    There are five datasets that may have some problems. In the cases of M91 and L22, the measurements for the line widths are highly uncertain, because of the lower spectral resolution in the former \citep{Metzroth2006} and the impact from the residuals of narrow lines in the latter \citep{Li2022}. For the cases of K96, S9 and S19, the time lags have lower confidence due to the relatively low amplitude of variation for both the continuum and H$\beta$ emission line or the lower $r_{\rm max}$ (see \citealt{Metzroth2006} and Section~\ref{sec:lag_measurements} for more details). We therefore discard these five datasets and check the virial relation again. As shown in Figure~\ref{virial_relationship} (the bottom panels), the best-fit slope for the relation between the lag and the line dispersion is $\alpha=-0.54\pm0.13$, which also is consistent with $\alpha=-0.5$.

    \begin{landscape}
        \begin{table}
            \renewcommand{\arraystretch}{1.3}
            \setlength{\tabcolsep}{7.5pt}
            \caption{Time Lags (in the Rest Frame), Line Widths and BH Masses from All Available Campaigns}
            \label{table:delays_widths_mass}
            \begin{tabular}{lcccccccccccccccc}
            \hline
              & & & & & & & \multicolumn{2}{c}{RMS Spectra} & & \multicolumn{2}{c}{Mean Spectra} & \\
              \cline{8-9}
		   \cline{11-12}
		   Campaign &Line & $r_{\rm max}$ & $\tau_{\rm cent}$ & $\tau_{\rm JAV}$ & $\tau_{\rm MICA}$ & $\tau_{\rm Mean}$ & $\sigma_{\rm line}$ & FWHM & & $\sigma_{\rm line}$ & FWHM & VP $^a$ & $M_{\bullet}$ $^b$ & Ref. &\\
		& &  & (day) & (day) & (day) & (day) & ($\rm km\ s^{-1}$) & ($\rm km\ s^{-1}$) & & ($\rm km\ s^{-1}$) & ($\rm km\ s^{-1}$) & ($\times 10^7\rm M_{\odot}$) & ($\times 10^7\rm M_{\odot}$) & \\
             \hline
             \multicolumn{4}{l}{The FAST Spectrograph Publicly Archived Programs}\\
             \hline
             S2 & H$\beta$ & 0.67 & $6.4_{-4.9}^{+3.9}$ & $15.4_{-2.1}^{+1.6}$ & $11.7_{-6.5}^{+4.2}$ & $10.8_{-4.5}^{+4.3}$& 2068 $\pm$ 72 & 4689 $\pm$ 170&& 2315 $\pm$ 3 & 6630 $\pm$ 29&$ 0.90_{-0.38}^{+0.37}$& $5.67_{-2.75}^{+2.68}$ & 1 \\
            S3 & H$\beta$ & 0.78 & $11.9_{-1.8}^{+2.4}$ & $9.6_{-0.3}^{+0.3}$ & $9.5_{-0.4}^{+0.4}$ & $10.4_{-1.3}^{+1.4}$& 2300 $\pm$ 31 & 6723 $\pm$ 184&& 2307 $\pm$ 2 & 6360 $\pm$ 21&$ 1.07_{-0.13}^{+0.14}$& $6.75_{-1.82}^{+1.85}$ & 1 \\
            S4 & H$\beta$ & 0.67 & $16.6_{-3.8}^{+3.8}$ & $18.8_{-5.5}^{+0.8}$ & $19.2_{-2.0}^{+2.1}$ & $17.7_{-2.3}^{+2.1}$& 2153 $\pm$ 81 & 4122 $\pm$ 1228&& 2093 $\pm$ 2 & 5741 $\pm$ 21&$ 1.60_{-0.24}^{+0.22}$& $10.10_{-2.83}^{+2.78}$ & 1 \\
            S5 & H$\beta$ & 0.80 & $9.9_{-2.7}^{+3.6}$ & $17.0_{-2.3}^{+0.9}$ & $11.1_{-2.9}^{+2.9}$ & $12.6_{-3.5}^{+3.5}$& 1971 $\pm$ 44 & 4990 $\pm$ 315&& 2062 $\pm$ 3 & 6069 $\pm$ 22&$ 0.96_{-0.27}^{+0.27}$& $6.02_{-2.20}^{+2.21}$ & 1 \\
            S6 & H$\beta$ & 0.80 & $7.6_{-4.6}^{+3.0}$ & $14.1_{-10.7}^{+2.2}$ & $14.6_{-1.0}^{+1.1}$ & $11.3_{-4.5}^{+3.7}$& 2181 $\pm$ 67 & 4936 $\pm$ 1001&& 2042 $\pm$ 4 & 5714 $\pm$ 21&$ 1.05_{-0.42}^{+0.35}$& $6.61_{-3.10}^{+2.73}$ & 1 \\
            S9$^c$ & H$\beta$ & 0.50 & $1.7_{-2.1}^{+8.7}$ & $1.4_{-1.0}^{+2.2}$ & $0.7_{-1.9}^{+2.2}$ & $1.9_{-1.5}^{+2.7}$& 2608 $\pm$ 70 & 6224 $\pm$ 793&& 2600 $\pm$ 6 & 7352 $\pm$ 34&$ 0.25_{-0.20}^{+0.35}$& $1.56_{-1.34}^{+2.26}$ & 1 \\
            S10 & H$\beta$ & 0.68 & $13.0_{-6.3}^{+6.6}$ & $13.8_{-4.2}^{+3.5}$ & $12.3_{-3.3}^{+4.7}$ & $13.2_{-2.9}^{+2.9}$& 2117 $\pm$ 66 & 5707 $\pm$ 543&& 2187 $\pm$ 4 & 6224 $\pm$ 24&$ 1.15_{-0.26}^{+0.27}$& $7.26_{-2.39}^{+2.41}$ & 1 \\
            S11 & H$\beta$ & 0.94 & $13.2_{-4.9}^{+3.3}$ & $6.9_{-1.8}^{+2.0}$ & $6.4_{-1.3}^{+1.7}$ & $8.8_{-3.5}^{+3.4}$& 2091 $\pm$ 14 & 4930 $\pm$ 69&& 2409 $\pm$ 10 & 6118 $\pm$ 42&$ 0.75_{-0.30}^{+0.29}$& $4.74_{-2.21}^{+2.16}$ & 1 \\
             \hline
             \multicolumn{4}{l}{MAHA Program}\\
             \hline
             S18 & H$\beta$ & 0.89 & $5.5_{-1.0}^{+1.1}$ & $6.6_{-0.4}^{+0.1}$ & $5.8_{-0.8}^{+0.8}$ & $5.9_{-0.6}^{+0.6}$& 1844 $\pm$ 44 & 3235 $\pm$ 243&& 2030 $\pm$ 2 & 5366 $\pm$ 18&$ 0.39_{-0.05}^{+0.05}$& $2.48_{-0.66}^{+0.66}$ & 1 \\
            S19$^c$ & H$\beta$ & 0.49 & $7.0_{-1.9}^{+1.9}$ & $11.5_{-1.6}^{+2.2}$ & $8.4_{-0.2}^{+0.3}$ & $9.0_{-2.1}^{+2.1}$& 2067 $\pm$ 61 & 1838 $\pm$ 227&& 1973 $\pm$ 1 & 5191 $\pm$ 12&$ 0.75_{-0.18}^{+0.18}$& $4.74_{-1.59}^{+1.60}$ & 1 \\
            S20 & H$\beta$ & 0.90 & $6.8_{-1.2}^{+0.9}$ & $5.9_{-0.8}^{+0.7}$ & $5.2_{-0.6}^{+0.8}$ & $5.9_{-0.8}^{+0.8}$& 1740 $\pm$ 66 & 2108 $\pm$ 302&& 1794 $\pm$ 2 & 4677 $\pm$ 19&$ 0.35_{-0.06}^{+0.06}$& $2.21_{-0.63}^{+0.63}$ & 1 \\
            S21 & H$\beta$ & 0.95 & $12.3_{-1.0}^{+0.8}$ & $12.3_{-5.9}^{+0.1}$ & $10.0_{-0.3}^{+0.3}$ & $11.2_{-2.0}^{+1.2}$& 1767 $\pm$ 10 & 2501 $\pm$ 53&& 1963 $\pm$ 2 & 5072 $\pm$ 17&$ 0.68_{-0.12}^{+0.08}$& $4.29_{-1.27}^{+1.13}$ & 1 \\
            \hline
             \multicolumn{4}{l}{Previous Campaigns}\\
             \hline
             C90 & \civ & 0.88 & $3.4_{-1.2}^{+1.4}$  & $2.7_{-0.7}^{+0.8}$  & $2.4_{-0.5}^{+0.6}$  & $2.9_{-0.7}^{+0.7}$  & 5698 $\pm$ 245 & 7797 $\pm$ 543 && ··· & ··· & $ 1.83_{-0.46}^{+0.48}$& $11.51_{-3.98}^{+4.07}$ & 2, 3 \\
            C90 & \heii$^d$ & 0.82 & $3.5_{-1.6}^{+2.0}$  & $2.5_{-1.0}^{+1.0}$  & $2.5_{-1.0}^{+0.9}$  & $2.8_{-0.9}^{+0.9}$  & 5013 $\pm$ 323 & 5356 $\pm$ 1270 && ··· & ··· & $ 1.39_{-0.46}^{+0.47}$& $8.78_{-3.57}^{+3.64}$ & 2, 3 \\
            C90 & \ciii] & 0.71 & $6.9_{-3.8}^{+4.6}$  & $7.2_{-3.7}^{+7.7}$  & $6.9_{-4.3}^{+8.4}$  & $7.9_{-2.9}^{+3.5}$  & 2553 $\pm$ 307 & 2646 $\pm$ 745 && ··· & ··· & $ 1.01_{-0.45}^{+0.51}$& $6.37_{-3.20}^{+3.53}$ & 2, 3 \\
            C90 & \mgii & 0.88 & $6.8_{-2.1}^{+1.7}$  & $6.4_{-1.4}^{+1.4}$  & $6.3_{-1.5}^{+1.7}$  & $6.5_{-1.0}^{+1.0}$  & 2581 $\pm$ 179 & 4823 $\pm$ 1105 && ··· & ··· & $ 0.85_{-0.17}^{+0.17}$& $5.35_{-1.68}^{+1.67}$ & 2, 3 \\
            M91$^e$  & H$\beta$ & 0.77 & $11.5_{-3.7}^{+3.7}$  & $9.3_{-2.0}^{+2.0}$  & $9.3_{-1.4}^{+1.7}$  & $10.1_{-1.8}^{+1.8}$  & 1958 $\pm$ 56 & 5713 $\pm$ 675 && ··· & ··· & $ 0.75_{-0.14}^{+0.14}$& $4.76_{-1.44}^{+1.45}$ & 4, 5 \\
            M91$^e$  & H$\alpha$ & 0.75 & $11.0_{-3.1}^{+4.1}$  & $9.8_{-1.8}^{+1.6}$  & $9.9_{-1.9}^{+1.4}$  & $10.2_{-1.5}^{+1.6}$  & 1721 $\pm$ 47 & 3724 $\pm$ 529 && ··· & ··· & $ 0.59_{-0.09}^{+0.10}$& $3.73_{-1.06}^{+1.08}$ & 4, 5 \\
            U96 & \civ & 0.97 & $3.3_{-0.9}^{+0.8}$  & $2.8_{-0.3}^{+0.3}$  & $2.8_{-0.3}^{+0.3}$  & $3.0_{-0.4}^{+0.4}$  & 5140 $\pm$ 113 & 4858 $\pm$ 149 && ··· & ··· & $ 1.54_{-0.21}^{+0.21}$& $9.69_{-2.67}^{+2.65}$ & 3, 6 \\
            U96 & \heii$^d$ & 0.93 & $2.6_{-1.2}^{+1.1}$  & $2.2_{-1.1}^{+1.3}$  & $2.2_{-1.3}^{+1.3}$  & $2.4_{-0.7}^{+0.7}$  & 4530 $\pm$ 92 & 4597 $\pm$ 659 && ··· & ··· & $ 0.94_{-0.29}^{+0.29}$& $5.94_{-2.32}^{+2.32}$ & 3, 6 \\
            U96 & \ciii] & 0.88 & $3.5_{-1.2}^{+1.5}$  & $4.4_{-1.4}^{+0.9}$  & $4.4_{-1.4}^{+0.9}$  & $4.0_{-0.8}^{+0.8}$  & 2817 $\pm$ 81 & 6997 $\pm$ 1366 && ··· & ··· & $ 0.62_{-0.14}^{+0.13}$ & $3.90_{-1.27}^{+1.25}$ & 3, 6 \\
            U96 & \mgii & 0.93 & $5.3_{-1.8}^{+1.9}$  & $3.8_{-0.4}^{+1.1}$  & $3.8_{-0.4}^{+0.4}$  & $4.4_{-1.0}^{+1.0}$  & 2721 $\pm$ 141 & 6458 $\pm$ 1850 && ··· & ··· & $ 0.64_{-0.16}^{+0.16}$ & $4.04_{-1.37}^{+1.39}$ & 3, 6 \\
            K96$^f$ & H$\beta$ & 0.84 & $3.1_{-1.3}^{+1.3}$  & $2.7_{-0.3}^{+0.9}$  & $1.0_{-0.7}^{+1.7}$  & $2.4_{-1.1}^{+1.1}$  & 1914 $\pm$ 42 & 4248 $\pm$ 516 && ··· & ··· & $ 0.17_{-0.08}^{+0.08}$ & $1.10_{-0.57}^{+0.58}$ & 5, 7 \\
            B06 & H$\beta$ & 0.94 & $6.6_{-0.8}^{+1.1}$  & $6.1_{-0.4}^{+0.4}$  & $5.7_{-0.4}^{+0.4}$  & $6.2_{-0.5}^{+0.5}$  & 2680 $\pm$ 64 & 4711 $\pm$ 750 && ··· & ··· & $ 0.87_{-0.08}^{+0.09}$ & $5.45_{-1.40}^{+1.40}$ & 8 \\
            R18 & H$\beta$ & 0.94 & $6.8_{-0.6}^{+0.5}$  & $6.6_{-1.0}^{+1.1}$  & $6.3_{-0.2}^{+0.2}$  & $6.6_{-0.4}^{+0.4}$  & 1940 $\pm$ 22 & 4393 $\pm$ 110 && 2078 $\pm$ 2 & 5174 $\pm$ 32 & $0.48_{-0.03}^{+0.03}$ & $3.05_{-0.76}^{+0.76}$ & 9 \\
             \hline
            \end{tabular}
        \end{table}
        \end{landscape}

        \begin{landscape}
        \begin{table}
        \renewcommand{\arraystretch}{1.3}
        \setlength{\tabcolsep}{9.5pt}
            \contcaption{Time Lags (in the Rest Frame), Line Widths and BH Masses from All Available Campaigns}
            \label{tab:continued}
            \begin{tabular}{lcccccccccccccccc}
            \hline
              & & & & & & & \multicolumn{2}{c}{RMS Spectra} & & \multicolumn{2}{c}{Mean Spectra} & \\
              \cline{8-9}
		   \cline{11-12}
		   Campaign &Line & $r_{\rm max}$ & $\tau_{\rm cent}$ & $\tau_{\rm JAV}$ & $\tau_{\rm MICA}$ & $\tau_{\rm Mean}$ & $\sigma_{\rm line}$ & FWHM & & $\sigma_{\rm line}$ & FWHM & VP $^a$ & $M_{\bullet}$ $^b$ & Ref. &\\
		& &  & (day) & (day) & (day) & (day) & ($\rm km\ s^{-1}$) & ($\rm km\ s^{-1}$) & & ($\rm km\ s^{-1}$) & ($\rm km\ s^{-1}$) & ($\times 10^7\rm M_{\odot}$) & ($\times 10^7\rm M_{\odot}$) & \\
             \hline
            L22 & H$\gamma$ & 0.81 & $4.9_{-2.6}^{+2.0}$  & $6.2_{-1.3}^{+1.3}$  & $6.0_{-1.9}^{+1.7}$  & $5.7_{-1.9}^{+1.6}$  & 2359 $\pm$ 73$^g$ & ··· && 2011 $\pm$ 8 & 4799 $\pm$ 20 & $ 0.62_{-0.21}^{+0.18}$ &  $3.88_{-1.64}^{+1.47}$ & 10 \\
            L22 & \heii & 0.90 & $0.2_{-1.4}^{+1.8}$  & $0.6_{-0.8}^{+1.0}$  & $0.6_{-0.9}^{+0.9}$  & $0.5_{-1.1}^{+1.2}$  & 3172 $\pm$ 82$^g$ & ··· && 2375 $\pm$ 6 & 7099 $\pm$ 52 & $ 0.09_{-0.21}^{+0.24}$ &  $0.57_{-1.32}^{+1.52}$ & 10 \\
            L22 & H$\beta$ & 0.88 & $5.0_{-1.2}^{+2.1}$  & $6.7_{-0.9}^{+0.9}$  & $6.9_{-1.3}^{+1.2}$  & $6.2_{-1.1}^{+1.4}$  & 2020 $\pm$ 76$^g$ & ··· && 2074 $\pm$ 5 & 5003 $\pm$ 13 & $ 0.50_{-0.10}^{+0.12}$ &  $3.12_{-0.96}^{+1.05}$ & 10 \\
            L22 & \hei & 0.91 & $1.5_{-1.5}^{+0.9}$  & $1.8_{-0.8}^{+0.7}$  & $1.5_{-1.0}^{+1.0}$  & $1.6_{-1.1}^{+0.9}$  & 2004 $\pm$ 48$^g$ & ··· && 2340 $\pm$ 11 & 6031 $\pm$ 50 & $ 0.12_{-0.09}^{+0.07}$ &  $0.79_{-0.58}^{+0.47}$ & 10 \\
            L22 & H$\alpha$ & 0.81 & $5.0_{-3.8}^{+0.8}$  & $8.6_{-1.6}^{+1.8}$  & $9.3_{-2.5}^{+2.9}$  & $7.6_{-2.6}^{+1.9}$  & 1772 $\pm$ 59$^g$ & ··· && 2069 $\pm$ 2 & 4897 $\pm$ 5 & $ 0.47_{-0.16}^{+0.12}$ &  $2.95_{-1.25}^{+1.02}$ & 10 \\
             \hline
            \end{tabular}
            \begin{list}{}{}
            \item[$^a$]{The virial products are calculated by using $\sigma_{\rm line}$ and the mean time lags.}
            \item[$^b$]{The BH masses are estimated adopting a viral factor of $f=6.3\pm1.5$~\citep{ho2014} for all RM campaigns.}
            \item[$^c$]{The lower $r_{\rm max}$ indicate the time lags is unreliable (the details are described in Section~\ref{sec:lag_measurements}).}
            \item[$^d$]{\heii~ refers to \heii$\lambda1640$ line.}
            \item[$^e$]{The line widths of the broad line may be unreliable (see more details in \citealt{Metzroth2006}).}
            \item[$^f$]{In this RM campaign, the weak response between the 5100~{\AA} continuum and the H$\alpha$ emission (even after subtracting a monotonic trending) makes the time lag unreliable, we therefore use only the measurement results of H$\beta$ line.}
            \item[Ref:] {(1) This work, (2)~\cite{clavel1990}, (3)~\cite{Metzroth2006}, (4)~\cite{maoz1991}, (5)~\cite{peterson2004}, (6)~\cite{ulrich1996}, (7)~\cite{kaspi1996}, (8)~\citet{bentz2006}, (9)~\citet{rosa2018}, (10)~\cite{Li2022}.}
            \end{list}
        \end{table}
        \end{landscape}

		\subsection{Black Hole Mass and Accretion Rates}
		\label{sec:bhmass}
		BH mass is estimated as $M_{\bullet}$=$f\rm \times VP$, where $f$ is the so-called virial factor and VP is the virial product. Adopting the line width and time lag, we determined the virial product
		\begin{equation}
			{\rm VP} = \frac{\ c\tau_{\rm  H\beta}\ \times (\Delta V)^2}{G} ,
		\end{equation}
		where $G$ is the gravitational constant, $c$ is the velocity of light, $\Delta V$ is either the FWHM or $\rm \sigma_{H\beta}$ determined by the H$\beta$ broad component in the mean or rms spectra. 
		
		Considering that the relation between the time lag and $\sigma_{\rm line}$ is consistent with the virial relation (Section~\ref{sec:virialtesting}), we use $\sigma_{\rm line}$ as the velocity widths to calculate VP. Table~\ref{table:delays_widths_mass} summarize VPs for the different campaigns. Because the standard deviation of the distribution of VP may represent the  systematic uncertainties of the measurements, we adopt the mean and standard deviation of the VP including previous results of measuring both the H$\beta$ line and UV lines as the final measurement. We obtained $\log\langle \rm VP \rangle=6.91\pm0.17$. In the above calculation, we have removed the possibly problematic measurements from M91, K96, L22, S9 and S19 (Section~\ref{sec:virialtesting}).
		
		For calculating the mass, a mean scaling factor $f\rm=6.3\pm1.5$ \citep{ho2014} is adopted.
		Combining this with the mean VP, we obtain $\log M_{\bullet}\approx 7.70\pm 0.20$. This is in reasonable agreement with the measurements given by stellar dynamical ($6.40 <  \log M_{\bullet} < 7.48$, \citealt{roberts2021}) and gas dynamical modeling ($\log M_{\bullet} \approx 7.56_{-0.31}^{+0.11}$, \citealt{{hicks2008}}). Additionally, \citet{Bentz2022} reanalyzed their RM data from 2006 by modeling the BLR and obtained a BH mass of $\log M_{\bullet} = 7.22_{-0.10}^{+0.11}$, which is similar to our measurement. Combined the values of BH mass obtained by the three methods independent of RM with the mean VP, we infer that the $f$-factors of about $1.1\pm1.4$, $4.5\pm2.8$, $2.0\pm1.0$ for the stellar/gas dynamical and BLR modeling, respectively. All of them are smaller than the scaling factor of $f\rm=6.3\pm1.5$, which is given by the $M_{\bullet}-\sigma_{\ast}$ relationship. 
		
		Based on the model of the standard accretion disk \citep{Shakura1973}, the dimensionless accretion rate defined as $\dot{\mathscr{M}}=\dot{M}_{\bullet}/L_{Edd}c^{-2}$, with the accretion rate $\dot{M}_{\bullet}$, the Eddington luminosity $L_{Edd}$ and the velocity of light $c$. Given the 5100~{\AA} luminosity and the BH mass, the $\dot{\mathscr{M}}$ can be estimated as \citep{du2016}
		\begin{equation}
			\dot{\mathscr{M}} = 20.1 \left(\frac{\ell_{44}}{\cos i}\right)^{3/2} m^{-2}_7,
		\end{equation}
		where $\ell_{44} = L_{5100} / 10^{44}\ {\rm erg\ s^{-1}}$, $m_7 = M_{\bullet} / 10^7 M_{\odot}$ and $i$ is the inclination of disk (we take $i=23^{\circ}$; \citealt{onken2007}). The average of BH masses ${M}_{\bullet}=(5.1\pm2.3) \times 10^7M_{\odot}$ is adopted. The dimensionless accretion rates are listed in Table~{\ref{table:summary}}.

        \begin{table*}
        \renewcommand{\arraystretch}{1.3}
        \setlength{\tabcolsep}{6.5pt}
        \caption{Summary of H$\beta$ RM Results}
        \label{table:summary}
        \begin{tabular}{lccccccccccc}
        \hline
        Campaign & Duration & $\tau$ & $\sigma_{\rm line}$ & VP & $M_{\bullet}$ & $F_{5100}$$^a$ & $L_{5100}$ & $\log\dot{\mathscr{M}}$$^b$ & Ref. \\
        & (yyyy/mm-yyyy/mm) & (day) & $(\rm km\ s^{-1})$ & $(10^7M_\odot)$ & $(10^7M_\odot)$ & & $(10^{42}\rm\ erg\ s^{-1})$ & &\\
        \hline
        S2 & 1994/12-1995/06 & $10.8_{-4.5}^{+4.3}$ & $2068\pm 72$ & $0.90_{-0.38}^{+0.37}$ &  $5.67_{-2.75}^{+2.68}$ & $7.47\pm0.80$ & $11.37\pm1.22$ & $-1.46\pm0.41$ & 1\\
        S3 & 1995/11-1996/07 & $10.4_{-1.3}^{+1.4}$ & $2300\pm 31$ & $1.07_{-0.13}^{+0.14}$ &  $6.75_{-1.82}^{+1.85}$ & $8.19\pm0.98$ & $12.46\pm1.49$ & $-1.40\pm0.41$ & 1\\
        S4 & 1996/12-1997/07 & $17.7_{-2.3}^{+2.1}$ & $2153\pm 81$ & $1.60_{-0.24}^{+0.22}$ &  $10.10_{-2.83}^{+2.78}$ & $7.07\pm0.45$ & $10.76\pm0.69$ & $-1.50\pm0.40$ & 1\\
        S5 & 1997/11-1998/07 & $12.6_{-3.5}^{+3.5}$ & $1971\pm 44$ & $0.96_{-0.27}^{+0.27}$ &  $6.02_{-2.20}^{+2.21}$ & $6.22\pm0.41$ & $9.47\pm0.63$ & $-1.58\pm0.40$ & 1\\
        S6 & 1998/12-1999/04 & $11.3_{-4.5}^{+3.7}$ & $2181\pm 67$ & $1.05_{-0.42}^{+0.35}$ &  $6.61_{-3.10}^{+2.73}$ & $5.36\pm0.47$ & $8.16\pm0.71$ & $-1.68\pm0.40$ & 1\\
        S9 & 2001/12-2002/07 & $1.9_{-1.5}^{+2.7}$$^c$ & $2608\pm 70$ & $0.25_{-0.20}^{+0.35}$ &  $1.56_{-1.34}^{+2.26}$ & $3.89\pm0.38$ & $5.92\pm0.57$ & $-1.88\pm0.40$ & 1\\
        S10 & 2002/12-2003/07 & $13.2_{-2.9}^{+2.9}$ & $2117\pm 66$ & $1.15_{-0.26}^{+0.27}$ &  $7.26_{-2.39}^{+2.41}$ & $5.22\pm0.52$ & $7.94\pm0.80$ & $-1.69\pm0.41$ & 1\\
        S11 & 2003/11-2004/07 & $8.8_{-3.5}^{+3.4}$ & $2091\pm 14$ & $0.75_{-0.30}^{+0.29}$ &  $4.74_{-2.21}^{+2.16}$ & $3.07\pm0.77$ & $4.67\pm1.18$ & $-2.04\pm0.43$ & 1\\
        S18 & 2018/11-2019/08 & $5.9_{-0.6}^{+0.6}$ & $1844\pm 44$ & $0.39_{-0.05}^{+0.05}$ &  $2.48_{-0.66}^{+0.66}$ & $0.81\pm0.18$ & $1.23\pm0.28$ & $-2.91\pm0.43$ & 1\\
        S19 & 2019/11-2020/05 & $9.0_{-2.1}^{+2.1}$$^c$ & $2067\pm 61$ & $0.75_{-0.18}^{+0.18}$ &  $4.74_{-1.59}^{+1.60}$ & $1.61\pm0.11$ & $2.46\pm0.17$ & $-2.46\pm0.40$ & 1\\
        S20 & 2020/11-2021/05 & $5.9_{-0.8}^{+0.8}$ & $1740\pm 66$ & $0.35_{-0.06}^{+0.06}$ &  $2.21_{-0.63}^{+0.63}$ & $2.66\pm0.31$ & $4.04\pm0.47$ & $-2.13\pm0.41$ & 1\\
        S21 & 2021/10-2022/08 & $11.2_{-2.0}^{+1.2}$ & $1767\pm 10$ & $0.68_{-0.12}^{+0.08}$ &  $4.29_{-1.27}^{+1.13}$ & $2.61\pm1.19$ & $3.97\pm1.81$ & $-2.14\pm0.50$ & 1\\
        M91 & 1987/12-1988/07 & $10.1_{-1.8}^{+1.8}$ & $1958\pm 56$$^d$ & $0.75_{-0.14}^{+0.14}$ &  $4.76_{-1.44}^{+1.45}$ & $2.74\pm0.18$ & $4.17\pm0.27$ & $-2.11\pm0.40$ & 2, 3\\
        K96 & 1993/11-1994/02 & $2.4_{-1.1}^{+1.1}$$^c$ & $1914\pm 42$ & $0.17_{-0.08}^{+0.08}$ &  $1.10_{-0.57}^{+0.58}$ & $6.63\pm0.48$ & $10.09\pm0.73$ & $-1.54\pm0.40$ & 4, 5\\
        B06 & 2005/03-2005/04 & $6.2_{-0.5}^{+0.5}$ & $2680\pm 64$ & $0.87_{-0.08}^{+0.09}$ &  $5.45_{-1.40}^{+1.40}$ & $2.41\pm0.35$ & $3.67\pm0.54$ & $-2.20\pm0.41$ & 6\\
        R18 & 2012/01-2012/05 & $6.6_{-0.4}^{+0.4}$ & $1940\pm 22$ & $0.48_{-0.03}^{+0.03}$ &  $3.05_{-0.76}^{+0.76}$ & $1.96\pm0.42$ & $2.98\pm0.65$ & $-2.33\pm0.42$ & 7\\
        L22 & 2020/11-2021/05 & $6.2_{-1.1}^{+1.4}$ & $2020\pm 76$$^d$ & $0.50_{-0.10}^{+0.12}$ &  $3.12_{-0.96}^{+1.05}$ & $2.80\pm0.30$ & $4.26\pm0.45$ & $-2.10\pm0.41$ & 8\\
        \hline
        \end{tabular}
        \begin{list}{}{}
        \item[$^a$] {$F_{5100}$ is the average flux of the 5100 ~{\AA} continuum in units of $10^{-15}~\rm erg\ s^{-1}\ cm^{-2}\ \text{\AA}^{-1}$ after correction in Section~\ref{sec:flux_correction}.}
        \item[$^b$] {Here we list the dimensionless accretion rates are calculated by using the average of BH masses ${M}_{\bullet}=(5.1\pm2.3) \times 10^7M_{\odot}$.}
        \item[$^c$] {The time lags are unreliable due to the weak response between the continuum and H$\beta$ emissions.}
        \item[$^d$] {The line width measurements are unconvinced for M91 campaigns (see more details in ~\citealt{Metzroth2006}). For L22 campaigns, the rms spectrum still has the narrow-line residuals, the value therefore is unreliable.}
        \item[Ref:] {(1) This work, (2)~\cite{maoz1991}, (3)~\cite{Metzroth2006}, (4)~\cite{kaspi1996}, (5)~\cite{peterson2004}, (6)~\citet{bentz2006}, (7)~\citet{rosa2018}, (8)~\cite{Li2022}}.
        \end{list}
        \end{table*}
        
        \section{Discussion}
		\label{discussions}

		\begin{figure}
			\centering
			\includegraphics[width=0.48\textwidth]{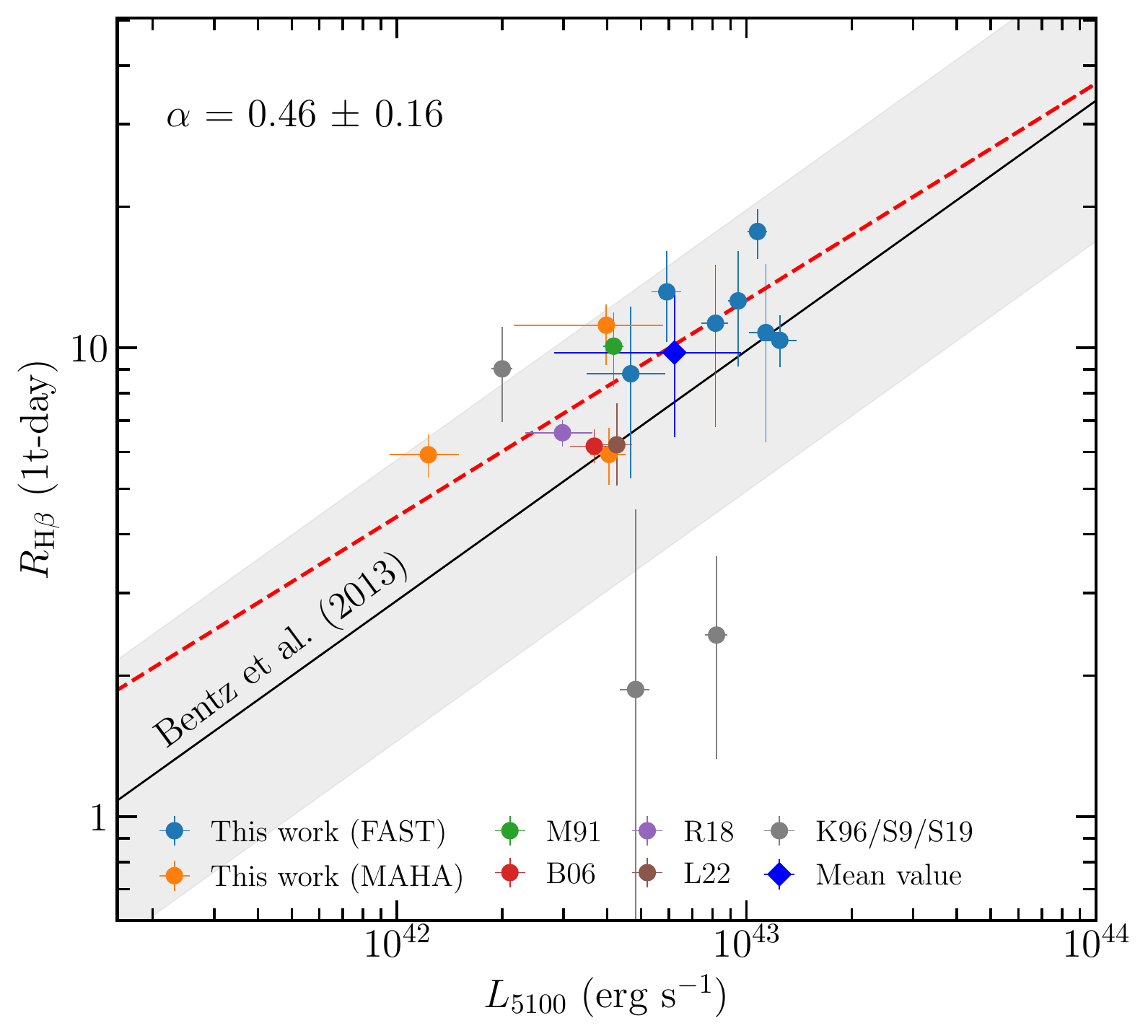}
			\caption{The $R_{\rm H\beta}$-$L_{5100}$ for NGC 4151. The red dashed line is the best-fit regression. The black solid line and the gray region are the traditional $R_{\rm H\beta}$-$L_{5100}$ relationship from \citet{bentz2013} and the corresponding scatter of 0.3 dex. The gray points are the problematic measurements, which we excluded in our fitting. $\alpha$ denotes the slope of our best-fit.}
			\label{figure:rl_relationship}
		\end{figure}

		\subsection{The $R_{\rm H\beta}-L_{5100}$ Relationship of NGC 4151}
		\label{sec:rl relationship}
		We examine the $R_{\rm H\beta}-L_{5100}$ relationship for NGC 4151 combining the measurements from all previous H$\beta$ RM campaigns with our results.
		Table~\ref{table:summary} summarizes the time lags (Section~\ref{sec:virialtesting}), the mean-corrected fluxes (Section~\ref{sec:flux_correction}) and their corresponding luminosities for each campaign. Meanwhile, Figure~\ref{figure:rl_relationship} shows the relation of the time delays versus the luminosities. We exclude the problematic measurements of K96, S9 and S19 (\citealt{kaspi1996,Metzroth2006} and Section~\ref{sec:lag_measurements}) in the following analysis. Adopting the technique \texttt{linmix}, we obtain the best-fitting regression line of
		\begin{equation}
			\log(R_{\rm H\beta}/\rm 1t-day)=(0.63\pm 0.13) + (0.46\pm 0.16)\log(\ell_{42})
		\end{equation}
		where $\ell_{42}=L_{5100}/10^{42}\, \rm erg\ s^{-1}$. We show the relationship between radius and 5100\,{\AA} luminosities in Figure~\ref{figure:rl_relationship}. The slope $0.46\pm 0.16$ is well consistent with the value of 0.5 expected from simple photoionization theory. 
		Compared to the traditional $R_{\rm H\beta}$-$L_{5100}$ relationship from \citet{bentz2013}, most measurements of NGC 4151 show slightly longer H$\beta$ lags than predicted, but they are consistent within a scatter of 0.3 dex. Also, the location of the mean value of time lags and luminosities is well consistent with the traditional $R_{\rm H\beta}$-$L_{5100}$ relationship. 
		
		\subsection{Evolution of BLR Kinematics: Radiation Pressure Driven?}
        \label{sec:kinematics}
        In several previous RM campaigns for NGC 4151, \citet{ulrich1996} and \cite{rosa2018} separately give velocity-resolved delays of \civ~ and H$\beta$ lines, both of them suggesting a virialized BLR with shorter lags for the high-velocity wings of the emission lines and the longer lags at the line centers. In the year 1991 data~\citep{ulrich1996}, the velocity-resolved delay of \civ~ line is asymmetric with a strong red wing showing $\tau\leqslant$ 2 days and a faint blue wing with $\tau\leqslant$ 10 days. This could be interpreted as a signal of inflow. While the velocity-resolved delay of H$\beta$ line in the 2012 data~\citep{rosa2018} demonstrates that the lags at the redshifted velocity-bins are longer than the ones at the blue. This suggests a signature of outflow~\citep{Denny2009}. Recently, \citet{Li2022} carried out RM observation using Lijiang for NGC 4151 at the same time as the MAHA campaign in S20. Their measurements of a velocity-resolved delay reveal that the BLR kinematics is a combined effect of both virial and inflow motions. In addition, \citet{Bentz2022} employed direct modeling of the H$\beta$-emitting BLR by using the data of B06. They found that the BLR kinematics prefers outflow.
        
        Our results of H$\beta$ velocity-resolved delays show two main kinds of structure in different seasons:
        
        1) $S3$, $S5$ and $S18$. The velocity-delay maps of the H$\beta$ line are measured in the data of years 1996, 1998 and 2018. They show the overall pattern of longer lags in the line centers and shorter lags in the wings, which are evidence for virial motion. For S3, the pattern displays lags at blue velocities longer than those at red velocities, which is similar to the velocity-resolved maps of the \civ~ line implying a virialized BLR with some potential contribution from inflow. Conversely, the velocity-resolved delays in S5 and S18 are the same as the ones measured by \cite{rosa2018} in 2012. The lags at blue velocities are shorter than those at red velocities, revealing the possibility of virial motion of BLR with a contribution from outflow.
        
        2) $S20$ and $S21$. The two velocity-resolved delays are measured from the data in 2021 and 2022. Both show a complex pattern of double peaks. For S20, the pattern, measured by WIRO, is similar to that measured by Lijiang \citep{Li2022}. The time lags at the range of velocity from $-$2000 to 0 $\rm km\ s^{-1}$ are longer than those at the velocity range 0 to 2000 $\rm km\ s^{-1}$. Considering that the velocity-resolved delays of H$\alpha$ and \hei~ lines measured by \citet{Li2022} show obviously longer lags for the blue side than the red side, we prefer interpreting the BLR kinematics as viral motion with a weak contribution from inflow. For the velocity-resolved time lag structure in S21, the blue and red side peaks are symmetric, while the lags at the blue wings are shorter than the ones at the red wings. This provides strong evidence that the BLR kinematics are outflow.
        
        In summary, the kinematics of BLR for NGC 4151 are complex and evolve over time. Considering the long-term photometric light curve (Figure~\ref{figure:lc_all}), we note that the BLR kinematics tends to virial and inflow motions during the AGN luminosity rising phase, and tends to virial and outflow motions when falling (we demonstrate the relationship between the variation of the light curve and the BLR kinematics in Figure~\ref{figure:lc_all}). Therefore, for NGC 4151, we propose that the rise of luminosity could be the result of an increasing accretion rate caused by the inflowing gas from BLR. After the luminosity reaches the higher state, the gravitational force is dominated by the radiation pressure, resulting outflow. Then the luminosity drops again. Clearly, more RM monitoring campaigns are needed to further verify this tentative result.
        \begin{figure*}
            \centering
            \includegraphics[width=0.8\textwidth]{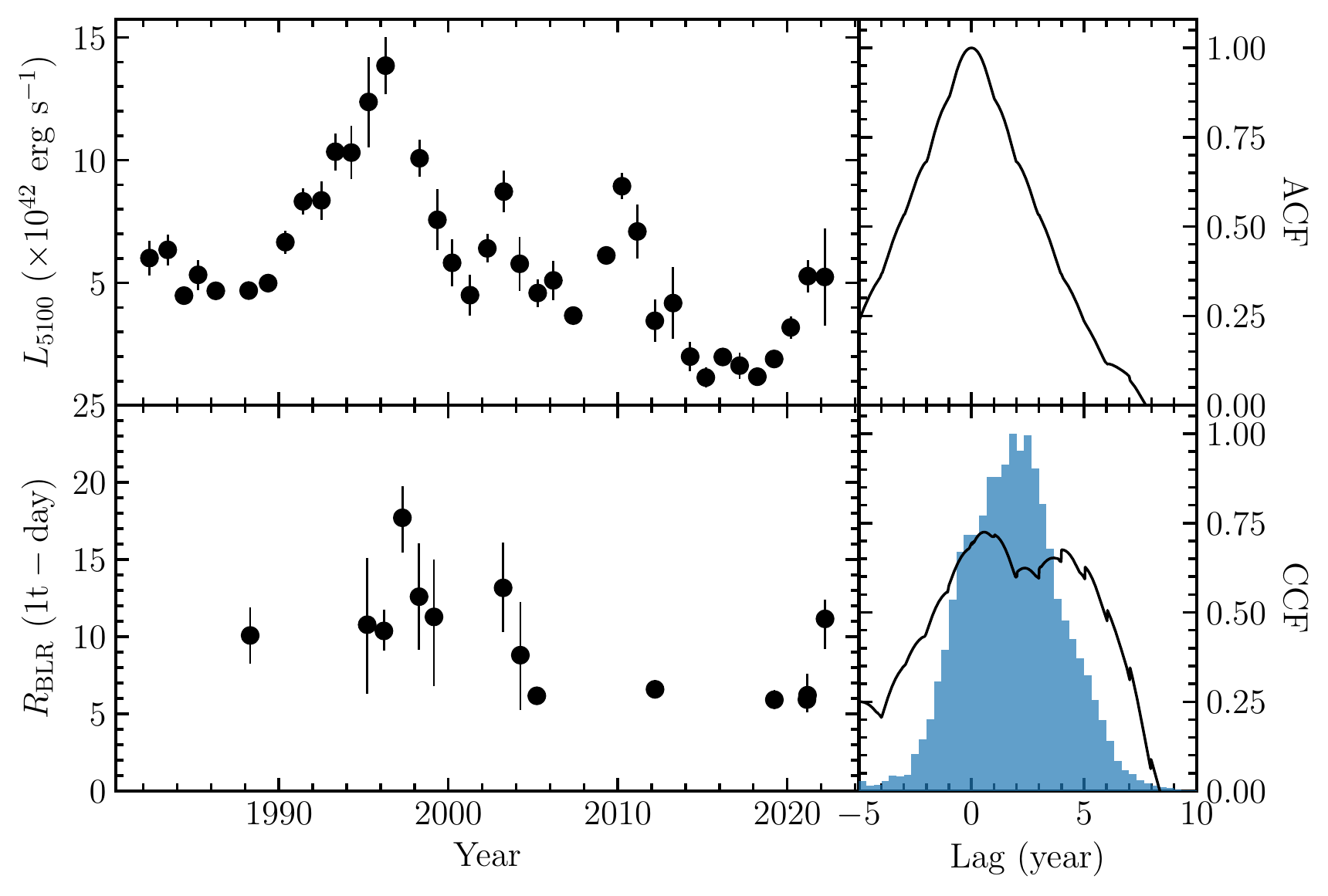}
            \caption{(Left) the variation over time of optical luminosity $L_{5100}$ and the BLR size $R_{\rm BLR}$. (Right) result of the correlation analysis. The top panel shows the ACF of luminosities. The bottom panel displays the ICCF of $R_{\rm BLR}$ with respect to $L_{5100}$. The blue histogram is the distribution of Monte Carlo simulations of the centroid.}
            \label{figure:lag_LR}
        \end{figure*}
	\subsection{Another Possible Evidence for Radiation Pressure}
        Using the previous 23 measurements for NGC 5548, \cite{lu2016,Lu2022} found that there is a $3.5_{-1.8}^{+2.0}$ years lag of $R_{\rm BLR}$ with respect to $L_{5100}$ (averaged luminosity within one individual season), which is consistent with the dynamical timescale of the BLR ($t_{\rm BLR}\approx2.1$ years). This suggests the possibility of the BLR being jointly controlled by the radiation pressure and SMBH gravity. For NGC 4151, as listed in Table~\ref{table:summary}, there is a total of 14 H$\beta$ RM campaigns, excluding three unreliable measurements. This allows us to examine whether there is a delay between $L_{5100}$ and $R_{\rm BLR}$. Figure~\ref{figure:lag_LR} shows the variation of $L_{5100}$ and $R_{\rm BLR}$, and the time lag between them ($\tau_{RL}$) is calculated by using ICCF. We obtain $\tau_{RL} = 1.86_{-2.24}^{+2.17}$ years with $r_{\rm max}=0.73$, which is in a good agreement with the dynamical timescale of the BLR determined by %
        \begin{equation}
            t_{\rm BLR} = \frac{c\tau_{\rm H\beta}}{V_{\rm FHWM}}=1.4\ \tau_{10}\ V^{-1}_{6000} \rm\ \text{years}
        \end{equation}
        where $V_{6000} = V_{\rm FWHM}/6000\ \rm km\ s^{-1}$ and and $\tau_{10}=\tau_{\rm H\beta}/10\ \rm days$.
        Here we take the typical values of $V_{\rm FWHM}$ and $\tau_{\rm H\beta}$.
        
        The result of $t_{\rm BLR}\approx \tau_{RL}$ is the same as that of NGC 5548, indicating that the BLR kinematics are perhaps the combined effect of radiation pressure and BH gravity. Luminosity variations lead to changes in radiation pressure acting on BLR clouds. It is thus expected to delay changes of the BLR with $\tau_{RL}$. The variations of luminosity and evolution of BLR kinematics are basically consistent with this scheme as we discuss in Section~\ref{sec:kinematics}. This needs more RM monitoring campaigns to verify.
		
		\section{Summary}
		\label{sec:summary}
        We present long-term RM measurements of NGC 4151 by using the spectra from the sixth project in the FAST Spectrograph Publicly Archive Programs (AGN Watch) and the MAHA program, and the main results are summarized as follows.
		\begin{enumerate}
			\item During the FAST projects and MAHA campaign, there are 21 seasons of spectroscopic data since 1994, of which reliable time lag measurements are obtained for 10 seasons. Meanwhile, we remeasured the lag in previously published data in a uniform way. Combining all of these measurements, we examine the relation between the time lag and line width and find a relationship that is well consistent with the expected $\Delta V=\tau^{-0.5}$.
			
			\item  Adopting the virial factor $f=6.3\pm1.5$ and $\sigma_{\rm line}$ from rms spectra as the line widths, we obtain a mean BH mass of ${M}_{\bullet}=(5.1\pm2.3) \times 10^7M_{\odot}$. This value is consistent with that measured by the stellar and gas dynamical modeling.
			
			\item Combining the calibrated photometric light curve and the starlight estimation using \texttt{GALFIT}, we conduct an absolute flux calibration for the fluxes in the AGN continuum both for this work and previous RM campaigns. We obtain $R_{\rm H\beta}\propto L_{5100}^{0.46\pm0.16}$, which follows simple photoionization theory $R_{\rm H\beta}\propto L_{5100}^{0.5}$ and is consistent with \citet{bentz2013}'s global $R_{\rm H\beta}$-$L_{5100}$ relationship within the dispersion of 0.3 dex.
			
			\item The high quality of spectroscopic data in S3, S5, S18, S20 and S21 allows us to recover their velocity-resolved time delays of the H$\beta$ broad line. With the previous measurements of \citet{ulrich1996}, \citet{rosa2018}, \citet{Li2022} and \citet{Bentz2022}, we note that BLR kinematics are tentatively related to the variation of the luminosity. The velocity-resolved delays show virial motion with the contribution from inflow/outflow in corresponding to the rising/declining phase of luminosity, indicating potential evidence of radiation pressure.
			
			\item We detected a delay of BLR secular changes ($\sim 1.86$ yrs) perhaps driven by radiation pressure. This is similar to that of NGC 5548. Details of modeling BLR secular evolution are necessary to include radiation pressure.
			
		\end{enumerate}
		
		NGC4151 is one of the AGNs with the largest number of spectroscopic observations. Combining the previous measurements, we present the possible relationship between dynamic evolution and luminosity in the BLR. Therefore, it is worthwhile to continuously monitor NGC 4151. In future works, we will model the BLR dynamics by utilizing \texttt{BRAINS} (\citealt{Li2013,Li2018}). Moreover, the profiles of the H$\beta$ broad line in different seasons significantly changed. We will explore the possible physical mechanism for the variation of the H$\beta$ line profile in Paper II.
		
		\section*{acknowledgements}
		This research is supported in part by the National Key R\&D Program of China (2021YFA1600404); by grant Nos. NSFC- 11833008, NSFC-11991051, and NSFC-11991054 from the National Natural Science Foundation of China; and by the China Manned Space Project with No. CMS-CSST-2021-A06 and CMS-CSST-2021-B11. 
		Y.-R.L. acknowledges financial support from the National Natural Science Foundation of China through grant Nos. 11922304 and 12273041 and from the Youth Innovation Promotion Association CAS.
		C.H. acknowledges support from the National Science Foundation of China (12122305). P.D. acknowledges financial support from NSFC through grant Nos. 12022301 and 11873048.
		
		Based on observations obtained with the Samuel Oschin Telescope 48-inch and the 60-inch Telescope at the Palomar Observatory as part of the Zwicky Transient Facility project. ZTF is supported by the National Science Foundation under Grants No. AST-1440341 and AST-2034437 and a collaboration including current partners Caltech, IPAC, the Weizmann Institute for Science, the Oskar Klein Center at Stockholm University, the University of Maryland, Deutsches Elektronen-Synchrotron and Humboldt University, the TANGO Consortium of Taiwan, the University of Wisconsin at Milwaukee, Trinity College Dublin, Lawrence Livermore National Laboratories, IN2P3, University of Warwick, Ruhr University Bochum, Northwestern University and former partners the University of Washington, Los Alamos National Laboratories, and Lawrence Berkeley National Laboratories. Operations are conducted by COO, IPAC, and UW. EB acknowledge the support of Serbian Ministry of Education, Science and Technological Development, through the contract number 451-03-68/2022-14/200002.

        We thank WIRO engineers James Weger, Conrad Vogel, and Andrew Hudson for their indispensable and invaluable assistance. 
        M.S. Brotherton enjoyed support from the Chinese Academy of Sciences Presidents International Fellowship Initiative, grant No. 2018VMA0005. 
        This work is supported by the National Science Foundation under REU grant AST 1852289.
        T.E. Zastrocky acknowledges support from NSF grant 1005444I.

        \section*{Data Availability}
	The ZTF photometric data are publicly available from IRSA at \url{https://irsa.ipac.caltech.edu/cgi-bin/Gator/nph-dd}. The ASAS-SN photometry could be obtained at \url{https://asas-sn.osu.edu}.
     \bibliographystyle{mnras}      
		\bibliography{reference}  

        \appendix
        \section{Log of the spectroscopic observations}
        \label{appendixa}
        In Table~\ref{table:log}, we summarize the observation logs for the spectroscopic data from FAST and MAHA programs.

        \begin{table*}
        \renewcommand{\arraystretch}{1.3}
        \setlength{\tabcolsep}{8.5pt}
        \caption{Log of the spectroscopic observations. The full version is available in a machine-readable form online.}
        \label{table:log}
        \begin{tabular}{cccccccccccc}
        \hline
        Date & JD & Exposure Time & Aperture & Position Angle & Seeing & $R([\text{\oiii}])$$^a$ & $F_{\rm gal}$ & \\
        (yyyy/mm/dd)& (-2,400,000) & (s) & (arcsec) & (deg) & (arcsec) & & ($10^{-14}\ \rm erg\ s^{-1}\ cm^{-2}\ \text{\AA}^{-1}$)\\
        \hline
        1994/Jan/16 & 49368.988 & 240.0 & 3.0 $\times$ 12.0 & 90 & 1-2 & 2.97 & 1.17 $\pm$ 0.07\\
        1994/Jan/17 & 49370.022 & 180.0 & 3.0 $\times$ 12.0 & 90 & 1-2 & 3.07 & 1.17 $\pm$ 0.07\\
        1994/Jan/17 & 49370.025 & 180.0 & 3.0 $\times$ 11.7 & 90 & 1-2 & 3.10 & 1.17 $\pm$ 0.07\\
        1994/Jan/18 & 49371.034 & 120.0 & 3.0 $\times$ 11.0 & 90 & 1-2 & 3.11 & 1.15 $\pm$ 0.07\\
        1994/Jan/19 & 49372.037 & 180.0 & 3.0 $\times$ 18.0 & 90 & 1-2 & 3.05 & 1.26 $\pm$ 0.08\\
        1994/Jan/22 & 49374.949 & 300.0 & 3.0 $\times$ 19.0 & 90 & 1-2 & 3.07 & 1.27 $\pm$ 0.08\\
        1994/Jan/22 & 49374.953 & 300.0 & 3.0 $\times$ 17.0 & 90 & 1-2 & 3.08 & 1.25 $\pm$ 0.08\\
        1994/Feb/15 & 49398.949 & 180.0 & 3.0 $\times$ 7.8 & 90 & 2 & 3.08 & 1.05 $\pm$ 0.07\\
        1994/Mar/05 & 49416.920 & 137.3 & 3.0 $\times$ 8.0 & 90 & 1-2 & 3.05 & 1.05 $\pm$ 0.07\\
        1994/Mar/06 & 49417.843 & 300.0 & 3.0 $\times$ 9.0 & 90 & 1-2 & 2.74 & 1.09 $\pm$ 0.07\\
        1994/Mar/06 & 49417.846 & 120.0 & 3.0 $\times$ 7.0 & 90 & 1-2 & 3.03 & 1.01 $\pm$ 0.06\\
        1994/Mar/06 & 49417.848 & 30.0 & 3.0 $\times$ 7.0 & 90 & 1-2 & 3.05 & 1.01 $\pm$ 0.06\\
        1994/Mar/07 & 49418.810 & 30.0 & 3.0 $\times$ 7.0 & 90 & 1-2 & 3.04 & 1.01 $\pm$ 0.06\\
        1994/Mar/21 & 49432.902 & 30.0 & 3.0 $\times$ 10.0 & 90 & 2-3 & 3.01 & 1.12 $\pm$ 0.07\\
        1994/Apr/01 & 49443.809 & 30.0 & 3.0 $\times$ 7.0 & 90 & 2-3 & 3.02 & 1.01 $\pm$ 0.06\\
        1994/Apr/02 & 49444.817 & 30.0 & 3.0 $\times$ 6.4 & 90 & 2 & 3.03 & 0.98 $\pm$ 0.06\\
        1994/Apr/03 & 49445.850 & 30.0 & 3.0 $\times$ 5.6 & 90 & 2 & 3.04 & 0.93 $\pm$ 0.06\\
        1994/Apr/05 & 49447.881 & 300.0 & 3.0 $\times$ 9.6 & 90 & 2-3 & 2.87 & 1.11 $\pm$ 0.07\\
        \hline
        \end{tabular}
        \begin{list}{}{}
        \item[$^a$] {$R( \text{[\oiii]})=F({\text{[\oiii]}\lambda5007})/F({\text{[\oiii]}\lambda4959})$.}
        \end{list}
        \end{table*}
        
        \section{[\oiii] Fluxes Measured in Different Spectral Extracted Aperture Sizes}
	\label{appendixb}
\begin{figure}
            \centering
            \includegraphics[width=0.45\textwidth]{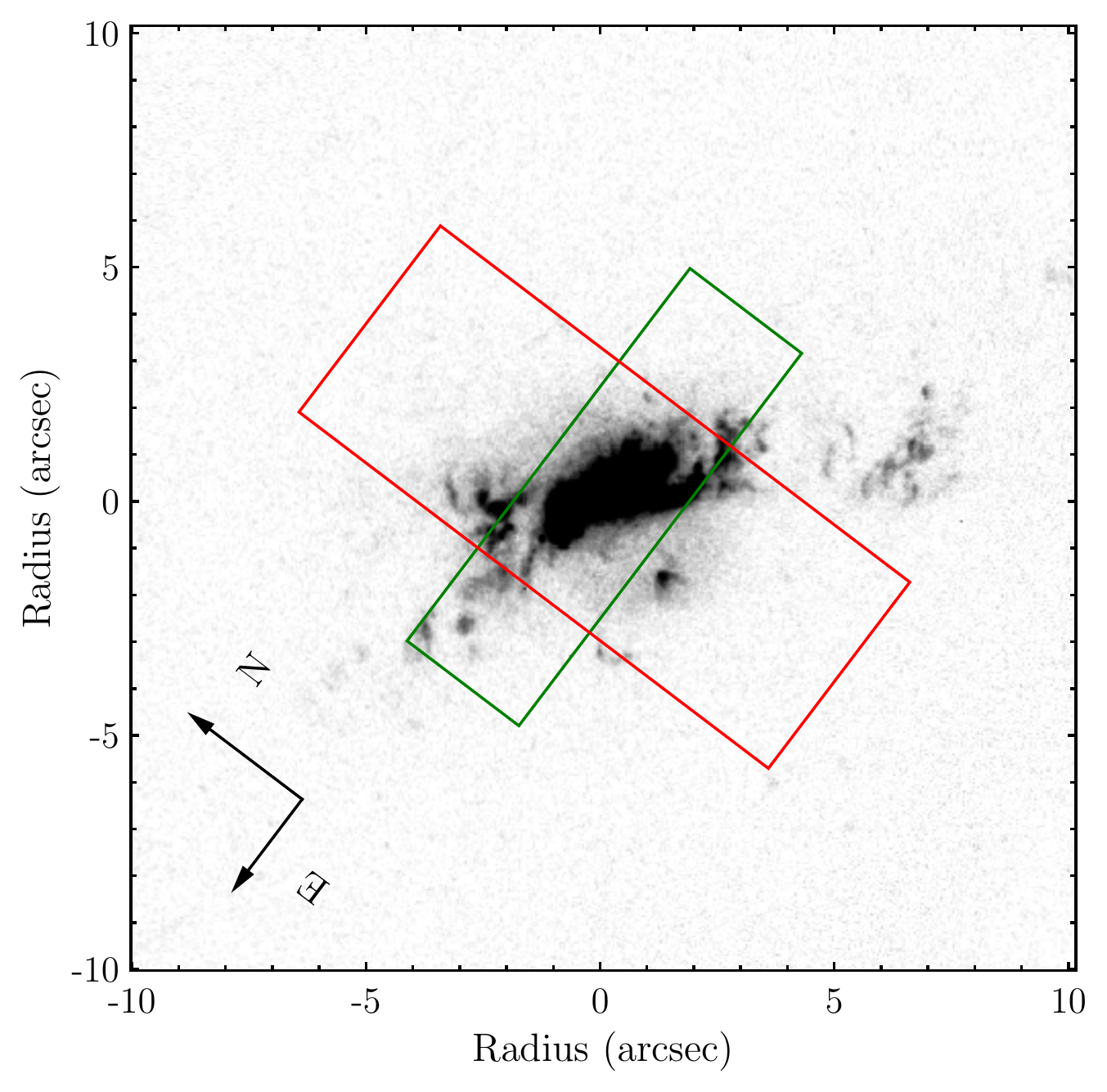}
            \caption{The continuum-subtracted [\oiii] image of NGC 4151. The green rectangle is the aperture of $3^{\prime\prime}\times 10^{\prime\prime}$ with a position angle of $90^{\circ}$. The red rectangle is the aperture of $5^{\prime\prime}\times 12^{\prime\prime}.6$ (WIRO, position angle is $0^{\circ}$).}
            \label{o3imag}
        \end{figure}

	For testing the influence of different spectral extraction apertures and position angles on the narrow lines flux, we retrieved [\oiii]$\lambda$5007 and continuum images from the HST archive. These images were observed with WFPC2 PC camera on HST. The [\oiii] and continuum images were separately observed with filters F502N and F547M, and their exposure times of them are 10s. After eliminating cosmic rays and correcting the background, we subtract the continuum image from the line image. The continuum-subtracted [\oiii] image is shown in Figure~\ref{o3imag}.
		
	In this work, the spectra with the extracting aperture $3\arcsec\times 10\arcsec$ were taken as the standard. We measure the [\oiii] emission from the continuum-subtracted [\oiii] image by integrating the flux within the apertures. The apertures and the position angles for the spectra from the FAST observing program are listed in Appendix~\ref{appendixa}. We show the result in Figure~\ref{oiiitesting}. When the sizes of spectral aperture are larger than $3\arcsec$, $F_{\rm 3^{\prime\prime}\times (Extracted\ Size)}/F_{3^{\prime\prime}\times 10^{\prime\prime}}$ of most spectra are more than 95\%. This indicates that the relative flux calibration described for the spectra from the FAST observing program in Section~\ref{sec:calibration} is not affected by the slit loss for more than 5\% after we discarded the spectra with extracting aperture less than $3\arcsec$. However, due to the existence of other effects, e.g. seeing, we do not correct the slit loss according to this simple test.

        \begin{figure}
            \centering
            \includegraphics[width=0.45\textwidth]{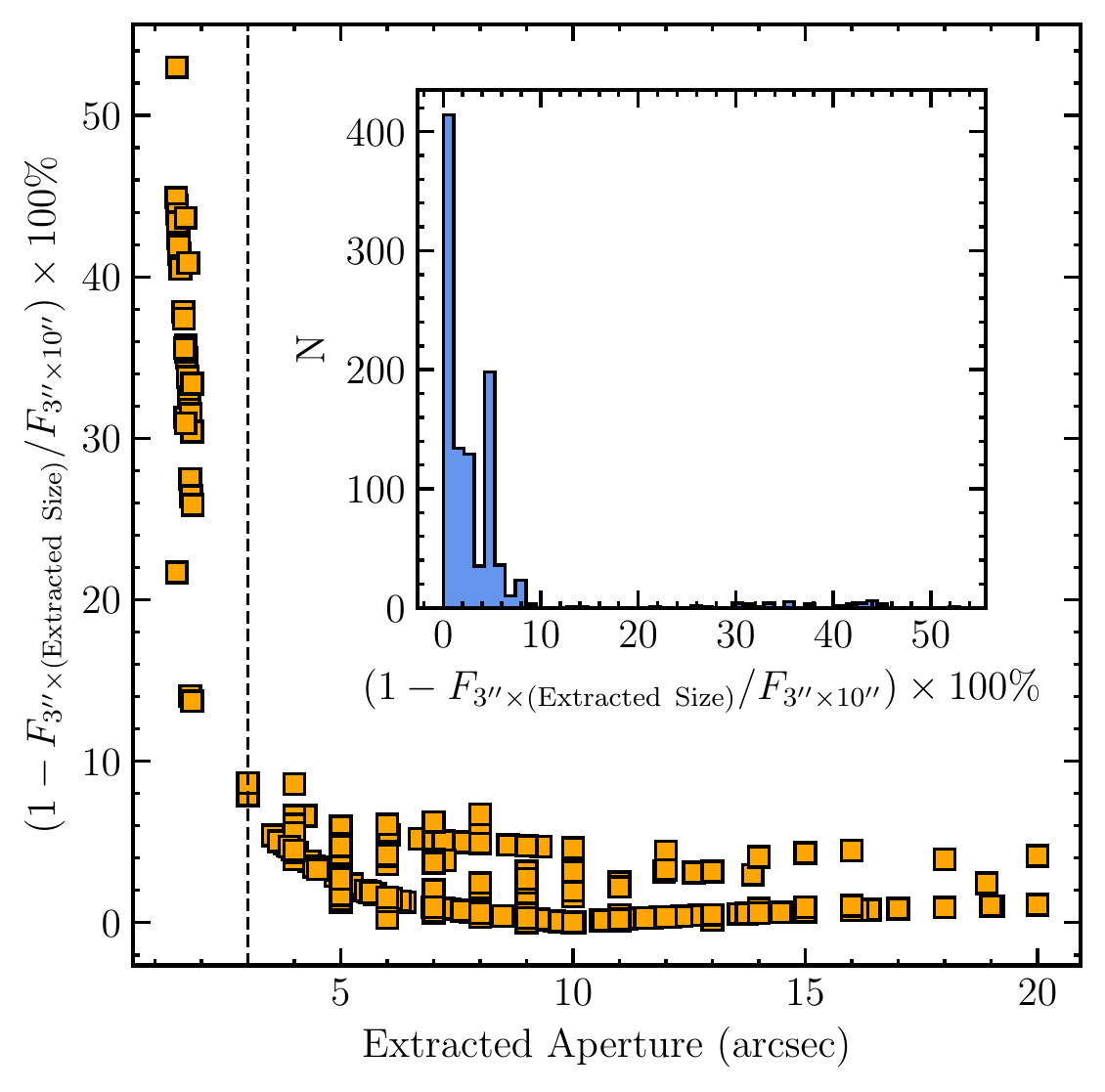}
            \caption{The spectral extracted aperture size vs. $1 - F_{\rm 3^{\prime\prime}\times (Extracted\ Size))}/F_{3^{\prime\prime}\times 10^{\prime\prime}}$. The inset plot shows the distribution of $1 - F_{\rm 3^{\prime\prime}\times (Extracted\ Size)}/F_{3^{\prime\prime}\times 10^{\prime\prime}}$. The vertical dashed line is the position of extracted aperture $3\arcsec$. }
            \label{oiiitesting}
		\end{figure}

    \section{S/N-weighted rms spectra}
    \label{appendixc}
    
\begin{figure*}
        \centering
        \includegraphics[width=0.95\textwidth]{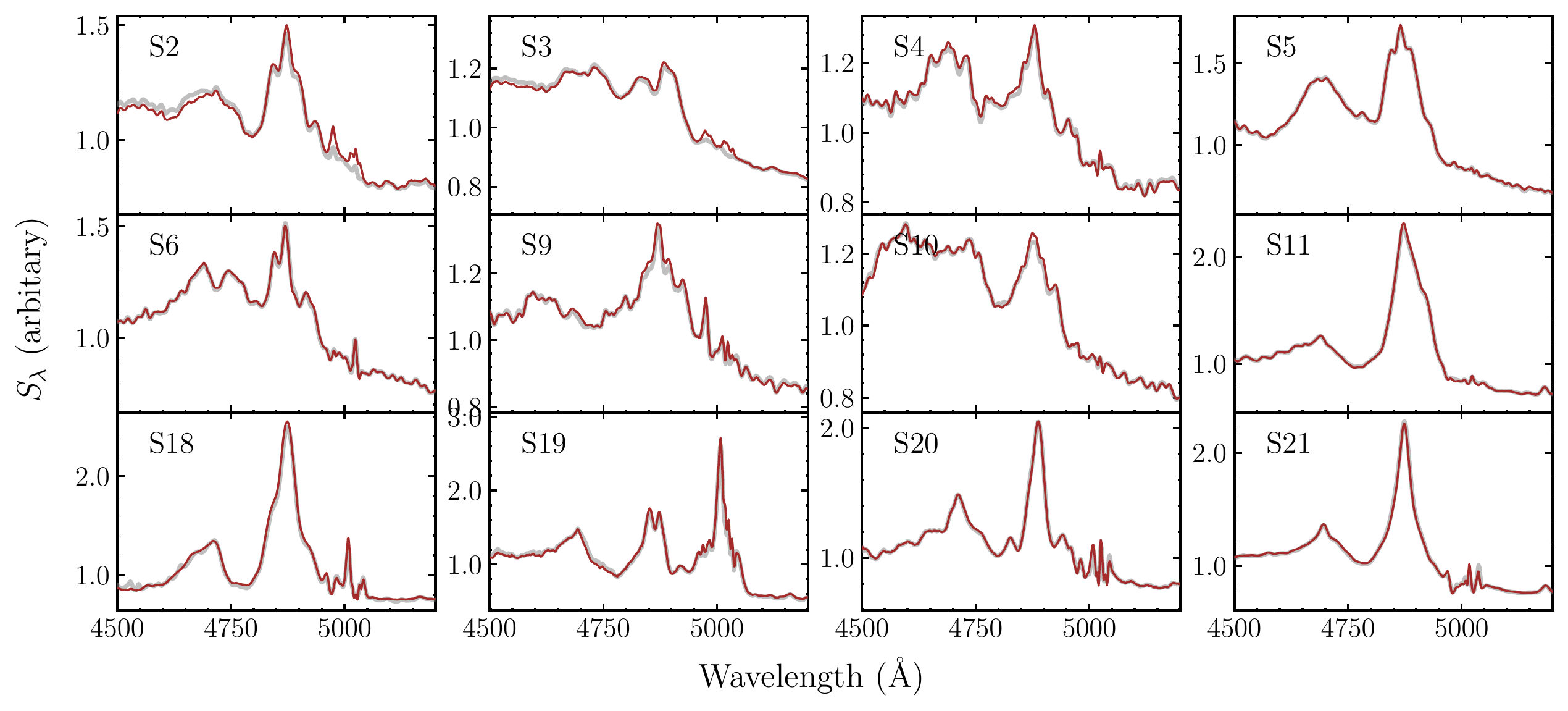}
        \caption{A comparison between the uniformly weighted (gray) and S/N-weighted (brown) rms spectra in the observed frame for the seasons used in RM analysis.}
        \label{fig:SN_rms}
    \end{figure*}

    In order to examine the influences of data errors on the rms spectra and line width measurements, we calculate the S/N-weighted rms spectrum as \citep{Park2012b, Barth2015}
    \begin{equation}
        S_{\lambda}^w = \left[\frac{1}{1-\sum_{i=1}^{N}w^2_i}\sum_{i=1}^{N}w_i\left(F_{\lambda}^{i}-\Bar{F}^w_\lambda\right)^2\right]^{1/2},
    \end{equation}
    where $w_i$ is the normalized S/N weight defined by
    \begin{equation}
        w_i = \frac{S_i/N_i}{\sum_{i=1}^{N}(S_i/N_i)},
    \end{equation}
    and $\Bar{F}^w_\lambda$ is the weighted mean spectrum defined by
    \begin{equation} 
    \Bar{F}^w_\lambda = \sum_{i=1}^{N}w_iF_{\lambda}^{i}.
    \end{equation} 
    Figure~\ref{fig:SN_rms} shows a comparsion between the uniformly weighted rms spectra defined in Equation~(\ref{eqn_rms}) and the S/N-weighted rms spectra for the seasons used for our RM analysis. There are no significant differences between the two types of rms spectra. We measured the line widths from the S/N-weighted rms spectra using the same procedure in Section~\ref{sec:width}. We find that the line width from the two types of rms spectra are well consistent within 1\% with each other. 

    \section{A Note on seasons S9 and S19}
    \label{appedixd}
      
    The seasons S9 and S19 show very low correlation coefficients between the 5100\,{\AA} continuum and H$\beta$ light curves ($r_{\rm max}\leqslant0.5$). There are several potential reasons for the low $r_{\rm max}$. (1) Poor sampling. The ICCF method linearly interpolates the light curves. When the sampling is uneven and very sparse, the interpolated light curves might be different from the true ones, leading to a poor correlation between the light curves; (2) There appear different long-term trends of variations in the 5100\,{\AA} continuum and H$\beta$ light curves. This phenomenon has been found in some previous RM campaigns (e.g., \citealt{grier2012,Fausnaugh2017, rosa2018}); (3) The H$\beta$ variations are decorrelated from the continuum variations like the ``BLR holiday'' phenomenon found in NGC 5548 (e.g., \citealt{Goad2016,pei2017}). All these three reasons may bias the time lag measurement.
    
\begin{figure}
        \centering
        \includegraphics[width=0.50\textwidth]{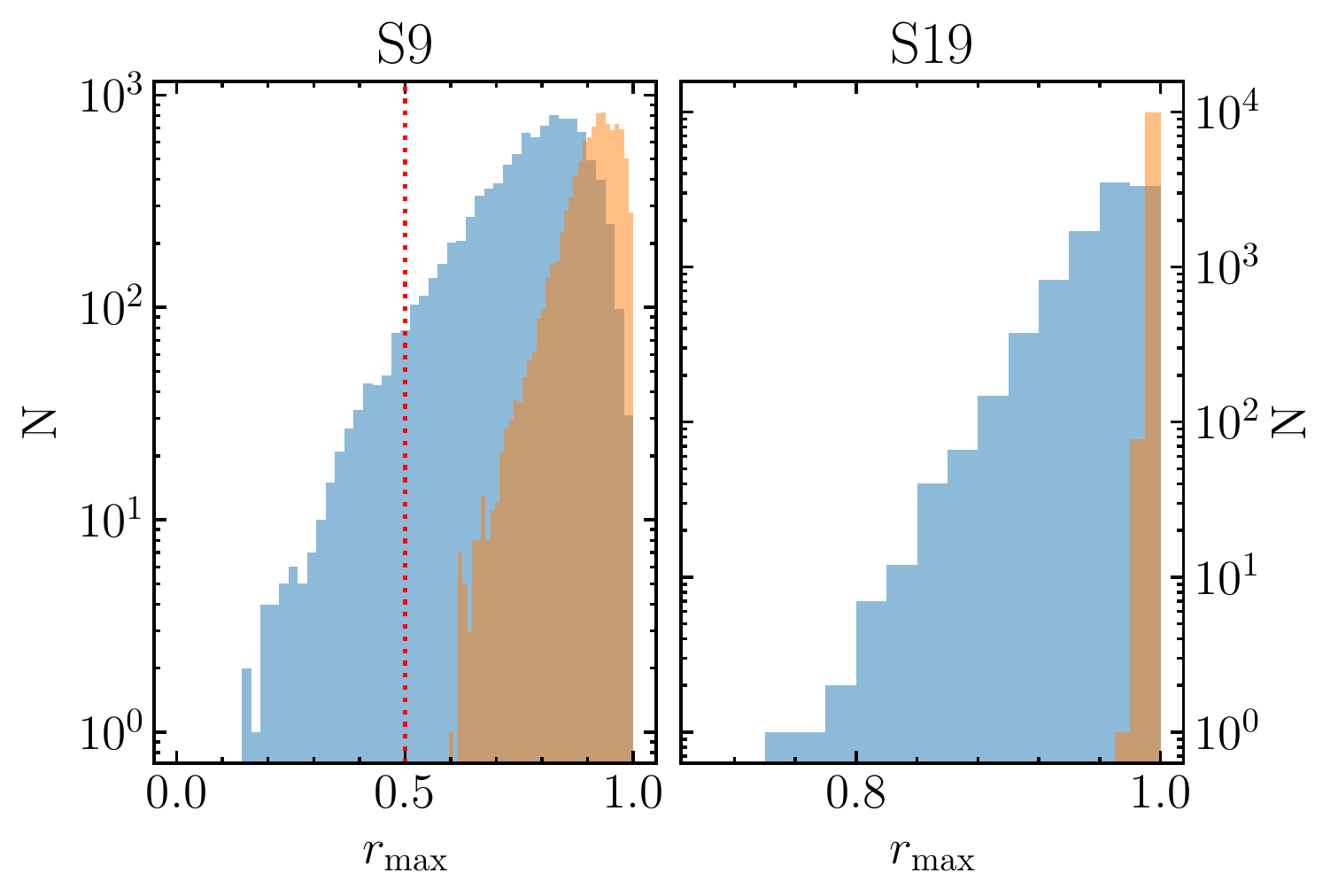}
        \caption{The $r_{\rm max}$ distribution of mock light curves for S9 and S19. The orange and blue histograms are for daily and observed sampling, respectively. The dotted line in the left panel marks the location of $r_{\rm max}=0.5$.}
        \label{fig:rmax_Simulation}
    \end{figure}

    We first test the influences of the sampling for the seasons S9 and S19 through simulations. We adopt the same procedure as \cite{lishasha2021}.  
    The parameters of the DRW process i.e., $\tau_d$ (the damped timescale) and $\sigma_d$ (the variability amplitude) as well as the center and width of the Gaussian transfer function are randomly chosen from their respective posterior samples obtained by MICA. We first generated mock light curves with a daily cadence and then linearly interpolated them onto the observed epochs. We use the same procedure as for the observed data to calculate $r_{\rm max}$ of the mock light curves. 
    In Figure~\ref{fig:rmax_Simulation}, we show the distributions of $r_{\rm max}$ and make a comparison between the daily and observed sampling intervals. 
    As can be seen, the observed sampling only mildly reduces the correlations.
    For S9 with the case of observed sampling, there is only a probability  of about 7\% for $r_{\rm max}<0.5$. For S19,  both the two types of sampling have $r_{\rm max}>0.7$. Those results indicate that the influence of data sampling is insignificant.
    
\begin{figure*}
        \centering
        \includegraphics[width=0.85\textwidth]{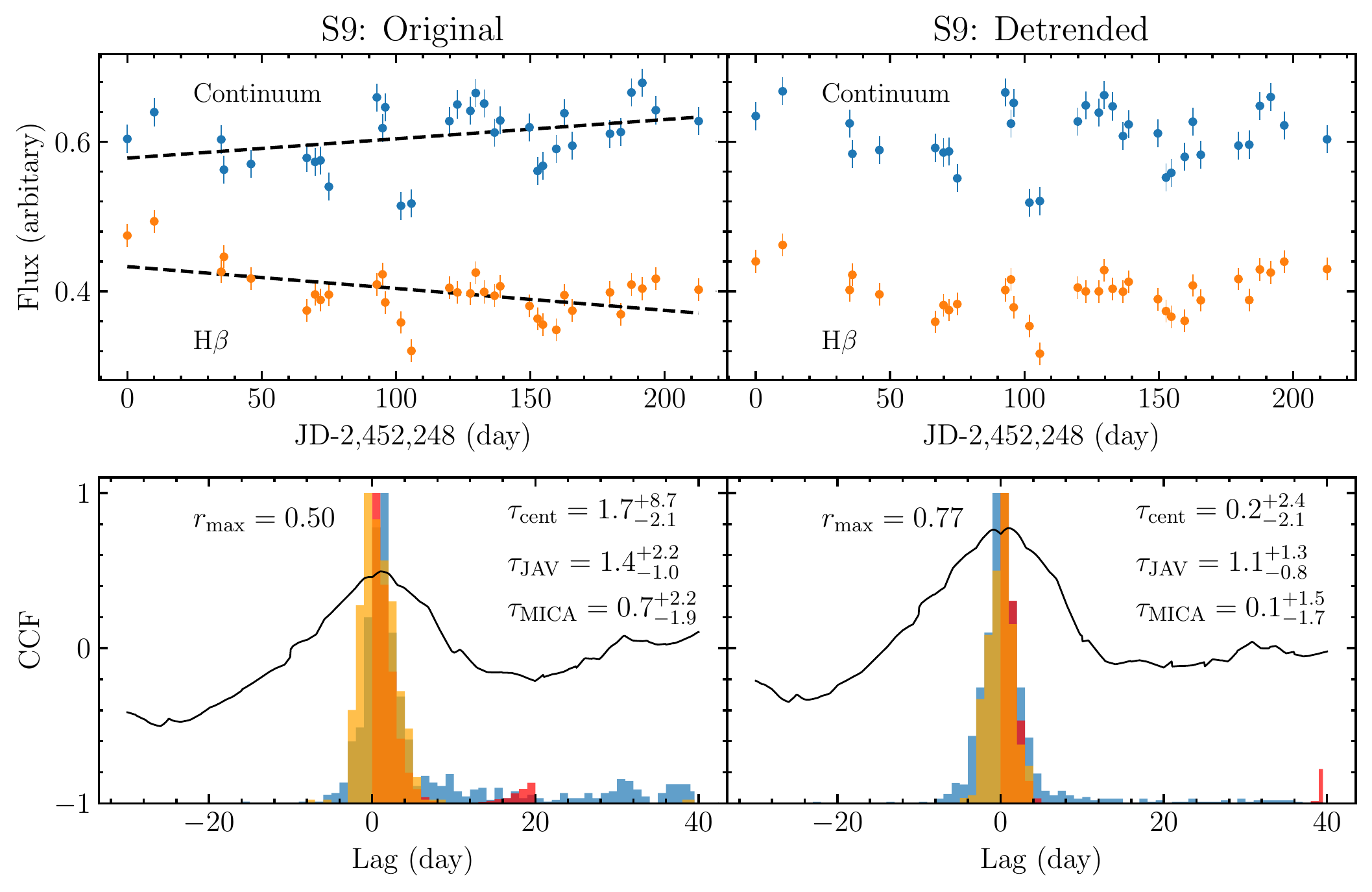}
        \caption{(Upper) the original (left) and detrended (right) light curves in S9. The dashed lines are the long-term linear trends. (Bottom) the CCFs between continuum and H$\beta$ light curves. The blue histogram is the cross-correlation centroid distribution calculated using the ICCF. The red and orange histograms show the posterior distributions of time lags from JAVELIN and MICA, respectively.}
        \label{fig:S9_Detrended}
    \end{figure*}

\begin{figure*}
        \centering
        \includegraphics[width=0.85\textwidth]{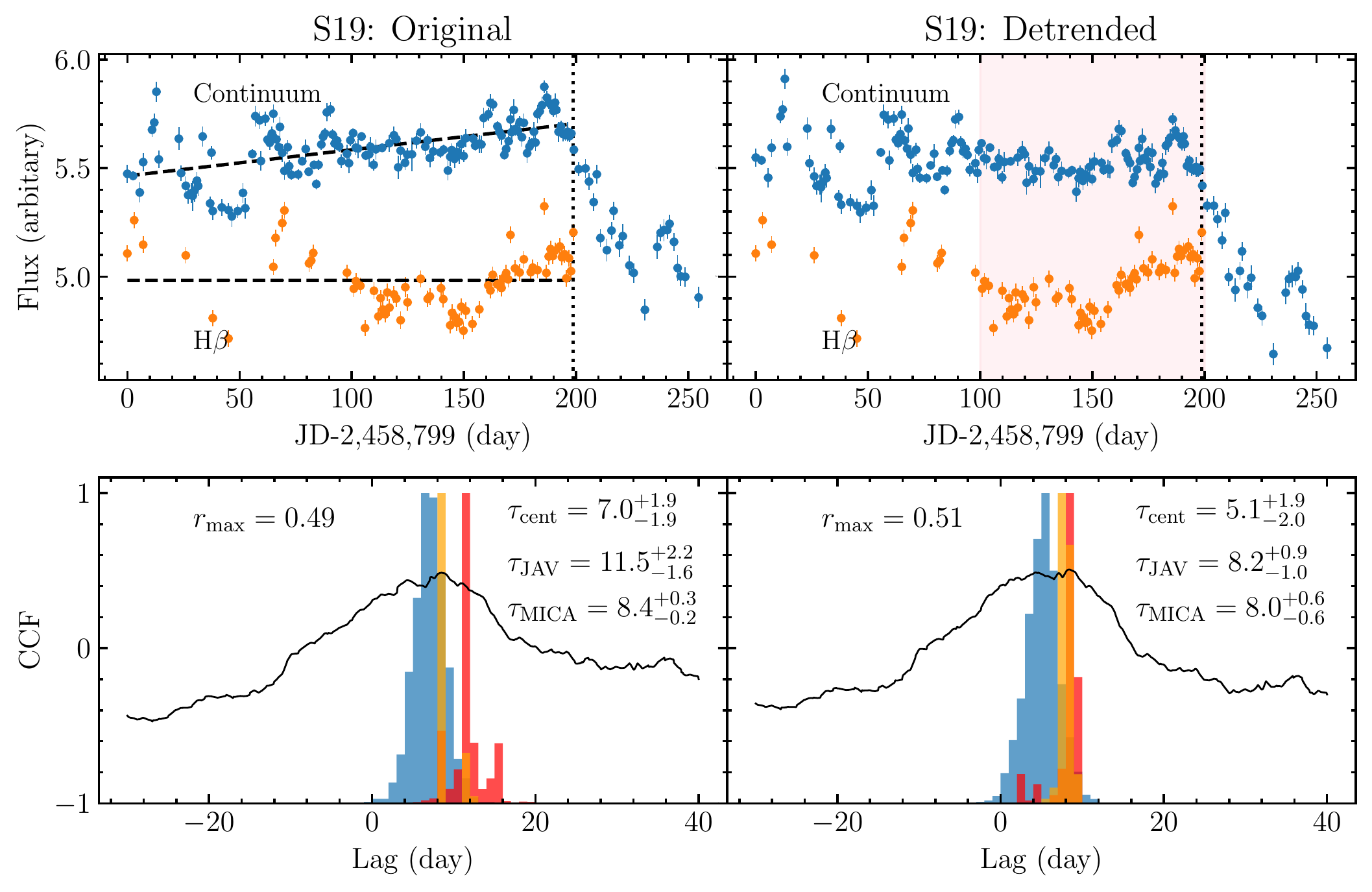}
        \caption{Same as Figure~\ref{fig:S9_Detrended} but for S19. The dotted lines show the linear fits to the long-term linear trends of the light curves for the beginning 200 days. The period of ``BLR holiday'' is marked with the pink band in the upper right panel.}
        \label{fig:S19_Detrended}
    \end{figure*}

    Next, we simply fit the continuum  and H$\beta$ light curves with straight lines to examine whether their long-term trends are different. The results are shown in the left panels of Figure~\ref{fig:S9_Detrended} and~\ref{fig:S19_Detrended}. In both S9 and S19, the long-term trends of the  continuum and H$\beta$ variations are different from each other. We subtract the best-fit trends and repeat the RM analysis in Section~\ref{sec:lag_measurements}. For S9, the correlation after detrending significantly improves, with $r_{\rm max}$ increasing from 0.50 to 0.77. This indicates that the different long-term trends might be responsible for the low $r_{\rm max}$ in S9.
    For S19, the $r_{\rm max}$ measured from the original ($r_{\rm max}=0.49$) and detrended ($r_{\rm max}=0.51$) light curves have no significant differences, implying that the simple linear long-term trends do not cause the low correlation. However, we note that there seems to appear the behavior of ``BLR holiday'' in the late part of the light curves in S19 (see the pink region in the right panel of Figure~\ref{fig:S19_Detrended}). We therefore tentatively ascribe the low correlation in S19 to the probable ``BLR holiday'' behavior.
    
    \end{document}